\documentclass[12pt]{article}
\pdfoutput=1
\usepackage{jheppub}
\usepackage{verbatim}

\newcommand{\be}{\begin{equation}}
\newcommand{\ee}{\end{equation}}

\newcommand{\HA}{\mathcal{H}_A}
\newcommand{\HB}{\mathcal{H}_B}
\newcommand{\HR}{\mathcal{H}_R}

\newcommand{\lan}{\langle}
\newcommand{\ran}{\rangle}
\newcommand{\tr}{\mathrm{tr}}

\newcommand{\mO}{\mathcal{O}}
\newcommand{\tO}{\widetilde{\mathcal{O}}}
\newcommand{\wt}{\widetilde}
\begin{document}
\title{Jerusalem Lectures on Black Holes and Quantum Information}
\author[a]{Daniel Harlow}
\affiliation[a]{Princeton Center for Theoretical Science, Princeton University, Princeton NJ 08540 USA}

\emailAdd{dharlow@princeton.edu}
\abstract{In these lectures I give an introduction to the quantum physics of black holes, including recent developments based on quantum information theory such as the firewall paradox and its various cousins.  I also give an introduction to holography and the AdS/CFT correspondence, focusing on those aspects which are relevant for the black hole information problem.}
\maketitle
\section{Introduction}
Black holes are fascinating objects in quantum gravity.  Starting from fairly mundane initial conditions (such as a collapsing star), nature is able to produce a geometry that amplifies short-distance fluctuations to macroscopic sizes.  This ``stretching'' of spacetime circumvents the Wilsonian decoupling of high energy physics from low energy physics, making deep questions of Planck-scale dynamics  relevant for low-energy (thought) experiments.\footnote{This amplification is also present in an expanding universe, and it seems likely that a complete understanding of black hole physics will lead to valuable lessons for quantum cosmology.}  Indeed in an extraordinary pair of classic papers \cite{Hawking:1974sw,Hawking:1976ra}, Stephen Hawking argued first that this stretching of fluctuations causes black holes to evaporate and second that the evaporation process is inconsistent with the quantum mechanical principle that pure states always evolve to other pure states.  This conclusion is usually called the \textit{Black Hole Information Problem}, and it has instigated a huge amount of research in the almost 40 years since Hawking's papers.  Is information indeed lost?  If not, then what is the nature of the Planckian interference that prevents it?  Significant progress has been made on these questions, but recent work has emphasized the extent to which we still do not have satisfactory answers to them.  

The goal of these lectures is first to give an introduction to as much as is reasonable of the techniques that go into formulating and analyzing these questions, and second to give an overview of the new paradoxes that have led to an explosion of recent work on the subject.  I will also discuss some ideas that have been proposed to resolve the paradoxes, but I will by no means aim at a comprehensive review; I have throughout done my best to prioritize pedagogy over completeness.  Of course in a field as chaotic as this one currently is, my views on which material should be included will be somewhat idiosyncratic.  As a general rule I have attempted to give, or at least sketch, the ``real'' arguments for things.  When the foundations of the subject are under as much doubt as they are here, it is my view that sloppy logic should be avoided as much as possible.

Occasionally some details of the material are new, but I will not try to call attention to this since it would awkward and tedious, and in any event my ``improvements'' are mostly cosmetic.  

Not all sections of the notes are equally important in getting to the paradoxes of section \ref{paradoxsec}.  Sections \ref{clbhsec} and \ref{qftentsec} are essential, as are sections \ref{2siderindsec}-\ref{infpsec}.  From there things get more flexible, sections \ref{bricksec} and \ref{eucblacksec} can be skipped on a first reading, as can sections \ref{testUsec}-\ref{scrambsec}.  When the lectures were actually presented I skipped all of section \ref{adssec}, although I wouldn't necessarily recommend that to the reader.

Finally a comment on the target audience for these notes: the lectures were given at the 31st winter school in theoretical physics at Hebrew University in Jerusalem to a diverse audience, including condensed matter, quantum information, and high-energy theorists.  The situation might be described by saying that the union of their background knowledge was maximal but the intersection was empty.  On behalf of the first two groups I have fairly extensively reviewed rather standard facts about general relativity, black holes, and AdS/CFT.  On behalf of the first and third groups I have done the same for some basic results in quantum information theory.  Quantum field theory is familiar to at least the first and third groups, but even they might not be comfortable with the aspects of it I use here so on behalf of all three I have reviewed that as well.  I hope that this will not cause the resulting size of these notes to deter potential readers from quickly jumping to whatever aspect they find most interesting.  My reason for writing up these notes is that there did not seem to be a convenient source for many of the things discussed here, some of which are dispersed throughout the literature and some of which are widely known but as far as I know do not appear in print anywhere.  The source with the most overlap is probably \cite{Susskind:2005js}, from which I learned many of the things in the earlier sections of these notes.  For some other (shorter) reviews see \cite{Preskill:1992tc,Giddings:1994pj,Mathur:2009hf}.  I have heard from many people that there is a high barrier of entry to this field, and I hope I have lowered it a bit.

\subsection{Conventions}
To simplify equations, starting in section \ref{krusksec} I work in units where the Schwarzschild radius $2GM$ is set to one.  This is a silly convention to use once we begin to consider situations where the black hole mass is time-dependent, so it will stop in section \ref{tevap}.  In section \ref{evapsec} I will switch to Planckian units where $8\pi G\equiv \ell_p^2$ is set to one, in section \ref{adssec} I will instead set the Anti- de Sitter radius to one, and from section \ref{paradoxsec} onwards the equations are simple enough that I will keep all three explicit.  Like any civilized person I will of course set $c=\hbar=k_B=1$ throughout.    

I will use the symbol $\Omega$ in two different contexts; I will denote ground state wave functions as $|\Omega\ran$, and I will refer to coordinates on the sphere $\mathbb{S}^d$ as $\Omega$.  $d\Omega_d^2$ is the standard ``round'' metric on $\mathbb{S}^d$, $d\Omega_d$ is the volume element for use in integrals, and $\Omega_d\equiv \frac{2\pi^{\frac{d+1}{2}}}{\Gamma\left(\frac{d+1}{2}\right)}$ is the volume.

In multipartite quantum mechanical systems I will refer to the subsystems by capital Roman letters such as $A$, $B$, $\ldots$, I will refer to their associated Hilbert spaces as $\HA$, $\HB$, $\ldots$, and I will call the dimensions of their Hilbert spaces $|A|$, $|B|$, $\ldots$.  

Except for section \ref{adssec} I will work almost exclusively in $3+1$ spacetime dimensions.  Results for asymptotically Minkowski black holes in other dimensions can be obtained from the AdS formulas in section \ref{adssec} by taking the limit $r_{ads}\to\infty$.  I will not discuss black holes with charge or angular momentum; these more general black holes provide interesting laboratories for the information problem and reconstructing the interior, but in the end they so far don't seem to add much conceptually. 

\section{Classical black holes}\label{clbhsec}

In this section I review the main properties of classical black holes in general relativity (GR).  For readers who are unfamiliar with this theory I give an extremely compact description in appendix \ref{GRapp}.
\subsection{The Schwarzschild geometry}\label{schgsec}
The Schwarzschild geometry is the unique source-free solution of Einstein's equation with spherical symmetry that approaches ordinary Minkowski space at large distances.  By itself the Schwarzschild geometry does not quite describe a black hole in the astrophysical sense, but understanding it is a necessary prerequisite for studying them.  Explicitly the spacetime metric for the Schwarzschild geometry is given by
\be\label{schmet}
ds^2=-\frac{r-2GM}{r}dt^2+\frac{r}{r-2GM} dr^2+r^2 (d\theta^2+\sin^2\theta d\phi^2).
\ee
Here $G$ is Newton's gravitational constant and $M$ is a parameter with units of mass.\footnote{To understand the physical meaning of the parameter $M$, we can observe that for $r\gg 2GM$ the geodesic equation \eqref{geodesiceq} for a massive non-relativistic test particle in this geometry reduces to the standard Newtonian equation
\be
m\ddot{\vec{x}}=-\frac{Gm M}{r^2}\hat{r}
\ee
for the motion of a particle about a point source of mass $M$.  Many experimental tests of general relativity are based on detecting the $O\left(\frac{r}{2GM}\right)$ corrections to this approximation.}
The quantity in brackets is the unit metric $d\Omega_2^2$ on the two sphere $\mathbb{S}^2$, so the coordinate $r$ parametrizes the proper size of this $\mathbb{S}^2$.  $r=0$ and $r=2GM$ are clearly special, most of the interesting physics of black holes lies in understanding what happens at these two radii.  

At $r=0$ the $\mathbb{S}^2$ shrinks to zero size and the metric diverges; this is called the \textit{singularity}.  This pathology can be described in a coordinate-invariant way as the divergence of the fully contracted Riemann tensor $R_{\alpha\beta\gamma \delta}R^{\alpha\beta\gamma\delta}$.  Since the Riemann tensor physically encodes the strength of tidal effects on freely falling objects, this divergence leads to formally infinite tidal forces that would destroy anyone unfortunate enough to find herself in the vicinity of $r=0$.  These divergences are probably regulated by Planck-scale physics, but that would be little consolation.   

The radius $r_s\equiv2GM$ is called the \textit{Schwarzschild radius}; the metric appears to also be singular here, but we will see in the next section that unlike the singularity at $r=0$ this singularity is a spurious artifact of our choice of coordinates. At least classically it does not lead to any locally detectable divergence.  Something important globally \textit{does} happen at $r=r_s$; in \eqref{schmet} the signs of the coefficients of $dr^2$ and $dt^2$ switch.  The coordinate $r$ has become timelike, and any test particle which falls into the region with $r<r_s$ will necessarily continue to evolve towards smaller $r$ until it approaches the singularity.  Somebody inside of the horizon cannot prevent this for the same reason you cannot prevent yourself moving forwards in ordinary time.  This is true even for massless particles, so the region behind the horizon is completely invisible to anybody who stays outside at $r>r_s$.\footnote{These statements can be justified more carefully by studying the geodesic equation in the Schwarzschild metric.  One also finds that any massive observer will always reach the singularity in a finite proper time.  In fact one survives the longest by not struggling; firing rockets to try to escape just causes you to die faster!}  

In any spacetime, if the set of points that can ever send a signal to a particular timelike geodesic has a boundary then that boundary is called the \textit{event horizon} of the geodesic.  In the Schwarzschild geometry we are especially interested in timelike geodesics that stay outside of the black hole for all times, and the surface $r=r_s$ is the event horizon for any such geodesic.  In this case this surface is usually just called the horizon of the black hole.

One very important feature of the Schwarzschild horizon is its \textit{gravitational redshift}; as we approach the horizon, a fixed unit of coordinate time counts for less and less proper time along a curve of fixed $r$.  This means that signals sent with a fixed energy from a point that gets closer and closer to the horizon have lower and lower energy when they reach $r\gg 2GM$.  Conversely, a signal propagating away from the horizon that has some fixed energy at $r\gg 2GM$ has higher and higher energy from the point of view of to a fixed $r$ observer as we move $r$ closer and closer to $2GM$.  This feature seems to allow an observer at $r\gg 2GM$ to be sensitive to very high energy physics, and it is at the heart of the connection between black hole thought experiments and quantum gravity.

\subsection{The Kruskal extension}\label{krusksec}
The $(t,r,\Omega)$ coordinate system we used in the previous subsection is convenient for thinking about experiments in the $r\gg 1$ region, but it is not well-suited for understanding near-horizon physics.\footnote{From here until section \ref{tevap} I work in units where $r_s=2GM=1$.}  A better choice is the Kruskal-Szekeres coordinates.  These are motivated by first observing that radial null geodesics in the Schwarzschild geometry can be parametrized as 
\be\label{nullg}
t=\pm r_*+C,
\ee
where $C$ is some constant of motion and $r_*$ is a new radial coordinate
\be
r_*\equiv r+\log(r-1).
\ee
$r_*$ is called the ``tortoise'' coordinate, presumably because it fits an infinite coordinate range into a finite geodesic distance.  The Kruskal-Szekeres coordinates are then defined as
\begin{align}\label{UVK}
U&\equiv -e^{\frac{r_*-t}{2}}\\
V&\equiv e^{\frac{r_*+t}{2}}.
\end{align}
By construction they have the property that lines of constant $U$ or $V$ are radial null geodesics.  These coordinates have the convenient feature that 
\be\label{uveq}
UV=(1-r)e^r,
\ee 
so the singularity is when $UV=1$ while the horizon is when either $U$ or $V$ is zero.  The metric is
\be
ds^2=-\frac{2}{r}e^{-r}\left(dUdV+dVdU\right)+r^2 d\Omega_2^2,
\ee
where $r$ is obtained implicitly from \eqref{uveq}.  The off-diagonal nature of the metric may appear unfamiliar, but it can be easily removed by defining yet another set of coordinates
\begin{align}\nonumber
U&=T-X\\
V&=T+X,
\end{align}
in terms of which the metric is
\be\label{TXmet}
ds^2=\frac{4}{r}e^{-r}\left(-dT^2+dX^2\right)+r^2 d\Omega_2^2.
\ee
Note that there is now no singularity of any kind at $r=1$.  

\begin{figure}
\begin{center}
\includegraphics[height=6cm]{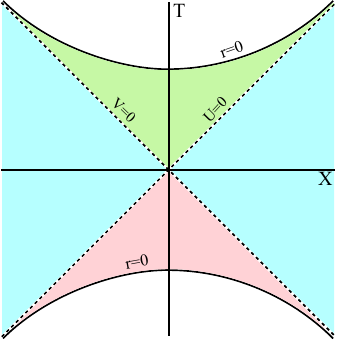}
\caption{The $XT$ plane of the Kruskal extension.  Lines of constant $U$ and $V$, or in other words radial null geodesics, are straight lines with slope $\pm \pi/4$.  Lines of constant $r$ are hyperboloids centered at the origin, with the blue regions having $r>1$ and the red/green regions having $r<1$.  The horizons are the dashed lines.  The original exterior region is the right light blue wedge, the new exterior is the left blue wedge, the future interior is in green, and the past interior is in red.  It is manifest that no radial null geodesic can escape the future interior into one of the blue regions, and it is also clear that no null geodesic connects the right and left blue wedges.}\label{kruskfig}
\end{center}
\end{figure}
The expression \eqref{TXmet} defines a geometry over the full $XT$ plane; it is interesting to understand which parts of it correspond to the regions discussed in the previous subsection.  This is illustrated in figure \ref{kruskfig}.  The region defined by $r>1$, $-\infty<t<\infty$ in the old Schwarzschild coordinates is sent to the right blue wedge.  In continuing to $r<1$ however there is a branch cut in the definition \eqref{UVK} which allows us to either reach the region $T>0$, $X^2-T^2<0$, shown in green, \textit{or} the region $T<0$, $X^2-T^2<0$, which is shown in red.  The singularity at $r=0$ is the hyperboloid $X^2-T^2=-1$, so it has not one but two connected components, one at the boundary of each of these regions.  I'll refer to these two regions as the future and past interiors respectively.  Finally there is a fourth region, the left blue wedge, which is a second asymptotically Minkowski region in which we can again have $r\gg 1$.  Combining all of the regions, we can interpret the full Schwarzschild geometry as a wormhole connecting two asymptotically-flat universes, each of them behaving at $r\gg 1$ as if there were a gravitational point source of mass $M$.  The wormhole is non-traversable in the sense that no signal can be sent from one blue region to the other, but observers who jump in from opposite sides are able to meet in the middle and compare notes!

\subsection{Penrose diagrams}\label{penrosesec}
In gravitational thought-experiments we are often uninterested in the details of the spacetime geometry; we might care only about the causal structure of the spacetime in the sense of which points can receive signals from which other points.  If this is the case, then we should be able to throw out some of the irrelevant information in the metric.  Indeed the following theorem gives us a very useful way to do this:
\begin{itemize}
\item THEOREM: Two spacetimes whose metrics differ only by multiplication by a positive scalar function on the spacetime, that is which are related as $g'_{\mu\nu}(x)=e^{2\omega(x)}g_{\mu\nu}(x)$ for some smooth real function $\omega(x)$, have the same null geodesics.  Timelike/spacelike geodesics in one metric will not necessarily be timelike/spacelike geodesics in the other, but timelike/spacelike curves in one metric \textit{will} be timelike/spacelike curves in the other.
\end{itemize}
Two metrics which are related in this way are said to be \textit{conformally equivalent}.  The proof of this theorem is not difficult, and is left to the interested reader as an exercise.  

It was realized long ago by Penrose that this theorem gives an elegant way to represent the asymptotic behavior of spacetimes at large distance (such as the $r\to \infty$ limit in Minkowski space).  The idea is to judiciously choose a function $\omega(x)$ that diverges as we approach infinity in just such a way that infinity is brought in to finite proper distance.  We may then include infinity as a boundary of the spacetime, turning it into a manifold with boundary.  This procedure is called \textit{conformal compactification}.  

\begin{figure}
\begin{center}
\includegraphics[height=6cm]{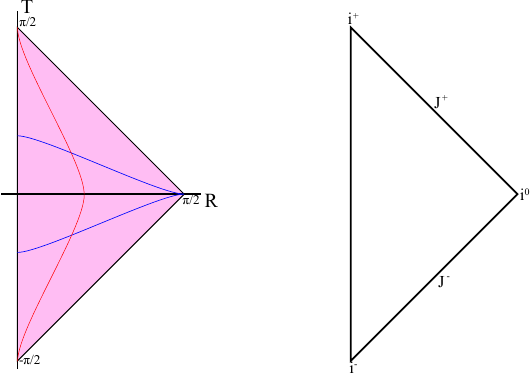}
\caption{On the left, the full Minkowski space is the pink wedge in the $RT$ plane.  Radial light rays move on lines of slope $\pm \pi/4$.  Some slices of constant $t$ are shown in blue and a slice of constant $r$ is shown in red.  On the right we formalize this into a genuine Penrose diagram.}\label{flatpenrose}
\end{center}
\end{figure}

Conformal compactification is perhaps best understood by studying an example, so let's consider ordinary flat Minkowski space.  The usual spacetime metric is 
\be
ds^2=-dt^2+dr^2+r^2 d\Omega_2^2.  
\ee
There is interesting asymptotic behavior as we take $r\to\infty$ and/or $|t|\to \infty$ that we would like to analyze.  The rough idea is to use the function $\arctan(x)$ to ``pull in'' the boundary to finite distance, but to preserve the simplicity of the causal structure we need to do this in light-like directions. We thus define
\begin{align}\nonumber
T+R&\equiv \arctan (t+r)\\
T-R&\equiv \arctan(t-r),\label{comptran}
\end{align}
which gives a metric
\be\label{penrosemet}
ds^2=\frac{1}{\cos^2 (T+R)\cos^2(T-R)}\left[-dT^2+dR^2+\left(\frac{\sin (2R)}{2}\right)^2 d\Omega_2^2\right].
\ee
These new coordinates have ranges $|T\pm R|<\pi/2$, $R\geq 0$, so we can now compactify the spacetime by including the points at the boundary $|T\pm R|=\pi/2$.  The prefactor diverges at this boundary, as it must since the boundary is infinitely far away, but we can now use the theorem to define a new spacetime with the same causal structure as Minkowski space by simply removing this prefactor.  This construction is illustrated in figure \ref{flatpenrose}.

The new boundary is naturally divided into five parts; \textit{past/future timelike infinity}, marked as $i^{\mp}$ in the figure, \textit{past/future null infinity}, marked as $J^\mp$ in the figure, and \textit{spatial infinity}, marked as $i^0$ in the figure.  $i^{\mp}$ are where timelike geodesics ``come from'' and ``go to'', $J^\mp$ are the same for null geodesics, and $i^0$ is where spatial geodesics end.  The scattering matrix maps states on $J^-\cup i^-$ to states on $J^+\cup i^+$, with massless particles entering/leaving at $J^{\mp}$ and massive particles entering/leaving at $i^{\mp}$.  Conserved charges in general relativity such as the total energy or electromagnetic charge are always written as boundary integrals at $i^0$.  The diagram also makes it very clear that there are no event horizons in Minkowski space; any timelike geodesic can eventually receive signals from everywhere in the space.  

The right-hand diagram in figure \ref{flatpenrose} is our first example of a \textit{Penrose diagram}; it is an extremely compact way of describing the causal structure of the spacetime without any extraneous details.  Just from the diagram we have already seen that this spacetime has no horizons, and that it should have a nice description in terms of an S-matrix.  

Let's now understand in more detail what happened to the other two dimensions.  At each point of the Penrose diagram there is an $\mathbb{S}^2$ which we have suppressed.  It is the spherical symmetry of the metric \eqref{penrosemet} which allows us to do this without losing much information about the spacetime, and indeed we can draw a similar diagram for any spacetime with $\mathbb{S}^2$ symmetry.  There are only two ways to have a boundary of a Penrose diagram; one is for the $\mathbb{S}^2$ to have infinite size, as happens at $|T\pm R|=\pi/2$ in our Minkowski diagram.  The other way is for the $\mathbb{S}^2$ to collapse to zero size, as happens at $R=0$.  In the interior of the diagram, the spacetime is locally just a product manifold where the radius of the $\mathbb{S}^2$ varies as we move around in the diagram; this is often called a \textit{warped product}.\footnote{More generally, in $d$ spacetime dimensions Penrose diagrams are useful anytime we can write the geometry as a warped product of a $d-2$ dimensional space over some subset of the two-dimensional plane.}

\begin{figure}
\begin{center}
\includegraphics[height=4cm]{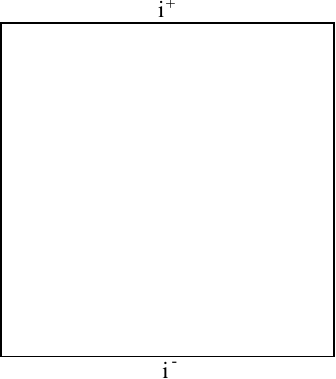}
\caption{The Penrose diagram of de Sitter space.  The $\mathbb{S}^2$ shrinks to zero size on both the left and the right boundaries, while it grows to infinite size at $i^{\pm}$.  Note that here $i^\pm$ are each spacelike surfaces instead of just points; it is this property that leads to the presence of horizons.}\label{dspenrose}
\end{center}
\end{figure}

Another useful example of a Penrose diagram is that of de Sitter space, which has metric
\be
ds^2=-d\tau^2+\cosh^2\tau d\Omega_3^2.
\ee
This is a solution of Einstein's equation with positive vacuum energy, and it is a good approximation to the geometry of our universe today at the largest scales, as well as during cosmological inflation in the past.  The spatial geometry is an $\mathbb{S}^3$ which first shrinks exponentially to some minimum size and then expands exponentially.  The Penrose diagram of de Sitter space is shown in figure \ref{dspenrose}.  
From the diagram it is easy to see two very important properties of de Sitter space:
\begin{itemize}
\item de Sitter space has event horizons.  Two observers moving on timelike geodesics, say vertical straight lines in the diagram, will eventually be unable to communicate.  Crudely this is because the accelerating expansion of the universe has caused them to be moving away from each other faster than light.  
\item de Sitter has no infinite spatial boundary $i^0$, nor any separate light-like infinity $J^\pm$ distinct from $i^\pm$.  This is a serious problem for attempts to formulate a quantum theory of de Sitter space, since existing well-defined theories of quantum gravity require at least one of these.  In particular there is no straightforward sense in which de Sitter space has an S-matrix.  
\end{itemize}

\begin{figure}
\begin{center}
\includegraphics[height=5cm]{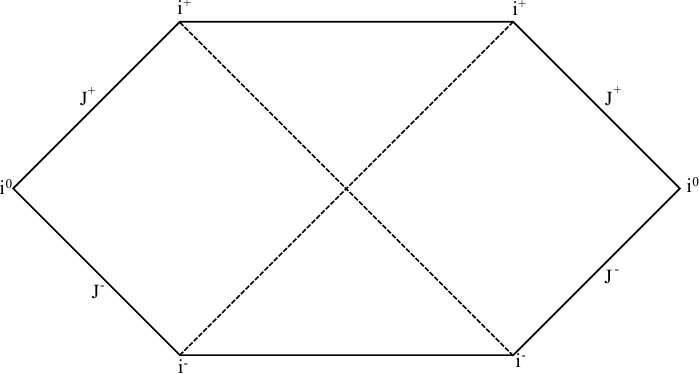}
\caption{The Penrose diagram for the Schwarzschild geometry.  The $\mathbb{S}^2$ shrinks to zero size only at the singularities at the top and bottom horizontal lines.  There are two copies of the asymptotic boundaries of Minkowski space, one on either side.  For convenience the horizons are marked with dashed lines.}\label{schpenrose}
\end{center}
\end{figure}

Finally we of course should understand the Penrose diagram of the Schwarzschild geometry.  The Kruskal-Szekeres coordinates have already done most of the work for us, since the metric is already in the form \eqref{TXmet}.  From a causal structure point of view the only difference between these $(T,X,\Omega)$ coordinates and the $(t,r,\Omega)$ coordinates in Minkowski space we just considered are their ranges.  In Minkowski space we had $-\infty <t<\infty$ and $r\geq0$, while for the Kruskal-Szekeres coordinates we have $X^2-T^2>1$.  We can use the same compactification transformation as for Minkowski space
\begin{align}\nonumber
T'+X'&\equiv \arctan (T+X)\\
T'-X'&\equiv \arctan (T-X),
\end{align} 
but now instead starting with the diamond $|R\pm T|<\pi/2$ and throwing out the region $R<0$, we instead start with the diamond $|X'\pm T'|<\pi/2$ and throw out the region $|T|>\pi/4$.  The resulting Penrose diagram is shown in figure \ref{schpenrose}.
This diagram is rather similar to the Kruskal diagram in figure \ref{kruskfig} above, but the spacetime boundaries are now explicitly shown.

\subsection{Real black holes}\label{realsec}
\begin{figure}
\begin{center}
\includegraphics[height=6cm]{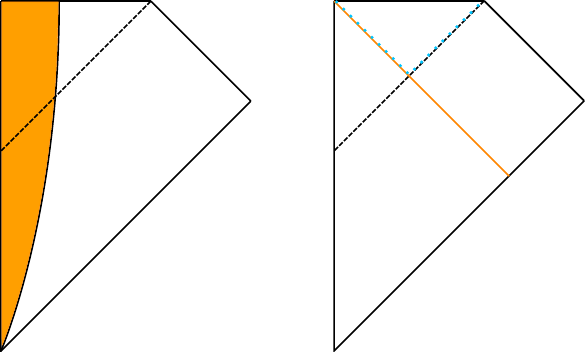}
\caption{Classical black hole formation.  On the left we have a black hole forming from the collapse of a cloud of massive particles, shown in orange.  On the right we have a black hole forming from the collapse of a spherical shell of photons.  In both cases the top boundary is the singularity, the left boundary is the origin of polar coordinates, and the other boundaries are the usual asymptotic ones for Minkowski space.  In the right-hand figure, the geometry above the orange line is exactly a piece from the upper right corner of the Schwarzschild geometry, while below it we have a piece of Minkowski space.  As usual the horizon is a dashed line.}\label{collapse}
\end{center}
\end{figure}

So far I haven't said much about black holes.  The reason is that the Schwarzschild geometry, with its two asymptotic infinities and no matter, is not a good description of the black holes that usually form in nature.  Real astrophysical black holes result from the gravitational collapse of ordinary matter either at the end of the life of a star or in the center of a galaxy.  These processes have some irritating limitations arising from details of of particle physics.  For example there is a lower bound on the mass of black holes that can form from stellar collapse: below a mass which is of order the solar mass $M_\odot$, gravitational collapse is halted by the formation of a neutron star and no black hole is created.\footnote{This bound is fairly easily understood as a consequence of the uncertainty principle.  A neutron star is a degenerate Fermi gas of neutrons, so the typical neutron momentum $k_n$ is of order the inverse spacing of the neutrons, which is of order $N^{1/3}/R$, where $N=M/m_n$ is the total number of neutrons, $R$ is the radius of the star, $M$ is its mass, and $m_n$ is the mass of a neutron.  We can relate $M$ and $R$ by equating the gravitational energy $GM^2/R$ and the kinetic energy $N_nk_n^2/m_n$.  Demanding that the neutrons stay non-relativistic, ie that $k_n\lesssim m_n$, gives $M\lesssim m_p \left(\frac{m_n}{m_p}\right)^2\approx M_\odot$ for the existence of a neutron star, where $m_p$ is the planck mass.  Determining the $O(1)$ coefficient requires more work, see for example \cite{bombaci1996maximum}, where a bound between $1.5 M_\odot$ and $3 M_\odot$ is quoted.}   To avoid having to worry about this kind of thing, it is convenient to instead imagine making a black hole out of a spherically symmetric infalling shell of photons.\footnote{If you don't want to assume the existence of photons then you can use gravitons instead.}  The infalling shell can be very diffuse at the moment when it passes through its own Schwarzschild radius, so there is no obstacle to forming a black hole in this way.  In fact we can explicitly construct the geometry in this case by sewing together a piece of Minkowski space and a piece of the Schwarzschild solution; the Penrose diagrams for these two methods of collapse are shown in figure \ref{collapse}.

It is interesting to note that in the case of the collapsing photon shell, the horizon extends into the region that is purely Minkowski space.  Somebody who was passing through the horizon down at that point would have absolutely no idea that his fate was sealed.  In fact we could currently be passing through the horizon of some gigantic yet-to-be-formed black hole, and we wouldn't know!  This illustrates the ``acausal'' nature of horizons; their locations depend on events that have not yet happened.

\section{Entanglement in quantum field theory}\label{qftentsec}
I'll now take a break from black holes to recall some basic facts about relativistic quantum field theory.

\subsection{Quantum field theory}\label{qftrevsec}

 A quantum field theory is a particular quantum mechanical system where the Hilbert space can be thought of as an infinite tensor product over all points in space of a finite number of degrees of freedom at each point.\footnote{The infinite number of points on a spatial slice is the source of the well-known UV (short-distance) and IR (long-distance) divergences of quantum field theory.  IR divergences can be regulated by working in finite volume, while UV divergences can be controlled by instead considering a theory with degrees of freedom only on some fine spatial lattice of points.  It is rather awkward to carry this around explicitly, so for the most part I will not attempt to.  I will bring back the UV cutoff occasionally when it is needed to regulate a divergence.}  The simplest example is a single degree of freedom at each spatial point; a scalar field $\phi(x)$.  The Hilbert space is spanned by states $|\phi\ran$ where the field has a definite value at each point in space.  Time evolution is generated by a Hamiltonian $H$ which is usually taken to be a single integral over space of some fairly simple function of the degrees of freedom and their derivatives.  For example a \textit{free scalar field of mass $m$} has a Hamiltonian
\be
H=\frac{1}{2}\int d^3 x \left(\pi(x)^2+\vec{\nabla}\phi(x)\cdot \vec{\nabla}\phi(x)+m^2\phi(x)^2\right).
\ee
Here $\pi(x)$ is the canonical momentum $-i\frac{\delta}{\delta \phi(x)}$ conjugate to $\phi$; together they obey
\begin{align}\nonumber
[\phi(x),\pi(y)]&=i\delta^3(x-y)\\\nonumber
[\phi(x),\phi(y)]&=0\\
[\pi(x),\pi(y)]&=0.
\end{align}
Note that these commutation relations are consistent with the statement that fields at different points act on different tensor factors of the Hilbert space.  This Hamiltonian follows from the Lorentz-invariant action
\be
S=-\frac{1}{2}\int d^4x \left(\partial_\mu \phi \partial^\mu \phi+m^2 \phi^2\right).
\ee

In conventional quantum mechanics we are often interested in explicitly describing the ground state wave function $|\Omega\ran$.\footnote{In relativistic quantum field theory one often refers to the ground state as the vacuum.  I will use the two terms interchangeably.}  This is possible for the free massive scalar field; indeed one can show \cite{Weinberg:1995mt} that
\be\label{freevac}
\lan\phi|\Omega\ran\propto e^{-\frac{1}{2}\int d^3x d^3y \phi(x)\phi(y)K(x,y)},
\ee
where 
\be
K(x,y)=\int \frac{d^3 k}{(2\pi)^3}e^{i \vec{k}\cdot (\vec{x}-\vec{y})}\sqrt{\vec{k}^2+m^2}=\frac{m}{2\pi^2r}\frac{d}{dr}\left(\frac{1}{r}K_{-1}(mr)\right).
\ee
Here $r\equiv|x-y|$ and $K_{-1}$ is a modified Bessel function \cite{abramowitz1964handbook}.  This expression is not particularly useful however, and it is rather inconvenient to generalize to theories with interactions.

Rather than trying to write down the vacuum wave functional explicitly it is usually more fruitful to instead study the vacuum expectation values of products of Heisenberg picture fields $\phi(t,x)\equiv e^{i H t} \phi(x)e^{-iHt}$.  For the free massive scalar field we can write a simple expression for the solution of this operator equation of motion:
\be\label{freeheis}
\phi(t,x)=\int \frac{d^3 k}{(2\pi)^3}\frac{1}{\sqrt{2\omega_k}}\left[e^{i (\vec{k} \cdot \vec{x}-\omega_k t)}a_{\vec{k}}+e^{-i (\vec{k} \cdot \vec{x}-\omega_k t)}a_{\vec{k}}^\dagger\right],
\ee
where I've defined $\omega_k\equiv \sqrt{\vec{k}^2+m^2}$.  The creation and annihilation operators $a_{\vec{k}}$, $a_{\vec{k}}^\dagger$ obey
\begin{align}\nonumber
[a_{\vec{k}},a_{\vec{k'}}^\dagger]&=(2\pi)^3 \delta^3(\vec{k}-\vec{k}')\\\nonumber
[a_{\vec{k}},a_{\vec{k'}}]&=0\\\nonumber
[a_{\vec{k}}^\dagger,a_{\vec{k'}}^\dagger]&=0\\\nonumber
[H,a_{\vec{k}}]&=-\omega_k a_{\vec{k}}.
\end{align}
The vacuum state $|\Omega\ran$ is annihilated by all of the $a_{\vec{k}}$, and low-lying excitations are created by acting on the vacuum with $a_{\vec{k}}^\dagger$'s.

More abstractly what this expression says is that we look for a complete basis of positive-frequency solutions\footnote{Somewhat confusingly, positive frequency means time dependence of the form $e^{-i\omega t}$ with $\omega>0$.} $f_n(x)$ to the wave equation 
\be
\left(\partial_\mu\partial^\mu-m^2\right)f(t,x)=0.
\ee
We do not want solutions that grow at infinity, so we want them to in some sense be normalizeable.  The best choice is to have the solutions $f_n$ be orthonormal in the Klein-Gordon norm
\be\label{KGnorm}
(f_1,f_2)_{KG}\equiv i\int d^3 x \left(f_1^*\dot{f}_2-\dot{f}_1^* f_2\right). 
\ee
To each $f_n$ we then associate an annihilation operator $a_n$, and we express the field as
\be\label{field}
\phi=\sum_n\left(f_n a_n+f_n^* a_n^\dagger\right).
\ee
The choice of normalization ensures that $a_n$ and $a_n^\dagger$ have the standard algebra.  The solutions $f_n$ are typically referred to as ``modes''.  In equation \ref{freeheis} we chose a plane-wave set of modes, which are delta-function normalized, but we could also have chosen some other set.  This more abstract formalism will be very useful in thinking about black holes.  

Finally let's look at some vacuum expectation values in the free massive theory.  These are typically called correlation functions.  The one-point functions vanish trivially:
\be
\lan\Omega|\phi(t,x)|\Omega\ran=0.
\ee
This follows from the action of the creation and annihilation operators on the vacuum, but it is also a consequence of the translation invariance of the vacuum.  The two-point function is more interesting, when the two points are at equal times it is given by:
\be\label{free2pt}
\lan \Omega|\phi(0,x)\phi(0,y)|\Omega\ran=\frac{1}{4\pi^2}\frac{m}{|x-y|}K_1(m |x-y|).
\ee
This correlation function scales like $1/|x-y|^2$ for $|x-y|\ll m^{-1}$, while for $|x-y|\gg m^{-1}$ it goes as $e^{-m|x-y|}$.  When a correlation function falls exponentially with separation, the decay constant, here $m^{-1}$, is often called the correlation length.  Note that if $m=0$, the correlation length is infinite and we have power law behavior all the way out.  In this case the theory is sometimes said to be ``gapless'', since there are excited states with energies arbitrarily close to the ground state energy.  

In fact a massless scalar field enjoys a symmetry larger than just the relativistic Poincare group; it is invariant under the \textit{conformal group}.  Among other things this bigger symmetry group includes rescalings of spacetime $x'^\mu=\lambda x^\mu$, so the theory is scale-invariant.  A quantum field theory with this larger symmetry group is called a \textit{conformal field theory}, or CFT.  

Although the correlation function \eqref{free2pt} is valid only in free scalar field theory, its basic structure of short distance power-law divergence and long distance decay is expected to be true in \textit{any} relativistic QFT.  

We will also eventually want to consider the two-point function when the two fields have different times, but since fields at timelike separation will not necessarily commute we need to be more careful about their ordering.  A particularly nice object is the ``time-ordered'' two-point function $\lan\Omega| T \phi(t,x)\phi(t',y)|\Omega\ran$, which is defined by ordering the fields in such a way that their time increases as we go to the left.  This can be interpreted as a ``transition amplitude'' where we act on the vacuum with a field $\phi$ at some time, evolve forward for a while, and then compute the projection onto the state we would have gotten by acting on the vacuum at the later time.  For our free massive scalar theory the time-ordered two-point function is
\be\label{to2pt}
\lan\Omega| T \phi(t,x)\phi(t',y)|\Omega\ran=\frac{1}{4\pi^2}\frac{m}{\sqrt{|x-y|^2-(t-t')^2+i\epsilon}}K_1\left(m \sqrt{|x-y|^2-(t-t')^2+i\epsilon}\right),
\ee
where $\epsilon$ is a positive infinitesimal quantity which is there to remind us which branch of the square root we should take when the points are timelike-separated; we should always take $\epsilon\to 0$ in any observable.

\subsection{Entanglement in the vacuum}
We saw in the last section that the ground state of a relativistic QFT has nonzero correlation between field operators at spatially separated points.\footnote{The presence of interesting correlation in the ground state is a special property of quantum field theories.  The ground state of a non-interacting non-relativistic gas of massive particles just has all the particles sitting on the floor!}  One convenient way to interpret this correlation is as an illustration of the \textit{entanglement} of different regions of spacetime in the vacuum of a relativistic QFT.  

For comparison recall the two-qubit state\footnote{Readers unfamiliar with qubit notation should read appendix \ref{qubitapp} now.  Also readers who would like a more detailed discussion of the definition and meaning of entanglement can look at appendix \ref{BPapp}.}
\be
|\Psi\ran=\frac{1}{\sqrt{2}}\left(|00\ran+|11\ran\right).
\ee
This state is clearly entangled in the sense that the full state is pure while the reduced state on either qubit is mixed, but we can also illustrate this using correlation functions.  Consider the Pauli operators $X_i$, $Y_i$, $Z_i$ acting on the spins.  Their one-point functions in the state $|\Psi\ran$ are all zero, but they have two-point functions
\begin{align}\nonumber
\lan\Psi|X_1 X_2|\Psi\ran&=1\\\nonumber
\lan\Psi|Y_1 Y_2|\Psi\ran&=-1\\\nonumber
\lan\Psi|Z_1 Z_2|\Psi\ran&=1.
\end{align}
This is the same qualitative behavior for one- and two- point functions that we saw in QFT, provided that we make the connection that the different qubits correspond to different regions in the QFT and the Pauli operators correspond to fields.  

\begin{figure}
\begin{center}
\includegraphics[height=3cm]{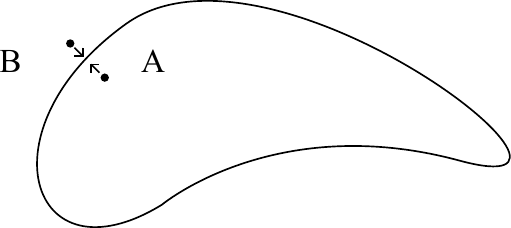}
\caption{Decomposing a QFT into a tensor product of the fields in region $A$ and the fields in region $B$.  The two-point function of fields at the indicated points diverges as they approach each other, indicating UV divergent entanglement between the two factors.  The boundary between the two regions is often called the \textit{entangling surface}.}\label{entangledcorr}
\end{center}
\end{figure}
More explicitly we could imagine decomposing the Hilbert space of the quantum field theory into a tensor product of the local field degrees of freedom in a region $A$ and its complement $B$.  By studying a two-point function with one field in $A$ and the other in $B$, we can learn about the entanglement between $A$ and $B$.  In fact we saw that at short distances the correlation functions are divergent, so if we allow our two points to approach each other at the interface between region $A$ and region $B$, as illustrated in figure \ref{entangledcorr}, the correlation becomes infinite.  This must mean that in some formal sense there is an infinite amount of entanglement between neighboring regions in the ground state of a relativistic quantum field theory.

Another fairly rigorous (and rather amusing) illustration of the entangled nature of the vacuum state in relativistic QFT is the Reeh-Schlieder theorem\cite{Streater:1989vi}, which says that for any region $A$, by acting on the vacuum $|\Omega\ran$ with operators located in that region we can produce a set of states which is dense in the full Hilbert space of the QFT.  In other words, by acting on the vacuum with some operators localized in this classroom we can create the moon.  Or the Andromeda galaxy!  This is possible because of the highly entangled nature of the vacuum; were the field theory degrees of freedom in the vicinity of the moon in a product state with the field degrees of freedom here, then no operators we act with here could do anything there.\footnote{To see the connection with entanglement more clearly, consider the state $|\Psi\ran=\sum_{ab}C_{ab}|a\ran|b\ran$ in a tensor product Hilbert space $\HA \otimes \HB$.  For simplicity say that the dimensionalities of $\HA$ and $\HB$ are equal.  Then if the matrix $C_{ab}$ is invertible, it is easy to see that we can produce any state in the Hilbert space by acting on $|\Psi\ran$ with an operator which acts nontrivially only on $\HA$.  That $C_{ab}$ is invertible is a statement about entanglement.  In quantum field theory both dimensionalities are infinite so one has to work harder to prove the theorem, but this example captures the basic point.  Note that a common confusion here is that if we were restricted to only acting with unitary operators on $\HA$ the theorem would be false, since they would not be able to change the amount of entanglement; it is essential that we are allowed to use non-unitary operators such as projections.}

\subsection{The Rindler decomposition}\label{rindsec1}
In the previous section we demonstrated vacuum entanglement in QFT in a rather general sense, but we'd of course like to have a more quantitative understanding of who exactly is entangled with who and by how much.  To this end, it is extremely useful to introduce the Rindler decomposition of Minkowski space.

\begin{figure}
\begin{center}
\includegraphics[height=6cm]{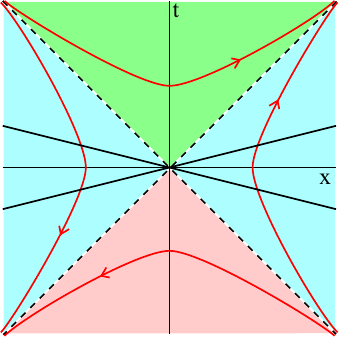}
\caption{The Rindler decomposition of Minkowski space.  The orbits of the boost operator are shown in red, and slices of Rindler time are shown as black straight lines.  The right and left Rindler wedges are shown in blue, the future wedge is shown in green, and the past wedge is shown in red.}\label{boost}
\end{center}
\end{figure}
The basic idea is to pick one of the three spatial coordinates, let's say $x$, and then divide the Hilbert space of the field theory into a factor $\mathcal{H}_R$ acted on by fields with $x>0$ and a factor $\mathcal{H}_L$ acted on by fields with $x<0$.  We'd then like to find a nice basis of states for each factor in which we can decompose the vacuum.  For this purpose it is extremely convenient to introduce the Lorentz boost operator $K_x$ that mixes $x$ and $t$, while acting trivially on $y$ and $z$.  This operator will exist in any relativistic QFT, for example in our free massive theory it is given by
\be\label{qftboost}
K_x=\frac{1}{2}\int d^3x \left[x (\dot{\phi}^2+\vec{\nabla}\phi \cdot \vec{\nabla}\phi+m^2\phi^2)+t\dot{\phi}\partial_x\phi\right].
\ee 
This operator appears time-dependent, but in the Heisenberg picture the time-dependence of the fields cancels the explicit time dependence.  The action of the boost operator in the $xt$ plane is shown in figure \ref{boost}.  As the figure makes clear, there are four different regions on which the boost operator has a well-defined action.  On the right Rindler wedge, shown in blue, it evolves forward in time from one black line to another.  On the left Rindler wedge, also shown in blue, it evolves \textit{backwards} in time along the same lines.  In the future and past wedges, shown in green and red respectively, its action is spacelike.  The resemblance of this figure to figure \ref{kruskfig} is not a coincidence; indeed an embarrassingly large percentage of the confusions one encounters about black holes are eventually resolved by understanding the Rindler decomposition better!

In order to understand how to express the vacuum state in the basis of boost eigenstates in the left and right Rindler wedges, it is extremely useful to introduce the Euclidean path integral.  Recall that in any quantum system, one way to find the ground state is simply to act on any generic state $|\chi\ran$ with $e^{-H T}$, where $T$ is some long time.  More precisely, we have
\be
\lan\phi|\Omega\ran=\frac{1}{\lan\Omega|\chi\ran}\lim_{T\to \infty}\lan\phi|e^{-H T} |\chi\ran.
\ee
In the Euclidean path integral formalism, this means that we can compute this wave functional as\footnote{In this expression $\hat{\phi}$ is a field history depending on both position and time, while $\phi$ is the field configuration at $t=0$ and is thus time-independent.}  
\be\label{epath}
\lan\phi|\Omega\ran\propto\int_{\hat{\phi}(t_E=-\infty)=0}^{\hat{\phi}(t_E=0)=\phi} \mathcal{D}\hat{\phi}e^{-I_E},
\ee
where $I_E$ is the Euclidean action, obtained from the usual one by analytic continuation $t \to -i t_E$.  For the free massive scalar field it is 
\be
I_E[\hat{\phi}]=\frac{1}{2}\int d^3x dt_E \left[(\partial_{t_E}\hat{\phi})^2+(\vec{\nabla}\hat{\phi})^2+m^2\hat{\phi}^2\right].
\ee
For simplicity I choose the early time boundary conditions to be $\phi=0$.  
\begin{figure}
\begin{center}
\includegraphics[height=5cm]{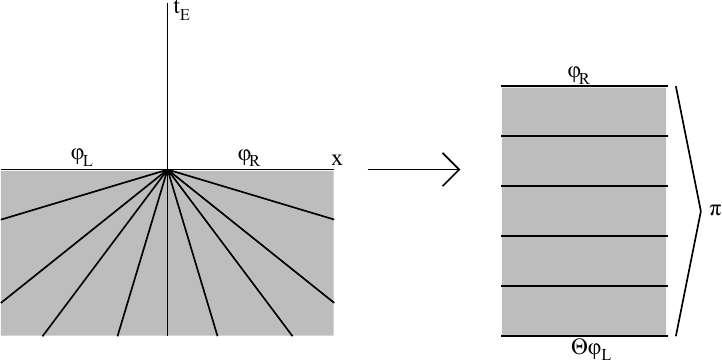}
\caption{Evaluating the Euclidean path integral representation of the vacuum wave functional.}\label{eucpath}
\end{center}
\end{figure}
In the free massive case it is possible to evaluate this path integral explicitly to obtain equation \eqref{freevac}, and you are invited to do so in the homework.  In understanding the entanglement of the vacuum however, it is more convenient to evaluate it in a way that continues to work when interactions are present.  The method is shown in figure \eqref{eucpath}.  It is based on the observation that the boost operator $K_x$ in the Euclidean plane generates rotations in the $xt_E$ plane, as can be seen from analytically continuing its action on $t$ and $x$.  So instead of evaluating the path integral from $t_E=-\infty$ to $0$, we instead evaluate it along the angular direction over an angle $\pi$.  From the point of view of the path integral this is just a trivial change of variables, but it makes clear that there is a nice alternative Hilbert space interpretation of the same path integral;
\be\label{rindvac}
\lan \phi_L\phi_R|\Omega\ran\propto \lan \phi_R|e^{-\pi K_R} \Theta |\phi_L\ran_L,
\ee
where $K_R$ is the restriction of $K_x$ to the right Rindler wedge and $\Theta$ is an antiunitary operator that exists in all quantum field theories and is usually called CPT.  Its action on a Heisenberg picture scalar field is\footnote{In even numbers of spacetime dimensions one usually combines this with a rotation of the $\vec{y}$ directions to invert them as well, but this is the definition that works in any dimension.}
\be
\Theta^\dagger \Phi(t,x,\vec{y})\Theta=\Phi^\dagger(-t,-x,\vec{y}),
\ee
although for a real scalar like we have been considering we can ignore the dagger since the field is hermitian.  It gives a natural map between $\mathcal{H}_L$ and $\mathcal{H}_R$.  It is needed in equation \eqref{rindvac} for two reasons; first to allow us to reinterpret the left-hand side of \eqref{rindvac} as a matrix element just in $\mathcal{H}_R$ and secondly because in the left-hand diagram of figure \ref{eucpath} $\phi_L$ is playing the role of a final state while in the right-hand diagram we need an initial state. We can then evaluate \eqref{rindvac} by inserting a complete set of $K_R$ eigenstates:
\begin{align}\nonumber
\lan \phi_L\phi_R|\Omega\ran&\propto \sum_i e^{-\pi \omega_i}\lan i|\Theta |\phi_L\ran\lan \phi_R | i\ran_R \\
&\propto\sum_i e^{-\pi \omega_i}\lan \phi_L|i^*\ran_L \lan \phi_R | i\ran_R ,
\end{align}
where I've used the antiunitarity of $\Theta$\footnote{Remember that for an antilinear operator $A$ the adjoint is defined as $\lan x|A^\dagger y\ran=\lan y|A x\ran$.  If $A$ is antiunitary then $\lan A x| Ay\ran=\lan y|x\ran$.} and defined 
\be
|i^*\ran_L=\Theta^\dagger|i\ran_R.
\ee
Thus we arrive at the remarkably simple expression for the vacuum state:
\be\label{rindvac1}
|\Omega\ran=\frac{1}{\sqrt{Z}}\sum_i e^{-\pi\omega_i}|i^*\ran_L |i\ran_R.
\ee

Notice that the entanglement between the left and right wedges is now completely manifest.  We can also compute the reduced density matrix for the right wedge:
\be
\rho_R=\frac{1}{Z}\sum_i e^{-2\pi \omega_i}|i\ran_R \lan i|.
\ee
This is nothing other than the thermal density matrix with temperature
\be\label{rindlertemp}
T=\frac{1}{2\pi}.
\ee
It may appear mysterious that the temperature \eqref{rindlertemp} is dimensionless, I will comment on the physical meaning of this soon.

\subsection{Free fields in Rindler space}\label{rindsec2}
To get some more intuition for the Rindler decomposition, we can study it in the free massive scalar theory.  The basic idea is to find a set of modes $f_n$ which are better suited to the Rindler decomposition than the usual plane waves, and then expand the field in creation and annihilation operators for them.  To accomplish this, it is convenient to introduce new coordinates for the left and right Rindler wedges:\footnote{In this transformation I have suppressed an arbitrary choice of length scale, which I'll call $\ell$, which is needed to make the units work out.  I'll restore it at the end of this section.}
\begin{align}\nonumber
x&=e^{\xi_R}\cosh \tau_R=-e^{-\xi_L}\cosh \tau_L\\
t&=e^{\xi_R}\sinh \tau_R=e^{-\xi_L}\sinh\tau_L,\label{rindcoord}
\end{align}
These coordinates have the property that evolving with the boost operator $K_x$ is just translation of $\tau_R$ forwards in time and translation of $\tau_L$ backwards in time.  The $\xi_{L,R}$ coordinates label hyperboloids that are orbits of $K_x$; they are also trajectories of constant proper acceleration.  The coordinate ranges are $-\infty<\xi_{L,R}<\infty$, $-\infty<\tau_{L,R}<\infty$, and they cover the left and right Rindler wedges respectively but not the future or past Rindler wedges.  The surfaces $\xi_R=-\infty$, $\xi_L=\infty$ are usually called the Rindler horizons, although they are not actually event horizons according to the definition given in section \ref{schgsec}.  The metric in these coordinates is
\be\label{rindmec}
ds^2=e^{2\xi_R}(-d\tau_R^2+d\xi_R^2)+d\vec{y}^2=e^{-2\xi_L}(-d\tau_L^2+d\xi_L^2)+d\vec{y}^2,
\ee
where for convenience I have combined $y$ and $z$ into vector $\vec{y}$.  
The idea is then to look for solutions of the massive wave equation
of the form
\be\label{rindlermode}
f_{R/L\omega k}=e^{-i\omega\tau_{R/L}}e^{i\vec{k}\cdot \vec{y}}\psi_{R/Lk\omega}(\xi_{R/L}),
\ee
where $\omega>0$ and the wave equation implies that $\psi_{Rk\omega}$ and $\psi_{Lk\omega}$ obey
\begin{align}\nonumber
\left[-\partial_{\xi_R}^2+(m^2+\vec{k}^2)e^{2\xi_R}-\omega^2\right]\psi_{Rk\omega}=&0\\
\left[-\partial_{\xi_L}^2+(m^2+\vec{k}^2)e^{-2\xi_L}-\omega^2\right]\psi_{Lk\omega}=&0.
\end{align}
Formally these equations are nothing but the Schrodinger equation for a non-relativistic particle in an exponential potential.  It can be solved explicitly in terms of Bessel functions, but for our purposes it is sufficient simply to observe that the normalizeable solutions (in the KG norm \eqref{KGnorm}) will oscillate at sufficiently negative $\xi_R$ (or sufficiently positive $\xi_L$), and will decay exponentially at sufficiently positive $\xi_R$ (or sufficiently negative $\xi_L$).  We can thus think of the modes as being ``confined'' to be near the horizon, with lower energy and/or higher tranverse momentum modes being confined more strongly.  

Of course the point of introducing these modes is that they have definite boost energy; $\omega$ in the right wedge and $-\omega$ in the left wedge.  As such, if we expand the field in terms of them as
\be
\phi=\sum_{k,\omega}\left(f_{R\omega k}a_{R\omega k}+f_{L\omega k}a_{L\omega k}+f_{R\omega k}^*a_{R\omega k}^\dagger+f^*_{L\omega k}a_{L\omega k}^\dagger\right),
\ee
then the creation operators $a_{L,R \omega k}^\dagger$ create states of definite boost energy on the Rindler vacuum $|0\ran$, which is defined as being annihilated by the annihilation operators $a_{L,R,\omega k}$.  We can thus rewrite equation \eqref{rindvac1} as a product state over all modes
\be\label{rindvacstate}
|\Omega\ran=\bigotimes_{\omega,k}\left[\sqrt{1-e^{-2\pi\omega}}\sum_n e^{-\pi \omega n}|n\ran_{L \omega (-k)}|n\ran_{R \omega k}\right],
\ee
where $n$ labels the number of particles on top of the Rindler vacuum in the each mode.  The sign flip for $k$ comes from the CPT conjugation in \eqref{rindvac1}.\footnote{More explicitly, $\Theta$ sends $\tau_R\to-\tau_L$, $\xi_R\to -\xi_L$, and $\vec{y}\to \vec{y}$.  Applying this to the right-hand modes \eqref{rindlermode} sends positive frequency modes to negative frequency modes, so to get the coefficient of an annihilation operator we need to take the complex conjugate.  This flips the sign of $\vec{k}$.}  As expected, in the reduced density matrix on either side each mode is thermally occupied with temperature \eqref{rindlertemp}.  

For future reference I note that the state \eqref{rindvacstate} is annihilated by the operators
\begin{align}\nonumber
c_{1\omega k}&\equiv \frac{1}{\sqrt{1-e^{-2\pi \omega}}}\left(a_{R\omega k}-e^{-\pi \omega} a_{L\omega (-k)}^\dagger\right)\\
c_{2\omega k}&\equiv \frac{1}{\sqrt{1-e^{-2\pi \omega}}}\left(a_{L\omega k}-e^{-\pi \omega} a_{R\omega (-k)}^\dagger\right),\label{cmodes}
\end{align}
so we can think of the vacuum state as the joint zero-eigenspace of the ``number'' operators $c_{1\omega k}^\dagger c_{1\omega k}$, $c_{2\omega k}^\dagger c_{2\omega k}$.  States where these oscillators are excited are excited states with respect to the Minkowski Hamiltonian $H$.\footnote{A more technical way of seeing this is to observe that the modes annihilated by the $c$'s have only positive frequency with respect to the Minkowski time $t$, so they must agree with the usual plane wave modes about the definition of the ground state.}

We now come back to the meaning of the temperature \eqref{rindlertemp}.  In defining the dimensionless Rindler coordinates $\xi_{R,L}$ and $\tau_{R,L}$, I suppressed an arbitrary choice of length scale $\ell$.  Had I defined them to have units of length, the temperature would have been $\frac{1}{2\pi \ell}$.  But what is the meaning of $\ell$?  Indeed it is straightforward to see that it is the inverse proper acceleration of an observer at $\xi_R=0$, who also happens to have $\tau_R$ as her local proper time.  Energy defined with respect to $\tau_R$ is thus the energy that such an observer would define in her locally inertial frame.  This strongly suggests that any observer with acceleration $a=\frac{1}{\ell}$  should actually perceive a temperature
\be\label{unruh}
T_{Unruh}=\frac{\hbar a}{2\pi k_B c},
\ee
where for amusement I have temporarily restored the dimensionful constants.  This is called the \textit{Unruh effect}, and is one of the more striking results of relativistic QFT.  Notice that both relativity and quantum mechanics are important; either $c\to\infty$ or $\hbar\to0$ would send $T_{Unruh}$ to zero.  You may wonder if this temperature is actually ``real''; in fact one can build a model of a detector that couple to the scalar field and set it on accelerating trajectories, and indeed one finds that it clicks just as if it were in the presence of thermal fluctuations at temperature $T_{Unruh}$.  For more details about this, as well as many other interesting observations about the physics of the Rindler decomposition, I encourage the reader to consult \cite{Unruh:1983ms}.

\subsection{Entanglement is important for horizon crossing: an introduction to firewalls}\label{firewallsec}
Before finally getting back to black holes, I need to make one final point about the Rindler decomposition.  So far I've mostly focused on the left and right Rindler wedges, but of course the future and past wedges are also interesting.  In particular, the entanglement of the Minkowski Vacuum $|\Omega\ran$ across the entangling surface $x=0$ is \textit{essential} for having a smooth transition from either the left or the right Rindler wedges to the future interior.  To see this, imagine that instead of the ground state $|\Omega\ran$, we put the system into the mixed state
\be\label{badstate}
\rho=\rho_L\otimes \rho_R,
\ee
where $\rho_{L,R}$ are the thermal density matrices obtained on either side by tracing out the other in the vacuum $|\Omega\ran$.  For any observer who stays in the left or right Rindler wedges, this state is indistinguishable from the vacuum.  But in fact it has infinite energy: the Hamiltonian includes a gradient term that is divergent if the field is discontinuous at $x=0$.  More precisely if the left and right wedges are completely uncorrelated, as in the state \ref{badstate}, then the typical difference between neighboring fields on either side is of order the typical field fluctuation, which is given by $\frac{1}{\epsilon}$ where $\epsilon$ is a UV length cutoff, so we have
\be
\partial_x\phi|_{x=0}  \propto \frac{1}{\epsilon^2}.
\ee
The gradient term in the Hamiltonian then produces a contribution 
\be
dx \int d^2 y (\partial_x\phi)^2\propto\epsilon\int d^2 y  \frac{1}{\epsilon^4}=\frac{A}{\epsilon^3},
\ee 
where I've replaced $dx\to\epsilon$.  There is thus a huge concentration of energy sitting at $x=0$, waiting to annihilate anybody who tries to jump through the Rindler horizon into the future wedge.  This type of thing has recently been given a new name: a \textit{firewall}.\footnote{We can also study this firewall by computing the expectation value of the number operator $c_{1\omega k}^\dagger c_{1\omega k}$ for the modes defined by \eqref{cmodes}.  In the state \eqref{badstate} a short calculation shows that $\lan c_{1\omega k}^\dagger c_{1\omega k}\ran=\frac{2}{\left(e^{\pi \omega}-e^{-\pi \omega}\right)^2}$, so for $\omega\lesssim 1$ the mode will have an $O(1)$ excitation.  Since there are many such modes, the state will be quite singular.}

One can study this more concretely using a model detector.  I will not describe this in detail here, but one can see without too much difficulty that even fairly mild decorrelation already leads to $O(1)$ probability that the detector clicks as it crosses the horizon.  For a related calculation see section 3.4 of \cite{Giddings:2006sj}. 

An important point here is that although entanglement is necessary for a smooth infalling experience through the Rindler horizon, it is not sufficient.  Heuristically, imagine that the states $\frac{1}{\sqrt{2}}\left(|00\ran\pm|11\ran\right)$ both had smooth horizons.  By the linearity of quantum mechanics this would mean that the product states $|00\ran$ and $|11\ran$ must as well, but we just saw that no product state possibly can have a smooth horizon in QFT.  Not only do we need entanglement to get a smooth horizon, it needs to be the \textit{right} entanglement!\footnote{One way to think about this is that the divergence of the two-point function of $\phi$ as we bring the points together is precisely the divergent correlation between the left and right which is necessary to avoid a large expectation value for the gradient.  If we correlate the fields in the wrong manner, for example if we anti-correlate them, then the gradient will still be large.} 

\section{Quantum field theory in a black hole background}\label{qftbhsec}
We've now discussed classical black holes and quantum field theory; it is time to combine them.  To really analyze the problem correctly, we would need to treat both the metric and the matter fields quantum mechanically in a full theory of quantum gravity.  I will discuss how this might be done in section \ref{adssec} below, but I'll first discuss the much simpler problem of quantum field theory in a fixed black hole background.  Physically this is the limit where we send the Newton constant $G$ to zero and the black hole mass $M$ to infinity in such a way that the Schwarzschild radius $r_s=2GM$ is fixed.  Said in terms of dimensionless quantities, we send $\frac{M}{m_p}\to\infty$, where $m_p\equiv \frac{1}{\sqrt{8\pi G}}$ is the Planck mass, and then study observables whose length scale is of order $r_s$ in this ratio.  In this limit there are no gravitational interactions, and the metric can be viewed as a fixed external field.\footnote{This is not quite true; free gravitons are still present in this limit, but we can just treat them as another matter field.}

This limit is quite reasonable from the point of view of astrophysical black holes: for a solar mass black hole we are working to leading order in
\be
\frac{m_p}{M}\approx 10^{-38},
\ee 
which is not a bad thing to base a perturbation theory on.  The Schwarzschild radius is of order kilometers, which is indeed the type of scale at which we could imagine doing experiments.  Of course detecting individual photons whose wavelength is a kilometer is no picnic, but nobody credible ever said life would be easy.

\subsection{Two-sided Schwarzschild and the Rindler decomposition}\label{2siderindsec}
Before proceeding further, we need to decide which of the two geometries discussed in section \ref{realsec} to study; the full two-sided Schwarzschild geometry or the one-sided collapse geometry that is only Schwarzschild after the infalling matter has gone in.  Both are interesting, but it is a bit easier to begin with the two-sided case to avoid the complications of the infalling shell.  

\begin{figure}
\begin{center}
\includegraphics[height=5cm]{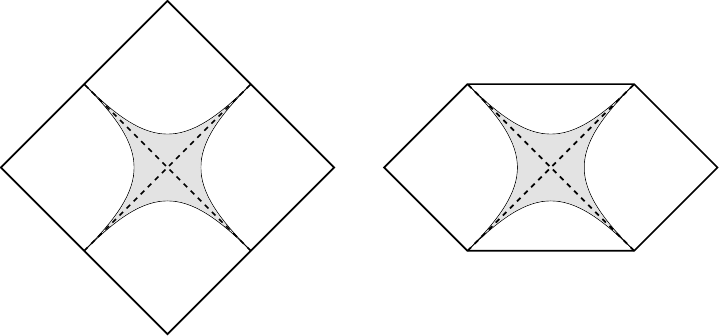}
\caption{The shaded region of the Rindler Penrose diagram well approximates the shaded region of the Schwarzschild Penrose diagram on the right.  The Rindler Penrose diagram is different from the Minkowski Penrose diagram in figure \ref{flatpenrose}, even though they represent the same spacetime.  The reason is that at each point I've here suppressed an $\mathbb{R}^2$ instead of an $\mathbb{S}^2$.}\label{rindsch}
\end{center}
\end{figure}

From the Kruskal/Szekeres expression for the metric \eqref{TXmet} it is clear that, for $r\approx 1$ and sufficiently small angular displacements, the Schwarzschild geometry resembles the region of Minkowski space that is near the Rindler horizon in the Rindler decomposition.  More explicitly, in the right exterior ($r>1$) and using the tortoise coordinate 
\be
r_*=r+\log (r-1),
\ee
we have
\be
ds^2=\frac{r-1}{r}\left(-dt^2+dr_*^2\right)+r^2 d\Omega_2^2.
\ee
If we define $y_1\equiv \theta_1$, $y_2\equiv \theta_2$, where $\theta_1$ and $\theta_2$ are two orthogonal angular coordinates on the sphere, then for $r\approx 1$ we have
\be
ds^2\approx e^{r_*-1}\left(-dt^2+dr_*^2\right)+d\vec{y}^2.
\ee
This is very reminiscent of the right Rindler wedge metric \eqref{rindmec}, and indeed if we define 
\begin{align}\nonumber
r_*&=2\xi_R+1-\log 4\\
t&=2\tau_R \label{rindschc}
\end{align}
they become equivalent.  We can extend this argument to the other three parts of the geometry; the upshot is illustrated in Penrose diagrams in figure \ref{rindsch}.

This observation makes it very plausible that, whatever initial state we pick for the quantum field in the Schwarzschild geometry, if it is to locally look like the Minkowski vacuum near the interface of the left and right exteriors, they will need to be thermally entangled just as in the Rindler decomposition.  We can determine the temperature with respect to Schwarzschild time, the natural time for obervations made at $r\gg1$, from the relationship \ref{rindschc} between Schwarzschild and Rindler time; apparently we must have
\be\label{Thawk}
T_{Hawking}=\frac{T_{Unruh}}{2}=\frac{1}{4\pi r_s}.
\ee
I have temporarily restored the Schwarzschild radius $r_s=2GM$; in fact for pedagogical purposes we can restore all of the dimensionful constants:
\be
T_{Hawking}=\frac{\hbar c^3}{8\pi k_B G M}.
\ee
Notice that the temperature decreases as the black hole increases in size; for example for a solar mass black hole this temperature is of order  $10^{-7}K$, which is fairly cold.  

There is a particularly simple global pure state for the Schwarzschild geometry, where the two exteriors are just thermally entangled as in Rindler space with no other excitations present, that is called the \textit{Hartle-Hawking State} (or the Hartle-Hawking-Israel State) \cite{Hartle:1976tp,Israel:1976ur}.  We will describe it more explicitly in section \ref{eucblacksec}. 

\subsection{Schwarzschild modes}\label{modesec}
Now let's study free fields in the Schwarzschild geometry.  To begin we need to find a simple set of modes $f$ that solve the free scalar equation of motion
\be
\frac{1}{\sqrt{-g}}\partial_\mu\left(\sqrt{-g}g^{\mu\nu}\partial_\nu \phi\right)=m^2\phi,
\ee
where $g_{\mu\nu}$ is the Schwarzschild metric and $g$ is its determinant.  We can then express the field in terms of these modes as in equation \eqref{field}, and study its properties in an appropriate quantum state such as the Hartle-Hawking state.   

I will focus on the right exterior of the Schwarzschild geometry, covered by the coordinates $(t,r,\Omega)$.  We will look for 
solutions of the form
\be\label{schmode}
f_{\omega \ell m}=\frac{1}{r}Y_{\ell m}(\Omega) e^{-i \omega t}\psi_{\omega\ell}(r).
\ee
As in the Rindler wedge, we can write an effective Schrodinger equation for $\psi_{\omega\ell}$.  It is again convenient to work in the tortoise coordinate, in terms of which we have
\be\label{modeeq}
-\frac{d^2}{dr_*^2}\Psi_{\omega\ell}+V(r)\Psi_{\omega\ell}=\omega^2\Psi_{\omega\ell},
\ee
with
\be\label{schpot}
V(r)=\frac{r-1}{r^3}\left(m^2r^2+\ell(\ell+1)+\frac{1}{r}\right).
\ee
In solving this equation we express $r$ implicitly in terms of $r_*$.

A considerable amount of the physics lies in the expression \eqref{schpot} for the effective potential.  Let's first consider the mass $m$; it is quite reasonable to assume for simplicity that the Compton wavelength $1/m$ is either much larger or much smaller than the Schwarzschild radius.  We will mostly be interested in the case where it is much larger, in which case we can just set the mass to zero, but I'll briefly comment on the massive case.  For $r\gg 1$ the potential just goes to a constant $m^2$, so massive modes will propagate near infinity only if $\omega \geq m$.  Since we are taking $m\gg 1$, this means that any modes whose energy $\omega$ is of order the Schwarzschild radius will be confined very near the horizon, having only exponentially small tails at infinity.  Since we have already concluded that the temperature of the black hole is of order the inverse Schwarzschild radius, we will indeed mostly be interested in modes with $\omega\approx 1$.  For this reason massive particles are usually not of much interest in black hole physics; we will from now on restrict to the case of $m^2=0$.  

\begin{figure}
\begin{center}
\includegraphics[height=4cm]{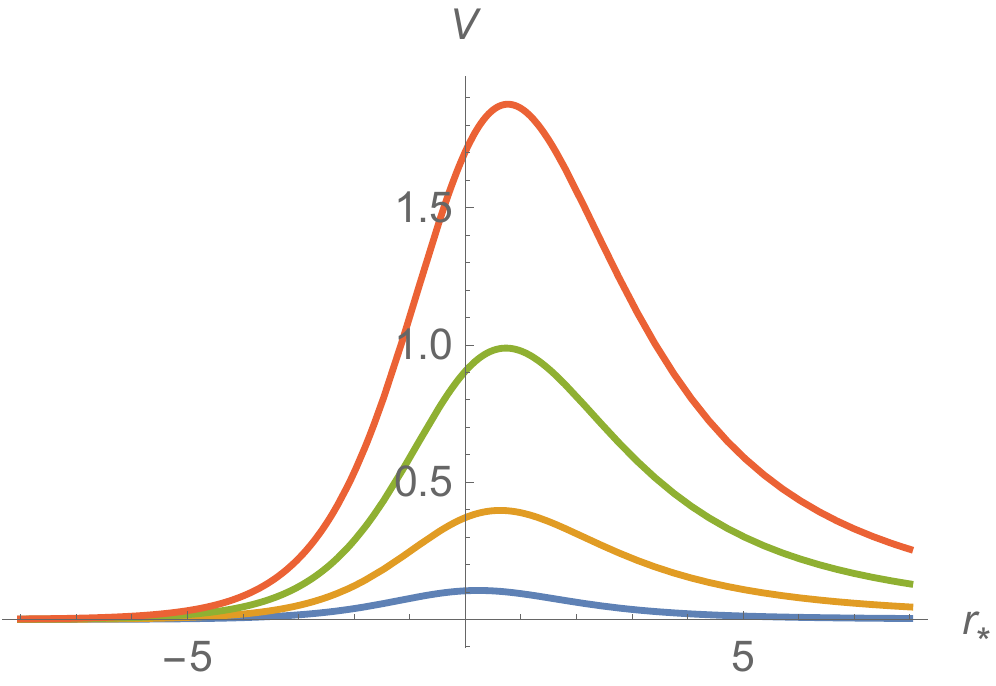}
\caption{Plots of $V$ as a function of $r_*$, for $\ell=\{0,1,2,3\}$.}\label{potentials} 
\end{center}
\end{figure}

In the massless case, the asymptotic behavior of the potential is
\be
V\approx
\begin{cases}
\frac{\ell(\ell+1)}{r_*^2} & r_*\to \infty\\
(\ell^2+\ell+1)e^{r_*-1} & r_* \to -\infty
\end{cases},
\ee
so it vanishes polynomially in $r_*$ at spatial infinity and exponentially in $r_*$ near the horizon.  In between the two regions there is a barrier.  For $\ell \gg 1$ the peak of the barrier is at $r=\frac{3}{2}$ and the height is of order $\ell^2$.  This potential is plotted for the first few $\ell$ in figure \ref{potentials}.

For modes whose energy is less than the height of the barrier, in particular those with energy of order the temperature, we can think of the barrier as dividing the black hole exterior into two regions.  For $r\gg \frac{3}{2}$ the geometry is weakly curved, and the propagating modes are just the usual ones of Minkowski space.  For $1<r<\frac{3}{2}$, there are also propagating modes but they are mostly confined to be near the horizon.  This region is sometimes called the ``thermal atmosphere'', since these modes will typically be occupied by a Boltzmann distribution with temperature $T_{Hawking}\approx 1$.  Recently people have also started to call the near-horizon region ``the zone''.  

\begin{figure}
\begin{center}
\includegraphics[height=5cm]{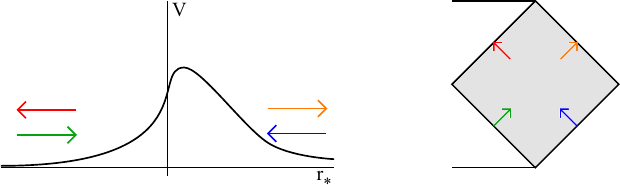}
\caption{The spacetime interpretation of scattering in the effective Schrodinger potential; on the right is a section of the Schwarzschild Penrose diagram, with the region $-\infty<r_*<r_*$ shaded in grey.}\label{schscat}
\end{center}
\end{figure}

From the effective Schrodinger equation point of view, it is clear that the relation between these two regions is a scattering problem. This is illustrated in figure \ref{schscat}.
Modes can come in either from the white hole horizon at $r_*\to -\infty$ or from $J_-$ at $r_*\to \infty$, and they can go out either through the black hole horizon at $r_*\to -\infty$ or through $J_+$ at $r_* \to \infty$.  To get a unique mode, we need to pick boundary conditions of some sort.  One obvious choice is send in particles from the right and then see if they are absorbed by the black hole; the absorption probability is just the transmission coefficient of this Schrodinger problem.  This type of mode has most of its support out in the asymptotic $r\gg 1$ region.  Another choice is to say that there is no flux coming in from the right, but allow some to come in from the left. This is a flux coming in out of the white hole and mostly being reflected off of the barrier back into the black hole, with a small amount tunneling through the barrier and transmitting out to infinity.  Modes of the latter type have most of their support in the near-horizon region, and are often called ``zone modes'' or ``modes in the zone''.  

To understand which modes are important, we clearly need some sort of prescription for the initial quantum state of the field we build out of creation and annihilation operators for them.  In the Hartle-Hawking state both are excited with a thermal distribution at temperature $T_{Hawking}$; there is a steady flux of thermal particles going in and coming out at each end of the wormhole.  The one-sided black hole made from collapse is more subtle, I'll turn to it now.

\subsection{Hawking's calculation of black hole radiation}\label{hawkrad}
\begin{figure}
\begin{center}
\includegraphics[height=5cm]{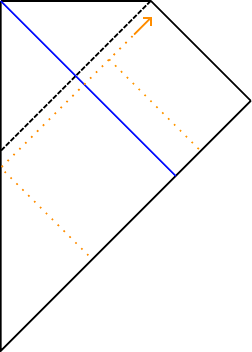}
\caption{The geometry for a black hole made from collapse.  The collapsing shell is now shown in blue, and the backwards evolution of an outgoing wavepacket at late times is shown in orange.  Part of it is scattered off of the potential barrier, and part of it goes back through the photon shell.}\label{collapse2}
\end{center}
\end{figure}

Let's now focus on a black hole created from the collapse of a matter; the Penrose diagram is shown in figure \ref{collapse2}.
 
The mode solutions we discussed in the previous section now only apply above the shell.  Below the shell they must be matched onto solutions of the ordinary Minkowski wave equation such as plane waves.  This matching will be nontrivial;  since there is no global time-translation symmetry, positive frequency modes with respect to Schwarzschild time above the shell will evolve back to mixtures of positive and negative frequency modes with respect to the Minkowski time below the shell.  In fact the negative frequency part comes entirely from the part of the solution that propagates back through the photon shell; the part that reflects off of the barrier while still in the Schwarzschild region will conserve energy and stay positive-frequency.  

We then define the quantum state in such a way that other than the shell which created the black hole, at early times the Minkowski modes below the shell are not excited.  In other words their annihilation operators will annihilate the state.  Because of the mixing between positive and negative frequency just described, the modes which are positive frequency in the Schwarzschild region \textit{will} be excited in this state.  To figure out how much one needs to figure out how to relate the creation and annihilation operators for the Minkowski modes below the shell to the creation and annihilation operators for the Schwarzshild modes above the shell.  The details are not hard but are slightly technical; they are beautifully explained in Hawking's original paper \cite{Hawking:1974sw} (and nicely reviewed in \cite{Wald:1984rg}), and I will not go through them here since there is a simpler way of getting to the final answer.  The idea is to simply guess the state of the ingoing and outgoing modes to the future of the blue shell in figure \ref{hawkrad} without worrying about what happened earlier (the fact that answer doesn't depend on what happened earlier is a warning sign, which we will return to soon).  

Indeed the state of the ingoing modes (the scattering states which are incident from the right in figure \ref{schscat}) at late times just describes any matter which is happens to be falling in, so if there is no such matter then these should be in their ground states.  The outgoing modes at the horizon however (the scattering states which are incident from the left in figure \ref{schscat}) must be thermally entangled across the horizon with their interior Rindler partners at the Hawking temperature $T_{Hawking}$, just as they would be in the Hartle-Hawking state, since otherwise there would be a large energy density at the horizon.  Thus the state of the exterior modes (which we can think of as the ``initial state'' for the scattering problem in \ref{schscat}) is
\be\label{BHmodestate}
\rho_{exterior}=|\Omega\ran\lan\Omega|_{\mathrm{ingoing}}\otimes \left(\bigotimes_{\omega \ell m}\left((1-e^{-\beta \omega})\sum_ne^{-\beta \omega n}|n\ran\lan n|_{\omega \ell m}\right)\right)_{\mathrm{outgoing}},
\ee
with $\beta=\frac{1}{T_{Hawking}}$ and $|n\ran_{\omega \ell m}$ indicates the state with $n$ particles in the mode $f_{\omega \ell m}$.  This state is sometimes called the Unruh state.  In the Unruh state the ingoing modes are in their ground states while the outging modes are thermal: this means that there is a net flux of energy away from the black hole \cite{Hawking:1974sw}!  We can compute the energy flux at infinity from the energy momentum tensor:
\begin{align}\nonumber
\frac{dE}{dt}&=-\lim_{r\to\infty}\int d\Omega r^2 \lan T^{0r}\ran\\
&=\lim_{r\to\infty}\int d\Omega r^2\lan\dot{\phi}\partial_r \phi\ran,
\end{align}
where in the second line we have used the energy-momentum tensor \eqref{scalarT} for a scalar field.  We can first consider this expectation value in the state $|1\ran_{\omega \ell m}$, meaning the state with one particle in the mode $f_{\omega\ell m}$, in which case we have (for simplicity treating $\omega$ as a discrete variable)\footnote{Here we assume that the energy-momentum tensor is ``normal-ordered'', meaning that all creation operators appear to the left of all annihilation operators, which can be accomplished by adding a (divergent) multiple of $g_{\mu\nu}$ to $T_{\mu\nu}$.}
\begin{align}\nonumber
\lan \dot{\phi}\partial_r \phi\ran &=\lan 0|a_{\omega \ell m}\left(\dot{f}_{\omega\ell m}^*\partial_r f_{\omega\ell m}+\dot{f}_{\omega\ell m}\partial_r f^*_{\omega\ell m}\right)a^\dagger_{\omega\ell m} a_{\omega\ell m}a^\dagger_{\omega\ell m}|0\ran_{\omega\ell m}\\
&=\dot{f}_{\omega\ell m}^*\partial_r f_{\omega\ell m}+\dot{f}_{\omega\ell m}\partial_r f^*_{\omega\ell m}
\end{align}
We are interested in evaluating this for $r\gg\ r_s$, so we can use the asymptotic form of the mode:
\be
f_{\omega\ell m}\approx \frac{1}{r}Y_{\ell m}(\Omega)\frac{T_{\omega\ell}}{\sqrt{2\omega}}e^{i\omega (r-t)},
\ee
where $T_{\omega\ell}$ is the transmission coefficient from the left for the scattering problem shown in figure \ref{schpot} (the factor of $1/\sqrt{2\omega}$ ensures the mode is normalized in the KG norm, and is the same factor as in \eqref{freeheis}).  In any one-dimensional potential scattering problem this transmission coefficient is the same whether the wave is incident from the left or the right (this is a consequence of time-reversal symmetry), and it is conventional to re-interpret his coefficient in terms of the absorption probability of an ingoing wave (from the right in figure \ref{schpot}) of frequency $\omega$ and angular momentum $\ell$
\be
P_{abs}(\omega,\ell)\equiv |T_{\omega\ell}|^2,
\ee
in terms of which have
\be
\dot{f}_{\omega\ell m}^*\partial_r f_{\omega\ell m}+\dot{f}_{\omega\ell m}\partial_r f^*_{\omega\ell m}\approx -\frac{\omega}{r^2}Y_{\ell m}Y_{\ell m}^*P_{abs}(\omega,\ell),
\ee
and therefore
\be
\lim_{r\to\infty}\int d\Omega r^2 \lan T^{0r}\ran=\omega P_{abs}(\omega,\ell).
\ee
This is the energy loss assuming there is one quanta in the mode $f_{\omega\ell m}$, but in the state \eqref{BHmodestate} the expected number is
\be
\lan n\ran=(1-e^{-\beta \omega})\sum_{n=0}^\infty n e^{-\beta \omega n}=\frac{1}{e^{\beta \omega}-1}
\ee
so the total energy flux is
\begin{align}\nonumber
\frac{dE}{dt}&=-\sum_{\omega\ell m}\frac{\omega P_{abs}(\omega,\ell)}{e^{\beta \omega}-1}\\
&=-\sum_{\ell, m}\int_0^\infty \frac{d\omega}{2\pi}\frac{\omega P_{abs}(\omega,\ell)}{e^{\beta \omega}-1},\label{hawkresult}
\end{align}
where in the second line we have gone back to continuum normalization for $\omega$.  Here $\beta=\frac{1}{T_{Hawking}}$, and we remind the reader that $P_{abs}(\omega,\ell)$ is the absorption probability for a ``blue'' mode of this frequency and angular momentum to be transmitted through the barrier from the right in figure \ref{schscat}.
%
$P_{abs}(\omega,\ell)$ is often called a \textit{greybody factor}, since other than this factor \eqref{hawkresult} is just the standard formula for radiation from a black body at temperature $T=\beta^{-1}$ into the vacuum.\footnote{Including this factor it is the standard formula for radiation from an imperfect black body that has some probability to absorb incoming modes.  It is derived by demanding that the absorption and emission probabilities are related in such a way that the body can be in thermal equilibrium with an external radiation field.}  We saw earlier that the height of the potential barrier grows like $\ell^2$, so the emission is dominated by the modes with the very lowest angular momentum.    

We are thus led to the following picture of the quantum state of a field near a black hole formed from collapse: after an initial ringdown period, the modes in the atmosphere are in a quasi-stationary state where they are excited thermally, with the low-$\ell$ modes gradually carrying energy out to infinity by tunneling through the barrier.  

Any massless field that happens to be around will carry away energy in this manner, but those with higher spin will carry less \cite{Page:1976df}; a simple way to understand this is that particles with spin can only radiate into modes with $\ell>0$ (dipole radiation for photons, quadrupole for gravitons, etc.), so they encounter higher potential barriers.  The largest fraction of the energy is thus carried away by whatever the lowest spin massless particles are; in our universe it would be photons.  A smaller, but still $O(1)$ fraction would be radiated into gravitons.    

There is a heuristic explanation of Hawking radiation that is occasionally brought up.  The idea is that entangled pairs of particles are constantly jumping into existence near the horizon via vacuum fluctuations, and sometimes one of them falls in and one of them gets out.  This cartoon has several problems if it is taken literally, among them that the ``particles'' involved have wavelengths comparable to the size of the black hole and the Hawking process isn't really stochastic (we are in a pure quantum state), so it should be used with care if at all.  

\subsection{Evaporation}\label{tevap}
In the QFT-in-curved-spacetime limit we have been considering so far, the mass of the black hole, and therefore its energy, is infinite.  The black hole can radiate away a constant energy flux forever without decreasing in size.  This is clearly unphysical, but to fix it we are now going to need to restore dynamical gravity.  Since the black hole mass will now be time dependent, it is obviously absurd to continue setting the Schwarszchild radius $r_s=2GM$ to one.  

Before discussing the decrease of the mass due to evaporation, I want to very briefly comment on the seemingly obvious fact that we should think of the mass $M$ of a black hole as its energy.  So far we have somewhat cavalierly defined the energy as the generator of $t$ translations in the Schwarzschild geometry, but in general relativity such a coordinate-dependent definition cannot really be correct.  There is a long and very interesting story about this which I unfortunately do not have time to get into, but the upshot is that in a proper Hamiltonian formulation of general relativity in asymptotically flat space, the energy is defined as a certain boundary integral on a two-sphere at $r\to\infty$, or more rigorously as a boundary integral on $i^0$ in the Penrose diagram \cite{Regge:1974zd}.  It is usually called the ADM energy, in honor of Arnowitt, Deser, and Misner, who were the first people to write it down and realize its relevance to a Hamiltonian formulation \cite{Arnowitt:1962hi}.  For our purposes though, it is enough just to know that the ADM energy is conserved and that a black hole of mass $M$ formed from the collapse of a shell \textit{does} have ADM energy $M$. 

We are now in a position to estimate the lifetime of a black hole of mass $M$.  The total energy flux leaving the black hole is
\be
\frac{dE}{dt}=-\sum_{\ell,m}\int_0^\infty \frac{d\omega}{2\pi}\frac{\omega P_{abs}(\omega,\ell)}{e^{\beta \omega}-1}.
\ee   
Unfortunately computing $P_{abs}$ involves solving the differential equation \eqref{modeeq}, which can't be done analytically.  Page has solved it numerically to compute exact lifetimes \cite{Page:1976df}, but for our purposes a simple estimate is enough.  For $\omega \lesssim \frac{1}{r_s}$ we can approximately solve the equation to find $P_{abs}(\omega,\ell=0)\sim (\omega r_s)^2$.  Higher $\ell$ modes have $P_{abs}$ proportional to some higher power of $(\omega r_S)$ in the same limit, so roughly we can neglect the $\ell>0$ terms in the sum and use this approximation for $\ell=0$ to find
\be
\frac{dE}{dt}\approx -\frac{C}{r_s^2},
\ee
where $C$ is some constant that this crude method is unable to compute.  Amusingly this is consistent with what a naive application of the Stefan-Boltzmann law $\frac{dE}{dA dt}=\sigma T^4$ would predict, although the constant factor is different.  The mass of the black hole as a function of time thus obeys the differential equation
\be
\frac{dM}{dt}=-\frac{C}{(GM)^2},
\ee
so a black hole of initial mass $M$ will evaporate in a time
\be
t_{evap}\sim G^2 M^3.
\ee
For a solar mass black hole this time is of order $10^{66}$ years.  This can be compared for example to the age of the universe, which is about $13.8$ billion years.\footnote{As a side comment, this should more accurately be called the time since the beginning of nucleosynthesis.  The actual age of the universe is unknown and could quite possibly be infinite.}  It seems that if we are ever to do experiments to test this prediction, we had better find a way to make black holes which are smaller than those produced astrophysically!
\begin{figure}
\begin{center}
\includegraphics[height=5cm]{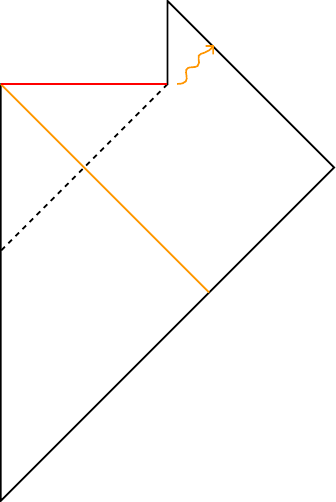}
\caption{The Penrose diagram for an evaporating black hole formed by a collapse of a shell of photons.  The Hawking radiation seems to all come out at once, but this is only an illusion arising from the conformal transformation.  The singularity is shown in red, and the two vertical black lines both represent a non-singular origin of polar coordinates.  The dashed line is the event horizon, which exists even though the black hole evaporates since there are still points that cannot send messages to $i^+$.}\label{evappen}
\end{center}
\end{figure}

Hawking evaporation means that our old Penrose diagram \ref{collapse} needs to be revisited, I show the improved diagram in figure \ref{evappen}.

\subsection{Entropy and thermodynamics}\label{entsec}
If a black hole has a temperature and an energy, it must also have an entropy.  Recall that the standard definition of temperature in statistical mechanics is
\be
\frac{dS}{dE}=\frac{1}{T}.  
\ee
For the black hole we have $T=\frac{1}{8\pi G M}$, so identifying $M=E$ and assuming that $S(E=0)=0$ we find
\be\label{bhent}
S=\frac{A}{4G}=2\pi \frac{A}{\ell_p^2},
\ee
where $A=4\pi r_s^2$ is the area of the horizon and $\ell_p=\sqrt{8\pi G}$ is the Planck length.  This entropy is enormous, of order $10^{78}$ for a solar mass black hole, which is much larger than the entropy of the sun, which is of order $10^{60}$.  In fact the entropy of the entire observable universe excluding black holes, which is dominated by the cosmic microwave background photons, is of order $\left(\frac{10^{10} pc}{1mm}\right)^3\approx 10^{87}$, while the entropy of a single $10^{6}$ solar mass black hole like that in the center of the milky way is of order $10^{88}$.  The largest supermassive black holes have masses of order $10^{10}$ solar masses, and thus entropies of order $10^{96}$.

Historically it was actually the black hole entropy which was discovered before the temperature. In classical general relativity, one can prove under rather general assumptions that the area of an event horizon can never decrease in time \cite{Hawking:1971tu}.  This property is reminiscent of the second law of thermodynamics, and if we formally define an entropy proportional to the horizon area and a temperature of order $\frac{1}{r_S}$, then a first law of thermodynamics $dM=TdS$ is also satisfied \cite{Bekenstein:1973ur,Bardeen:1973gs}.  Bardeen, Carter and Hawking viewed this as only a mathematical analogy \cite{Bardeen:1973gs}, but it was the point of view of Hebrew University's own Jacob Bekenstein that this entropy should actually represent the statistical entropy of the black hole in the sense of counting the number of ways the black hole could have formed \cite{Bekenstein:1973ur,Bekenstein:1974ax}.  Bekenstein argued the entropy should be given by some constant multiple of the horizon area in Planck units, and provided evidence for this by considering various thought-experiments where an entropic system is thrown into a black hole.  He saw that in each case the black hole entropy defined in this way always increased more than the exterior entropy decreased from losing the system.  Bekenstein called this observation the \textit{Generalized Second Law}, and conjectured that it was true in general.\footnote{The name is a bit misleading.  If we take seriously Bekenstein's suggestion that black hole entropy is real, then his ``generalized'' second law is really just the ordinary second law.}  It was at this point that Hawking's paper on the temperature appeared, closing the circle.  For these reasons the quantity \eqref{bhent} is usually called the Bekenstein-Hawking entropy.

The idea that the Bekenstein-Hawking entropy counts microstates has found quite strong support in string theory, a set of ideas that has produced many insights into quantum gravity in the last few decades \cite{Polchinski:1998rq,Polchinski:1998rr}.  General arguments based on counting the states of a long vibrating string are able to produce the area-scaling of \eqref{bhent} in a wide variety of situations \cite{Susskind:1993ws,Horowitz:1996nw}, and in certain special supersymmetric cases \cite{Strominger:1996sh} one is actually able to sharpen these arguments enough to compute the numerical coefficient analogous to the $\frac{1}{4}$ in equation \eqref{bhent}.  

\subsection{The information problem}\label{infpsec}
We have now seen that black holes behave like thermodynamic objects in many respects, and it seems almost irresistible, following Bekenstein, to take the point of view that Bekenstein-Hawking formula indeed counts the logarithm of the number of microstates of a black hole of a given size.  Let us say for a moment however that we nonetheless insist that in fact black holes are really only distinguished by their mass (and charge and angular momentum if we had included these), as suggested by general relativity.  We then find ourselves in tension with a basic principle of physics: if we know the state of a system at some time, we should be able to infer its initial state by running the dynamics backwards.  This proposal thus is really saying that by creating a black hole, we destroy most of the information about how we made it.  Since we would like to preserve this principle, we are naturally led to assume that indeed black holes have microstates.  In one of the most remarkable papers in the last 50 years however, Hawking brilliantly argued that, once we allow the black hole to evaporate, this assumption is \textit{not sufficient} to avoid the destruction of information.  

Before explaining Hawking's argument, it is worth saying a bit more about the principle of information conservation.  In classical mechanics time evolution is generated by Hamiltonian flow on phase space, which can always be reversed by changing the sign of the Hamiltonian.  More prosaically we can just solve the equations of motion backwards.  Similarly in quantum mechanics, time evolution is described as unitary evolution in Hilbert space, which can again be reversed by switching the sign of the Hamiltonian.  In both cases the principle applies only if the system is isolated; otherwise information can leak out.  The quantum case is slightly counter-intuitive since the measurement process is non-deterministic, but measurement always involves coupling the system being measured to some external apparatus; the evolution of the joint system remains unitary and deterministic.  



Now consider a black hole that was formed by a shell of matter in some pure quantum state $|\Psi\ran$.  After the initial collapse settles down, the state of the outgoing modes is givey by \eqref{BHmodestate}.  As time goes on, the quantum state of the radiation field outside becomes more and more mixed, which we can quantify by saying that its entanglement entropy is increasing.\footnote{For more details on the basic properties of pure and mixed states, as well as entanglement entropy, see appendix \ref{BPapp}.}  This may not seem so bad at first, since after all in looking at the late radiation we are looking just at the part of the state which is outside of the black hole horizon.  As the black hole evaporates however, it decreases in size until at some point it becomes Planckian.  \eqref{BHmodestate} is supposed to be accurate up to corrections of order $\frac{m_p}{M}$, so until this point the entanglement entropy of the radiation field outside continues to increase.  At this point one of two things must happen:
\begin{itemize}
\item[(1)] The evaporation stops, and the Planck-sized object just sits around.  This possibility is called a \textit{remnant}.  For the total state to remain pure, as required if the evolution of the system is to be unitary, the remnant must have an extraordinary amount of entanglement entropy. Even before it becomes Planck-sized its entanglement entropy would need to exceed the Bekenstein-Hawking value, which would violate the state-counting interpretation of \eqref{bhent}.
\item[(2)] The black hole finishes evaporating into ordinary quanta such as photons and gravitons.  Energy conservation prevents the final burst of quanta from containing nearly enough entanglement entropy to purify the earlier radiation, so the end result of the evaporation process is a mixed state of the radiation field whose entropy is of order the initial black hole horizon area in Planck units. 
\end{itemize}
Hawking argued for option (2), claiming that the process of black hole formation and evaporation cannot be described by a unitary map from the ingoing shell to the outgoing radiation, since this would have resulted in a pure quantum state of the radiation.  Moreover since different initial states result in the same final state, he claimed that black holes violate the principle of information conservation.  Option (2) is thus usually referred to as \textit{information loss}.   

Basic physical principles are not often discarded, and this should only be done once it is clear that we have no other choice.  How might information loss be avoided?  Option (1) is in principle possible, but is rather vile since it requires objects with finite energy but an infinite number of states, and has rarely been taken seriously \cite{Preskill:1992tc,Giddings:1994qt,Susskind:1995da}.  
  Most people who are unwilling to accept information loss have instead gone for a third option:
\begin{itemize}
\item[(3)] Equation \eqref{BHmodestate} is correct only in a coarse-grained sense; the Hawking radiation does not actually come out in a mixed state.  The information is carried out in subtle correlations between the Hawking photons, and the final state of the evaporation is a pure state of the radiation field.  Because it is a complicated state any small subsystem looks thermal, justifying the approximate validity of \eqref{BHmodestate} if we don't look at too many photons at once.  There is a complete basis of such pure states whose dimensionality is the exponential of the Bekenstein-Hawking formula, which thus can indeed be interpreted as counting microstates.
\end{itemize}
Option (3) may obviously seem like the best of the three, but it is more radical than it appears at first.  Equation \eqref{BHmodestate} seems to follow from very widely held assumptions about the validity of quantum field theory on scales that are large compared to the Planck scale.  If it is wrong, then shouldn't this violation of quantum field theory be detectable in other ways?  Since all three options thus have unappealing features, this state of affairs is referred to as the \textit{black hole information problem}.

There is however at least one reason why one might expect substantial corrections to \eqref{BHmodestate} \cite{'tHooft:1984re,Unruh:1994zw,Corley:1996ar}.  Returning to figure \ref{collapse2}, consider the limit where we have the orange outgoing mode coming out later and later.  If it comes out more than a time\footnote{In this name $scr$ stands for ``scrambling''. The reason is that this is the timescale it takes for perturbations of the black hole to die down to Planckian size; it is the time it takes them to be ``scrambled'' by the black hole horizon.  I'll discuss this further in section \ref{scrambsec} below.}
\be\label{tscrambling}
t_{scr}\equiv r_s\log \frac{r_s}{\ell_p}
\ee
after the initial shell falls in, then as we evolve it back in time its collision with any of the photons that make up the initial shell happens at a center of mass energy that is greater than the Planck scale.  Intuitively this is because near the horizon in tortoise coordinates the proper distance to the horizon behaves like $e^{r_*}$, so from \eqref{nullg} we see that wave packets which come out of the potential barrier at times that are later than \eqref{tscrambling} after the formation of the black hole will have come out from within a Planckian distance of the horizon.  In Hawking's calculation these modes are being ``pulled out of the vacuum'' near the horizon from an imaginary reservoir of trans-Planckian degrees of freedom;  this is called the \textit{Trans-Planckian problem}.

To see the trans-Planckian collision with the initial shell more explicitly,  we can note from \eqref{nullg} that an ingoing null geodesic that crosses the potential barrier at time $t=0$ intersects an outgoing geodesic that crosses the potential barrier at time $t_{out}>>r_s$ at
\be
r_{collision}\approx 1+e^{-\frac{t_{out}}{2r_s}}.
\ee
To compute the energy of the collision we need the four-momenta of the two photons in Schwarzschild coordinates, which are given by\footnote{This is determined by finding an appropriate affine parametrization of the null geodesics in \eqref{nullg}, as discussed in appendix \ref{GRapp}.}
\be
p^\mu_\pm=E_\pm\left(\frac{r}{r-r_s},\pm 1,0,0\right),  
\ee
where $E_\pm$ are the energies of the massless particles.  The center of mass energy of the collision is then
\be
E_{COM}=\sqrt{-\frac{1}{2}p_+^\mu p_-^\nu g_{\mu\nu}}\approx\sqrt{E_+ E_-}e^{\frac{t_{out}}{4r_s}}.
\ee
The outgoing photon will always have $E_+\sim 1/r_s$, which is also a lower bound for $E_-$ since otherwise the infalling photon wouldn't have fit into the black hole the first place, so we see that indeed after a time \eqref{tscrambling} the center of mass collision energy is Planckian.  

There has been much debate over whether or not the trans-Planckian problem is a serious criticism of Hawking's argument for information loss.  There is a well-known candidate rebuttal called the ``nice-slice'' argument \cite{Polchinski:1995ta}; basically one argues that we can just do Hawking's calculation in Kruskal coordinates where the trans-Planckian origin of the Schwarzschild modes is absorbed into the standard renormalization of quantum field theory in curved space time.  Although this prescription gives a well-defined procedure that reproduces Hawking's calculation for the late time state of the radiation, in my view it is not completely satisfactory since it does not get rid of the fact that projecting onto possible final states of the late-time Hawking radiation produces states with a genuine high energy collision in the past. The renormalization procedure used in the nice-slice argument does involve making an arbitrary choice about physics at high energy scales, and, unlike in most situations where quantum field theory is used, the redshifting of the black hole geometry allows low-energy properties of the state at later times to depend on this choice.\footnote{It is sometimes argued that the adiabatic theorem of quantum mechanics prevents us from making other choices that would lead to different results for the Hawking radiation \cite{Polchinski:1995ta}, but the adiabatic theorem applies only to the global conserved energy, not to the center of mass energy of localized excitations.  Projections on the late Hawking radiation will not appreciably change the energy, since any localized excitations that are created this way are redshifted by the horizon.}  I personally believe that the trans-Planckian problem gives a plausible excuse for why Hawking's calculation might not be completely correct for times longer than $t_{scr}$.  

You might ask why we don't just conclude from this that \eqref{BHmodestate} is \textit{completely} wrong for $t>t_{scr}$, for example the black hole could just explode at $t=t_{scr}$.  This is actually ruled out experimentally, the black hole in the center of our galaxy would have already exploded, but we could imagine something more mild.\footnote{For another observational argument, the trans-Planckian problem also exists in the early univere during inflation, but nothing too terrible seems to have come out of it.}  Anything along these lines besides option (3) however would mean that we should not really think of the black hole as a complex thermal system with entropy $S_{BH}$; the apparent successes of black hole thermodynamics would be a mirage.  Aesthetically this is rather unappealing, since so far all cases of systems that behave thermodynamically ultimately have their explanations in statistical mechanics.  This would also be in tension with the string theory microstate counting arguments mentioned in the previous subsection, and we will see powerful evidence in favor of unitary evaporation with approximate thermality from the AdS/CFT correspondence in section \eqref{adssec} below.  For now I will thus provisionally adopt the point of view that option (3) is correct.  
 
In the final two parts of this section I discuss two important aspects of QFT in black hole backgrounds which don't quite fit into the main flow of these notes but are in my opinion too important to leave out.  Casual or first-time readers may wish to skip these final two subsections and proceed to section \ref{evapsec}

\subsection{The brick wall model and the stretched horizon} \label{bricksec}
To get some intuition for black hole thermodynamics, it is interesting to study the thermodynamics of a scalar quantum field in a black hole background.\footnote{In this subsection I will again set $r_s=1$ to simplify formulas.}  We've already seen that the region outside the horizon can be understood in terms of the modes $f_{\omega \ell m}$, and Hawking's analysis says that all of these modes are excited thermally.  We then roughly have total energy
\be\label{bwE}
E=\sum_{\omega \ell m}\frac{\omega}{e^{\beta\omega}-1}
\ee
and entropy
\be\label{bwS}
S=\sum_{\omega \ell m}\left[\frac{\beta\omega}{e^{\beta\omega}-1}-\log\left(1-e^{-\beta\omega}\right)\right],
\ee
where $\beta$ is the inverse temperature.  These expressions are not really well-defined; it is not clear what is meant by $\sum_{\omega}$.  It can't just be $\int d\omega$, since this wouldn't have the right units.  This ambiguity reflects something physical; these quantities are both UV and IR divergent.  The IR divergence arises from the infinite volume of flat space in the region where $r\to \infty$, and the UV divergence arises from the near-horizon region.  The former can be regulated by putting the black hole in a large box, while the latter are presumably regulated by some sort of Planckian physics close to the horizon.  

As a simple model, 't Hooft has suggested implementing the near-horizon cutoff by simply putting Dirichlet boundary conditions $\phi=0$ on a surface a Planckian distance from the horizon \cite{'tHooft:1984re}.  More explicitly one demands the field vanishes at $r=r_{min}$, which we can express in terms of the proper distance $\epsilon$ from the horizon as
\be
r_{min}\equiv 1+\frac{\epsilon^2}{4}.
\ee
This is called the ``brick wall'' model.  It clearly is not a good model from the point of view of the full black hole geometry; we don't think the infalling observer will actually encounter a brick wall!  The actual cutoff imposed by quantum gravity is undoubtably more subtle.  Nonetheless the brick wall does give a physically motivated way to discretize the sum over $\omega$, allowing estimates for \eqref{bwE} and \eqref{bwS}.  

The IR box is a little awkward to deal with explicitly, but it can be dispensed by instead including only the modes which have no incoming piece outside of the barrier; these are the ``zone modes'' of section \ref{modesec}.  By doing this we are throwing out the contributions to the thermal energy and entropy of the radiation field in the region $r\gg 1$, but these should not be considered as part of the black hole anyway.  The quantization of $\omega$ is also messy to derive in detail, but there is a simple way to estimate it.  Returning for a moment to tortoise coordinates, the brick wall is at
\be
r_{*min}\approx 2\log\frac{\epsilon}{2}.
\ee
I will focus on modes with $\ell\gg 1$ and $\omega\lesssim 1$, since these will dominate the thermodynamic ensemble.\footnote{Modes with larger $\omega$ are Boltzmann suppressed since we will take $\beta=\frac{1}{T_{Hawking}}\sim 1$, and modes with low $\ell$ are entropically suppressed.}  The turning point where the potential barrier becomes important and the mode begins to decay exponentially is at
\be
r_{*turn}\approx 2\log \frac{\omega}{\ell}.
\ee
We can thus approximate this mode problem as the Schrodinger problem of a particle in a box of size
\be
\Delta r_*\approx 2\log \frac{2\omega}{\ell \epsilon},
\ee
where I have neglected various order one factors.  We then have a quantization condition
\be
\omega_n\approx \frac{\pi n}{2\log \frac{2\omega_n}{\ell\epsilon}},
\ee
which allows the replacement
\begin{align}\nonumber
\sum_{\omega \ell m}f(\omega)&\approx 4\int_0^\infty \frac{d\omega}{2\pi} f(\omega)\int_0^{\frac{2\omega}{\epsilon}} d\ell\,(2\ell+1) \log\frac{2\omega}{\ell\epsilon}\\
&\approx \frac{8}{\epsilon^2}\int_0^\infty \frac{\omega^2 d\omega}{2\pi}f(\omega).
\end{align}
We may finally then compute the energy and entropy
\begin{align}\nonumber
E&\approx \frac{r_s}{960\pi \epsilon^2}\\
S&\approx\frac{r_s^2}{180\epsilon^2},
\end{align}
where I've set $\beta=\frac{1}{T_{Hawking}}=4\pi r_s$ and restored $r_s$.  These results are manifestly divergent as we remove the short distance cutoff $\epsilon$, but on physical grounds we should probably not take $\epsilon$ to be any smaller than the Planck length $\ell_p=\sqrt{8\pi G}$. Indeed if we choose $180\pi \epsilon^2=G\sim \ell_p^2$ we then have
\begin{align}\nonumber
E=\frac{3}{8} M\\
S=\frac{A}{4G}.
\end{align}
With this choice the entropy of the field is the full Bekenstein-Hawking entropy $S_{BH}$ of the black hole, and its energy is comparable to the full black hole energy $M$.  Taking $\epsilon\gtrsim \ell_p$ is thus essential in ensuring that the fields do not carry \textit{more} energy and entropy than the black hole itself.  

In fact in this model we have essentially \textit{replaced} the black hole with the field theory degrees of freedom near the horizon, and there is nothing left for the black hole to do.  Of course this choice of $\epsilon$ was rather arbitrary, and by making it larger we can imagine a separation of degrees of freedom into QFT degrees of freedom at distances greater than $\epsilon$ from the horizon and ``quantum gravity'' degrees of freedom closer in.  In this more general model, one imagines a dynamical membrane a Planck distance away from the horizon, which is usually called the \textit{stretched horizon}.  The stretched horizon then carries most of the black hole entropy and energy, and in a unitary theory it absorbs infalling matter and re-emits it later in scrambled form.  In this picture the stretched horizon is in thermal equilibrium with the QFT modes in the atmosphere, and evaporation happens because these modes occasionally tunnel out to infinity.    

Clearly any such model cannot be taken too seriously; a real brick wall or stretched horizon would be detectable by an infalling observer who crosses the horizon.  This may seem like a trivial comment, but as we will see in section \ref{paradoxsec} below, harmonizing a unitary description of black hole evaporation with a smooth experience for an infalling observer is proving more difficult than many people expected.  In any event these models make it clear that there is a somewhat arbitrary distinction between which degrees of freedom are counted as being ``part of the black hole'' and which are ``part of the atmosphere''.

\subsection{The Euclidean black hole}\label{eucblacksec}
I now return to the full two-sided Schwarzschild geometry of figure \ref{kruskfig}.\footnote{In this subsection evaporation is unimportant, so I will continue to use units where $r_s=1$.}  In the Rindler decomposition of Minkowski space, we found the ground state by evaluating the Euclidean path integral \eqref{epath}.  We'd like to do the same for quantum fields in the Schwarzschild geometry, but to do this we need to identify an appropriate Euclidean version of the geometry to evaluate the path integral on.  This geometry should be a spherically symmetric solution of the Euclidean Einstein equation $R_{\mu\nu}=0$ with asympotically flat boundary conditions; since we already know that the Schwarzschild geometry is the unique such geometry in Lorentzian signature it is natural to find its Euclidean version by simple analytic continuation $t\to t_E$:
\be
ds^2=\frac{r-1}{r}dt_E^2+\frac{r}{r-1}dr^2+r^2 d\Omega_2^2.
\ee  

\begin{figure}
\begin{center}
\includegraphics[height=6cm]{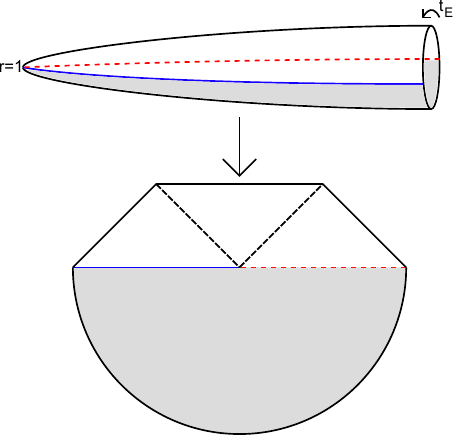}
\caption{The Hartle-Hawking construction.  The upper diagram is the Euclidean Schwarzschild ``cigar'', with the two-sphere suppressed.  The wave functional of the quantum fields in the Lorentzian Schwarzschild geometry is constructed by evaluating the Euclidean path integral on the lower half of the cigar and then using this as initial data, as shown in the lower diagram.}\label{cigar}
\end{center}
\end{figure}

This geometry still appears to be singular at $r=0$ and $r=1$, but here a surprise is in order.  If we define a new coordinate 
\be
d\rho=\sqrt{\frac{r}{r-1}}dr,
\ee
then near the horizon ($\rho\to 0$) we have
\begin{align}\nonumber
r&\approx 1+\frac{\rho^2}{4}\\
ds^2&\approx d\rho^2+\frac{1}{4}\rho^2 dt_E^2+d\Omega_2^2. 
\end{align}
The first two terms look very much like the origin of polar coordinates in $\mathbb{R}^2$, and in fact if we were to decide to have $t_E$ be an angular variable with period $4\pi$ then the apparent singularity at $\rho=0$ would be resolved by the geometry capping off smoothly.  Remarkably, this means that the curvature singularity at $r=0$ has been completely excised, and we are left with an entirely nonsingular geometry!  This is illustrated in the upper part of figure \ref{cigar}.

By cutting this Euclidean geometry in half and evaluating the Euclidean path integral over the lower half, we can define the Hartle-Hawking wave functional for the Lorentzian Schwarzschild geometry \cite{Hartle:1976tp,Israel:1976ur}.  By the same arguments we used for the Rindler decomposition, this allows a simple explicit expression for the quantum state of the fields:
\be\label{hhs}
|\Psi\ran_{HH}\propto\sum_ie^{-\beta E_i/2}|i^*\ran_L |i\ran_R. 
\ee
Here $i$ labels eigenstates of the Schwarzschild Hamiltonian in the left and right exteriors respectively, and $*$ indicates CPT conjugation.\footnote{CPT here acts as $\Theta^\dagger \Phi_L(t,r,\Omega)\Theta=\Phi_R^\dagger(-t,r,\Omega)$.  In the free case the modes which are entangled are related by $m\to -m$.}  As in the Rindler case, this derivation does not assume that the quantum fields are free. 

Tracing out one of the two sides we immediately see that in the Hartle-Hawking state the reduced density matrix on either side is the thermal density  matrix, with temperature
\be
T_{Hawking}=\frac{1}{4\pi}
\ee
as expected.  Indeed this type of state was suggested outside of the context of black holes as a method for doing thermal field theory calculations \cite{Takahashi:1996zn}, and in this broader context it is usually called the \textit{Thermofield Double State}.

This derivation of the black hole temperature is much more compact than that presented in section \ref{hawkrad}, and one might expect that a similar derivation is possible for the entropy.  This is indeed the case \cite{Gibbons:1976ue}, but the analysis is more subtle and involves some handwaving.  It also involves more advanced geometry than the rest of these notes, and the reader who is unfamiliar with general relativity may wish to skip the remainder of this section.  The basic idea is to interpret the path integral over the \textit{full} Euclidean Schwarzschild geometry as a partition function.  In ordinary field theory this is the standard observation that
\be\label{partf}
Z(\beta)\equiv \mathrm{Tr} e^{-\beta H}=\int \mathcal{D}\phi \int_{\hat{\phi}(t_E=0)=\phi}^{\hat{\phi}(t_E=\beta)=\phi} \mathcal{D}\hat{\phi}e^{-I_E},
\ee 
or in other words that the partition function can be computed from a Euclidean path integral that is periodic in Euclidean time.  Once we have computed the partition function it is straightforward to compute the energy and entropy via the standard formulae
\begin{align}\nonumber
E&=-\frac{Z'}{Z}\\
S&=\beta E+\log Z.\label{statmech}
\end{align}

We'd like to compute the black hole entropy along similar lines, but it is clear that in order to get the Bekenstein Hawking entropy $S=\frac{A}{4G}$ it is insufficient to only integrate over a single scalar field; there is nowhere for the $G$ to come from.  This isn't a surprise; in computing the partition function we must take into account \textit{all} dynamical fields, including the metric.  Once we start integrating over metrics, it is not immediately clear what family of geometries to integrate over.  Roughly we want them to be be asymptotically flat with periodic Euclidean time.  To make this condition precise it is convenient to introduce an explicit boundary in the spacetime by cutting off the geometry at some finite two-sphere radius $r_c$.  We then integrate over compact geometries whose only boundary has topology $\mathbb{S}^1\times \mathbb{S}^2$ and induced metric
\be\label{minkbmet}
ds^2=dt_E^2+r_c^2 d\Omega_2^2.
\ee
Here the periodicity of $t_E$ is taken to be $\beta$.  Two such geometries are the piece of the Euclidean Schwarzschild geometry with $r\leq r_c$, which I'll call $g_{sch}$, and ordinary Euclidean $\mathbb{S}^1\times \mathbb{R}^3$ in cylindrical coordinates
\be
ds^2=dt_E^2+dr^2+r^2d \Omega_2^2,
\ee
again with $r\leq r_c$ and $t_E$ periodicity $\beta$, which I'll call $g_{flat}$.  

It so happens that the leading order behavior in $\frac{M}{m_p}$ of the partition function is not affected by matter fields, so we just want to compute the path integral 
\be
Z(\beta)=\int \mathcal{D}g e^{-I_E}
\ee 
over the set of Euclidean compact geometries with boundary metric \eqref{minkbmet}.  The Euclidean action is
\be\label{eucact}
I_E=-\frac{1}{16\pi G}\int_{\mathcal{M}} d^4x \sqrt{g}R-\frac{1}{8\pi G}\int_{\partial \mathcal{M}} d^{3}x \sqrt{\gamma}K,
\ee
where $\mathcal{M}$ denotes some compact manifold with a boundary $\partial \mathcal{M}$.  The topology of this manifold is not fixed, we should sum over different topologies.  In particular $g_{sch}$ has the topology of a disk times a two-sphere while $g_{flat}$ has the topology of a circle times a three-dimensional ball.  $\gamma_{\mu\nu}$ is the induced metric on the boundary, and $K$ is the trace $\gamma^{\mu\nu}K_{\mu\nu}$ of the extrinsic curvature tensor $K_{\mu\nu}=\nabla_\mu n_\nu\equiv \partial_\mu n_\nu-\Gamma^\alpha_{\phantom{\alpha}\nu\mu} n_\alpha$, where $n_\nu$ is the outward-pointing normal vector at the boundary.  This extra term is necessary to include in the gravitational action when a boundary is present if we want to fix the induced geometry on the boundary \cite{Gibbons:1976ue}.  To leading order in the $\frac{m_p}{M}$ expansion we can compute this path integral by saddle-point techniques: 
\be\label{saddlept}
Z[\beta]\approx \sum_{g_{cl}} e^{-I_E[g_{cl}]},
\ee
where $g_{cl}$ correspond to geometries which solve the classical equations of motion. Here there are two contributions; $g_{sch}$ and $g_{flat}$,  and their Euclidean actions are
\begin{align}\nonumber
I_E[g_{flat}]&=-\frac{\beta r_c}{G}\\
I_E[g_{sch}]&=I_E[g_{flat}]+\frac{\beta^2}{16\pi G}.
\end{align}

The dominant contribution to \eqref{saddlept} thus comes from $g_{flat}$.  This is not surprising; we already know that black holes in asymptotically flat space evaporate, so it must be that a gas of radiation in flat space dominates the thermal ensemble.  If we are nonetheless interested in understanding the subleading contribution of the black hole to the ensemble, it is natural to look at the contribution to the partition function from $g_{sch}$.  Even in the black hole geometry however there is a large contribution from thermal excitations of gravitons far away from the black hole, which we do not want to include as part of the black hole.  We can remove both of these effects by including only $g_{sch}$ in the sum over solutions and subtracting from its action the action of $g_{flat}$, so we find that the partition function just of the black hole is \cite{Gibbons:1976ue}\footnote{This argument is admittedly rather vague; we will see in section \ref{adssec} below that for black holes in Anti-de Sitter space an analogous argument can be justified more rigorously.}
\be
Z_{BH}(\beta)\approx e^{-\frac{\beta^2}{16\pi G}}.
\ee
From \eqref{statmech} we then have
\begin{align}\nonumber
E&=M\\
S&=\frac{A}{4G},
\end{align}
as expected.  

It is instructive to compare what happened here to the brick wall model of the previous section.  We could have also included a scalar field here, whose saddle point would have been $\phi=0$, and its one-loop determinant would have produced a UV-divergent contribution to the partition function that would match the brick wall model result.  In the Euclidean formalism however we interpret this contribution (and a similar one from gravitons) as a \textit{renormalization} of $G$, which combines with the ``bare'' contribution from the gravitational action in such a way that the entropy becomes $\frac{A}{4G}$ with the renormalized $G$ \cite{Susskind:1994sm,Demers:1995dq}.  This is a nontrivial statement since the renormalization of $G$ in a given cutoff scheme can also be computed by using Feynman diagrams for gravitational scattering \cite{Demers:1995dq}; it guarantees that the Bekenstein-Hawking formula for the entropy is independent of the arbitrariness of how we divide up the black hole and the atmosphere.  Although it is somewhat mysterious, the Euclidean gravity path integral is apparently a quite powerful method for extracting the thermodynamic properties of gravitational systems.  

\section{Unitary evaporation}\label{evapsec}
In this section I will explore some consequences of the assumption that black hole evaporation is unitary.  In this section I will work in Planckian units, where $8\pi G=1$.  Up to order one constants, this means that we can collect the results of the previous section as
\begin{align}
T&\sim\frac{1}{M}\\
S&\sim M^2\\
t_{evap}&\sim M^3.
\end{align}

Starting in this section I will begin to use more techniques from quantum information theory, so the reader who is unfamiliar with these techniques may wish to consult the appendices as needed.  Readers who are interested in getting to firewalls as quickly as possible may want to skip subsections \ref{testUsec}-\ref{scrambsec} on a first reading, but subsection \ref{compsec1} is essential.

\subsection{The S-matrix}\label{smatsec}
So far we have mostly been working in the limit of quantum field theory in curved spacetime, with the gravitational coupling $G$ taken to zero compared to any other scale in the problem.  In this limit we are able to make precise sense out of local ideas such as a field operator at a particular point in the spacetime.  Operationally this is possible because in this limit one can imagine arbitrarily precise rods and clocks, which can be used to determine the background as accurately as one would like.  Once we allow nonzero $G$ however, we immediately run into the issue that any apparatus we might like to use will necessarily backreact on the geometry.  This is a consequence of the ``universal'' nature of gravitational interactions; it does not seem to be possible to invent matter that does not couple to gravity.\footnote{This is actually not quite true; there exist things called topological field theories where one indeed has fields that do not couple to the metric.  The simplest example is the Chern-Simons theory in 2+1 dimensions.  This only seems to be possible however if \textit{none} of the fields which are present interact with the metric; one cannot have fields which don't interact with the metric interacting with fields that do.  The reason is that any such interaction would induce interaction with the metric even if it wasn't there before.}  It thus seems likely that in an exact theory of quantum gravity we are going to need some formulation in which local field operators are emergent and approximate notions.\footnote{I'll discuss some more thoughts along these lines in section \ref{holosec} below.}

\begin{figure}
\begin{center}
\includegraphics[height=6cm]{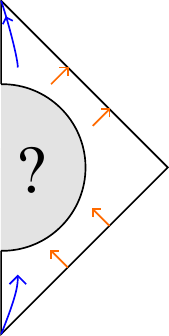}
\caption{The S-matrix; we send in massless (orange) particles and massive particles (blue) from past infinity, they interact in some complicated way, and then more particles come back out at future infinity.  The S-matrix encodes the transition probabilities of this process via equation \eqref{SP}.}\label{smat}
\end{center}
\end{figure}

In asymptotically flat space times, such as those relevant for the formation and evaporation of black holes we have considered so far, there is a very natural candidate for an exact quantity to study in place of the correlation functions of local operators that one naturally studies in quantum field theory.  This quantity is the S-matrix.  It is defined as a linear map from initial states on $i_-\cup J_-$ to final states on $i_+\cup J_+$ with the property that
\be\label{SP}
P(\chi|\psi)=|\lan\chi|S|\psi\ran|^2.
\ee
In other words the probability of finding an ``out'' state $|\chi\ran$ given an ``in'' state $|\psi\ran$ is given by the absolute value squared of the matrix elements of the S-matrix.  Formally we can think of the S-matrix as something like
\be\label{Sh}
S = e^{-i\infty H},
\ee
although this expression obviously needs to be taken with a grain of salt.  It is not hard to see that the S-matrix must be a unitary map for \eqref{SP} to make sense, which is consistent with \eqref{Sh}.\footnote{In Hawking's attempts to study information loss he introduced the idea of a ``$\$$''-matrix, which is a linear map from density operators to density operators.  Today this kind of thing is called a superoperator or a quantum channel, and it often appears in discussions of noise and communications.}  I illustrate the basic idea of the S-matrix in figure \ref{smat}.

Before the S-matrix can really have any nontrivial content, we of course need to specify the meanings of its labels.  In other words, are there natural bases for the spaces of in and out states respectively that have simple physical interpretations?  Here we are in luck; as we approach infinity any incoming/outgoing particles which are around become more and more widely separated, so gravitational (and any other) interactions between them become irrelevant and we can really make sense out of them as exact quantum states.  So in fact we expect the Hilbert space of states at past or future infinity to simply have the structure of a free quantum field theory!  States are labeled by how many particles there are of such-and-such type and such-and-such momentum/spin, making sure to include the appropriate boson/fermion statistics.\footnote{It is clear that this is a subtle claim to make precise, even in quantum field theory one has to make sure that one picks the right set of asymptotic particles, and the choice isn't always obvious.}

That the unitary S-matrix should be an exact observable of asymptotically-flat quantum gravity has been confirmed in the one real example we know so far of such a theory (in a number of dimensions large enough to have black holes); the BFSS matrix model \cite{Banks:1996vh}.  Among other things this model provided the first example of a unitary theory which is expected to have black holes; I won't discuss it in any detail because AdS/CFT gives a broader and easier set of examples.

\subsection{The Page curve}\label{pcurvesec}
\begin{figure}
\begin{center}
\includegraphics[height=6cm]{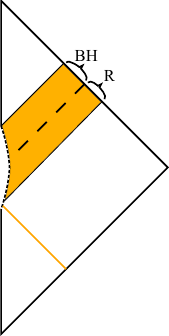}
\caption{The S-matrix for a black hole that is made from an initial orange shell of photons and then decays into a Hawking cloud.  The horizon is shown as the short-dashed line; it forms and then dissipates.  It is often convenient to split the Hawking radiation into an early part, $R$, and a late part, $BH$, which I do with the long-dashed line.  The names are meant to suggest that at some time the photons in $R$ were already out in the radiation while the photons in $BH$ had yet to be emitted from the black hole.}\label{bhscat}
\end{center}
\end{figure}

For our purposes we are interested in a particular type of scattering process; one where we create a black hole from infalling matter and then watch it evaporate.  This is illustrated in figure \ref{bhscat}.

In this setting we would like to get a sense of how the quantum state of the black hole and its radiation evolves with time; one fairly rigorous way to do this is by making a split of the Hawking radiation into ``late'' and ``early'', as shown in figure \ref{bhscat}.  This gives a tensor product decomposition of the out state of the Hawking radiation into a bipartite system
\be
\mathcal{H}_{out}=\mathcal{H}_R\otimes \mathcal{H}_{BH},
\ee 
and we can study how the reduced density matrix on $R$ or on $BH$ depends on when we do the decomposition.   

\begin{figure}
\begin{center}
\includegraphics[height=5cm]{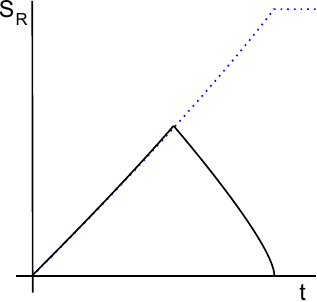}
\caption{Page's suggestion for a Page curve, in black.  $S_R$ increases as the black hole evaporates until a time of order $t_{evap}/2$, at which point we have $S_R\approx S_0/2$, where $S_0$ is the initial coarse-grained entropy of the black hole.  It then decreases back to zero as the black hole evaporates.  A more detailed analysis \cite{Page:2013dx} based on the arguments  in the next section puts the turnover at $t=.54 t_{evap}$ and at entropy $S_R=.6 S_0$.  For comparison I show Hawking's proposal as the dotted blue line, where $S_R$ continues to increase until at the end of the evaporation it reaches the full initial coarse-grained entropy, violating unitarity.}\label{pagecurve}
\end{center}
\end{figure}

One thing that is particularly interesting to compute is the entanglement entropy $S_R$ as a function of time; the plot of this function is called the ``Page curve'', in honor of its inventor \cite{Page:1993wv,Page:2013dx}.\footnote{In fact because of the UV divergences of quantum field theory, what one really plots is the \textit{renormalized} entanglement entropy $S_{R}^{\{ren\}}\equiv S_R-S_R^{\{vac\}}$, where $S_R^{\{vac\}}$ is the entanglement entropy of the vacuum.  For notational simplicity I will ignore this subtlety in what follows, but if you are worried about it an excellent laboratory for convincing yourself it is ok is the ``moving mirror'' model of black hole evaporation \cite{Holzhey:1994we}.  In this model the Page curve is computable analytically, and one can see that it has the basic features suggested here.}  Let's think about what to expect; at the beginning of the experiment the black hole is in a pure state, so the radiation field is trivial and has $S_R=0$  As the black hole begins to radiate $S_R$ will start increasing. At some point however it must turn over and come back to zero, since once all the radiation is out it must again be in a pure state.  I show a plot of one possibility in figure \ref{pagecurve}.

In fact Andy Strominger has argued that being able to compute the Page curve in some particular theory is what it \textit{means} to have solved the black hole information problem; even in AdS/CFT or the BFSS model we are far from being able to really do this.  Nonetheless there are some fairly compelling arguments, perhaps unsurprisingly due to Don Page, about what its basic form should be.  I will now explain these arguments, but first we need a bit of technology from quantum information theory.
 
\subsection{Page's theorem}\label{pagesec}
Consider a bipartite system
\be
\mathcal{H_{AB}}=\mathcal{H}_A\otimes \mathcal{H}_B.
\ee
Without loss of generality I will take $|A|\leq |B|$, where by $|\cdot|$ I mean the dimensionality of the indicated system.  We say that the system is \textit{maximally entangled} if the state $\rho_{AB}$ is pure but the state $\rho_A$ obtained from it by partial trace is proportional to the identity operator on $\mathcal{H}_A$.  

Page's theorem \cite{Page:1993df} then says that a randomly chosen pure state in $\mathcal{H}_{AB}$ is likely to be very close to maximally entangled as long as $\frac{|A|}{|B|}\ll1$  In order to state the theorem precisely, I need to describe first how to choose a random pure state and second how to quantify what ``close'' means.  

Choosing a random pure state is accomplished by acting on any particular state $|\psi_0\ran$ with a random unitary matrix:
\be
|\psi(U)\ran\equiv U|\psi_0\ran.  
\ee
$U$ is chosen from the group-invariant Haar measure.  We can compute the reduced density matrix $\rho_A(U)$ by tracing out $B$.  

Closeness of states is defined using the operator trace norm, also called the $L_1$ norm, which is defined for any operator $M$ as
\be
||M||_1\equiv \mathrm{tr} \sqrt{M^\dagger M}.
\ee
The trace norm distance between two density matrices $\rho$ and $\sigma$ is then defined as $||\rho-\sigma||_1$.\footnote{The motivation for this definition is as follows; say that $||\rho-\sigma||_1<\epsilon$.  Then for any projection operator $\Pi$ we also have $\mathrm{tr} \left(\Pi(\rho-\sigma)\right)<\epsilon$, so the probability for any experimental outcome differs between $\rho$ and $\sigma$ by at most $\epsilon$.  Also beware that it is somewhat common to define a trace distance $D(\rho,\sigma)\equiv \frac{1}{2}||\rho-\sigma||_1$, but this factor of two just would make our lives more difficult.}  We will also be interested in the $L_2$ norm, defined as
\be
||M||_2\equiv \sqrt{\mathrm{tr}M^\dagger M}.
\ee
It isn't too hard to show that these obey
\be\label{norms}
||M||_2\leq ||M||_1 \leq \sqrt{N} ||M||_2,
\ee
for any operator $M$, where $N$ is the dimensionality of the Hilbert space.

We are now in a position to state (a version of) Page's theorem:

THEOREM: For any bipartite Hilbert space $\mathcal{H}_A\otimes\mathcal{H}_B$, we have
\be\label{Pagethm}
\int dU ||\rho_A(U)-\frac{I_A}{|A|}||_1\leq \sqrt{\frac{|A|^2-1}{|A||B|+1}}.
\ee

For intuition we can slightly weaken the bound by just writing $\sqrt{\frac{|A|}{|B|}}$ on the right hand side.  Page's theorem then says that once $|B|$ is significantly larger than $|A|$, the typical deviation of $\rho_A$ from the maximally mixed state is extremely small.  For example say that the two systems are sets of qubits; then if $B$ has 10 more qubits than $A$ then the typical deviation from maximal entanglement is bounded by $2^{-5}$.  

The proof goes as follows; we first have
\begin{align}\nonumber
\left(\int dU ||\rho_A(U)-\frac{I_A}{|A|}||_1\right)^2&\leq \int dU \left(||\rho_A(U)-\frac{I_A}{|A|}||_1\right)^2\\
&\leq |A|\int dU \left(||\rho_A(U)-\frac{I_A}{|A|}||_2\right)^2,
\end{align}
with the first inequality following from Jensen's inequality and the second following from \eqref{norms}.  We can then evaluate the integral over $U$ exactly using unitary matrix technology which is developed in appendix \ref{Uintsec} and take the square root of both sides to find the right hand side of \eqref{Pagethm}.  Note that this version of the theorem does not have any assumptions about either $|A|$ or $|B|$ being large, although clearly to get $\frac{|A|}{|B|}\ll 1$ we will want to take $|B|\gg 1$.  

Page actually stated his theorem in terms of the entanglement entropy $S_A$ instead of the trace norm; this version is derived by a similar argument: First defining 
\be
\Delta\rho_A\equiv \rho_A-\frac{I_A}{|A|},
\ee
we have
\begin{align}\nonumber
\int dU \, S_A&=-\int dU \,\mathrm{Tr}\rho_A \log \rho_A\\\nonumber
&=\mathrm{Tr}\left[\left(\frac{I_A}{|A|}+\Delta\rho_A\right)\left(\log |A|-|A|\Delta\rho_A+\frac{1}{2}|A|^2 \Delta \rho_A^2+\ldots\right)\right]\\\nonumber
&=\log |A|-\frac{|A|}{2}\int dU \, \mathrm{Tr} \Delta \rho_A^2+\ldots\\
&=\log|A|-\frac{1}{2}\frac{|A|}{|B|}+\ldots,
\end{align}
where $\ldots$ indicate terms that are smaller in the limit $|A|,|B|\gg 1$.  Abstractly I prefer the trace norm version, both because the trace norm is a better measure of the distance between states and because no limit is necessary, but the entropy version is also useful, as we will now see.

Indeed we can now use Page's theorem to justify the proposed form of the Page Curve shown in figure \ref{pagecurve}.  The idea is that black hole evaporation is such a complex process that it is plausible to assume that the pure state of $R$ and $BH$ together is random, up to the basic constraints imposed by energy conservation and causality.  At early times the coarse grained entropy $S_R^{\{coarse\}}=\log|R|$ of the Hawking radiation is small compared to the coarse-grained entropy $S_{BH}^{\{coarse\}}=\log|BH|$ of the black hole.  More explicitly, most of the Hawking radiation is emitted into photons in the lowest $\ell$ modes, which we can approximately think of as a (1+1) dimensional photon gas.  Its coarse-grained entropy at early times is then roughly
\be
S_R^{\{coarse\}}\propto t T,
\ee  
where $t$ is the time since the black hole began evaporating and $T\sim 1/M$ is the temperature of the black hole.  The coarse-grained entropy of the black hole is just the Bekenstein-Hawking entropy, which is of order $M^2$, so we see that for $t\ll M^3$ we have $S_R^{\{coarse\}}\ll S_{BH}^{\{coarse\}}$.  By Page's theorem we then have\footnote{A subtlety in applying Page's theorem here is that the subspace of states of fixed energy in $\mathcal{H}_{BH}\otimes \HR$ does not have a tensor product form, so to respect energy conservation we shouldn't really average over all states in the product Hilbert space.  In the limit of weak interactions between $A$ and $B$ one can modify the proof of Page's theorem to deal this, the basic steps are the same but the notation is a bit heavier since we need to ensure that the random unitary acts only within the subspace of fixed total energy.  As you might expect, the result is  that the reduced density matrix for the smaller system $A$ is very close to the \textit{thermal} density matrix $\frac{1}{Z_A}e^{-\beta H_A}$, with the inverse temperature $\beta$ chosen so that $A$ has the right expectation value for its remaining energy.  The question of how weak the interactions need to be is a subtle one, but a good rule of thumb is that the unperturbed thermal expectation value of the perturbation to the Hamiltonian should be small compared to the unperturbed expectation value of the unperturbed Hamiltonian.}
\be
S_R\approx S_R^{\{coarse\}}\qquad \qquad t\ll M^3,
\ee
so for a while the Page curve grows linearly in $t$.  Eventually we have $S_R^{\{coarse\}}\approx S_{BH}^{\{coarse\}}$, which is defined to happen at the ``Page time'' $t_{Page}$.  At this point $S_R$ is some order one fraction of the original coarse-grained entropy $S_0$ of the black hole.  After the Page time we can apply Page's theorem in the other direction, so we expect to have $S_{BH}\approx S_{BH}^{\{coarse\}}$. Since the total state is pure we must have $S_{BH}=S_{R}$ at all times (this follows from the Schmidt decomposition described in appendix \ref{BPapp}), so we now roughly have 
\be
S_R\approx S_{BH}^{\{coarse\}}\propto S_0\left(1-\frac{t}{t_{evap}}\right)^{2/3} \qquad t_{page}<t<t_{evap}.
\ee
Thus by using Page's theorem we have reproduced the qualitative features of figure \ref{pagecurve}.  By being more careful about details of the evaporation, such as greybody factors and the number and helicities of the available massless particles, one can work out more of the quantitative details about exactly when the curve turns over and at what value of the entropy \cite{Page:2013dx}.  The results for a four-dimensional Schwarzschild black hole radiating into photons and gravitons are quoted in the caption of figure \ref{pagecurve}.

Intuitively we can think of what is going on as follows: at the beginning of the evaporation process the radiation that comes out is entangled with the remaining black hole.  But eventually it must start coming out entangled with the earlier radiation, since eventually the final state of the radiation must be pure.  It is only once we are past the Page time that we can think of the quantum information about the initial state as having started to come out.

\subsection{How hard is it to test unitarity?}\label{testUsec}
I now consider a different question; say that we are convinced theoretically that black hole evaporation is unitary.  Is there a good way to test this experimentally?  There are a host of practical difficulties that would need to be dealt with; first of all we would need to figure out how to reliably make black holes in a laboratory setting.  They will need to be considerably smaller than the black holes that are produced astrophysically, for example a black hole that evaporates in a year has a mass of order $10^9 kg$ and a radius of order $10^{-18}m$.  Secondly we would need to be able to directly manipulate individual photons (and possibly gravitons!) in the Hawking radiation; the energies of these photons and gravitons for this black hole would be of order $100GeV$. Most worrying, the entropy is of order $10^{34}$, so we have $10^{34}$ particles to deal with.  We can improve this by making the black hole even smaller, but that comes with its own technical challenges.  These obstacles are worrisome to say the least, but even if we were able to surmount them there is still a purely quantum mechanics question of how we should go about verifying the unitarity of black hole evaporation, say for a black hole whose entropy is only of order $10^{10}$.     

Let's thus imagine that we are successfully able to create a black hole and capture the full quantum state of the Hawking radiation, which we can then transfer to the memory of a quantum computer.\footnote{The reader who is unfamiliar with quantum computation and quantum circuits should now look at appendix \ref{compsec} for a brief introduction.}  We could of course just determine the quantum state explicitly by sampling from an ensemble of exponentially many (in the entropy!) identically-prepared black holes, but this is rather impractical.\footnote{People with a particle-physics background sometimes talk about ``measuring high-point correlation functions'' to test unitarity; that is another version of this brute-force algorithm.}  If we grant ourselves the ability to simulate the black hole formation and evaporation process on a quantum computer in polynomial time,\footnote{This is probably possible for the quantum gravity theories we know about like the BFSS matrix model or AdS/CFT \cite{feynman1982simulating,lloyd1996universal,Jordan:2011ne}, and if you believe in what Scott Aaronson and Stephen Jordan called in their lectures the ``Extended Church-Turing-Deutsch hypothesis'' then it must be possible.} then a simpler option is to transfer the final state of the Hawking radiation to a quantum computer and then apply the time-reverse of this simulation.  What emerges should then be the initial state of the black hole.  We could easily check this, for example by entangling the first few qubits in the collapse with some reference system and then checking if the output of the computation still possesses that entanglement.  A downside of this method is that it requires us to know the S-matrix, but an upside is that we can therefore check if we got it right.  

\begin{figure}
\begin{center}
\includegraphics[height=4cm]{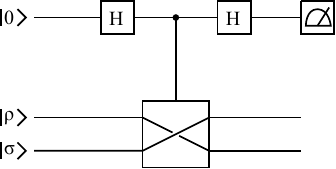}
\caption{The swap test \cite{PhysRevLett.87.167902}.  This circuit measures the expectation value of the swap operator in the state $\rho\otimes\sigma$.  The circuit works by starting with an extra qubit in the state $|0\ran$, applying the Hadamard transformation that exchanges the $Z$ and $X$ eigenbases, applying a controlled swap which exchanges the two lower states only if the upper qubit is $|1\ran$, reapplying the Hadamard, and then measuring the $Z$ operator for extra qubit.  You can convince yourself (homework!) that the expectation value for this last measurement is $\mathrm{Tr} \rho \sigma$.}\label{swap} 
\end{center}
\end{figure}

Alternatively Hayden and Preskill have suggested a clever experiment that does not require a time-reversed simulation \cite{Hayden:2007cs}.  The idea is to prepare two identical black holes and then capture their radiation in the memory of a quantum computer.  We then pass these two states through the quantum algorithm shown in figure \ref{swap}.  This algorithm measures the unitary ``swap'' operator that exchanges the two systems, so it returns $\pm 1$ with expectation value $\mathrm{Tr} \rho \sigma$, where $\rho$ and $\sigma$ are the (possibly mixed) quantum states of the two Hawking clouds.  This expectation value is close to zero unless the two states are both equal and pure.\footnote{One way to see this is to observe that $\mathrm{Tr}\rho \sigma$ defines an inner product on the space of Hermitian operators.  We can then apply the Cauchy-Schwarz inequality to see that
\be
(\mathrm{Tr}\rho \sigma)^2\leq \mathrm{Tr} (\rho^2) \mathrm{Tr}(\sigma^2).
\ee
The right-hand side will be small unless both $\rho$ and $\sigma$ are close to being projection operators, or in other words are close to being pure.  Moreover the inequality is saturated only if they are equal.}  This is thus like trying to distinguish a fair coin from a weighted one that almost gives heads; an O(1) number of trials is sufficient to determine if the states are pure and equal to high probability, which would confirm the unitarity of the evaporation.  Somewhat miraculously this is possible without knowing either the final state of the Hawking radiation or the dynamics of the black hole!  This protocol is called the \textit{swap test} \cite{PhysRevLett.87.167902}.  

These algorithms inconveniently both require us to really capture the full state of the evaporation.  As soon as we start making mistakes, for example because it is rather difficult to detect gravitons, they don't work.  In particular if we try to use the swap test to compare subsystems of the two states then the states would be mixed and as just argued the test would fail.  To deal with missing (or corrupted) radiation we need to do some kind of quantum error correction.  A brief introduction to quantum error correction and its computational complexity is given in section 4 of \cite{Harlow:2013tf}.  The upshot is that quantum error correction is possible provided that we don't lose too much of the radiation; we can lose or corrupt up to almost half if we know which part is affected, while if we don't know we can lose or corrupt up to a quarter.  Recalling our observation in section \ref{hawkrad} that black holes radiate less into higher spin particles, losing gravitons is thus not really a principled obstruction: even if we don't measure any of them we will be able to correct for the loss.  Unfortunately it seems likely that this error correction procedure will take exponential time in the number of affected qubits, so if an $O(1)$ fraction of the radiation is lost then the correction procedure will probably take a time which is exponential in the black hole entropy.  To deal with this exponential we thus need to take the entropy to be at most $O(10-20)$, which corresponds to a black hole whose radius is only slightly larger than the Planck scale.  

At present it thus seems that testing the unitarity of Hawking radiation experimentally will be outside of our technological capability for the foreseeable future, even though there seems to be no in principle obstruction preventing it.  Does this mean we should not worry about it?  Not necessarily; remember that thought-experiments about trains moving close to the speed of light were very helpful to Einstein in understanding relativity, even though we still do not have trains that move at almost the speed of light.  The black hole information problem is a litmus test for theories of quantum gravity; any consistent theory of quantum gravity must either give us an answer or explain why the question does not make sense, and once we have found such a theory it may be experimentally testable in other ways which today are unimaginable.  
  
\subsection{What is a typical microstate?}\label{typicalsec}
So far I've been discussing black holes made out of a fairly rapid collapse of a matter or photon shell.  But what fraction of the total number of microstates can we make this way?  We can estimate this as follows; a 3+1 dimensional photon gas at temperature $T$ and in volume $V$ has energy and entropy
\begin{align}\nonumber
E&\sim V T^4\\
S&\sim VT^3,
\end{align}
so if we imagine forming the black hole by compressing a gas of energy $E=M$ into a volume $V\approx r_s^3\sim M^3$, then apparently the entropy is
\be
S\sim M^{\frac{3}{2}}\sim A^{\frac{3}{4}},  
\ee
where $A$ is the area of the horizon.  Comparing this to the full black hole entropy $S\sim A$, we see that we can make only a small fraction of the microstates this way.  How do we make the rest?  In statistical mechanics there is a standard way of answering this question; all known laws of physics are $CPT$ invariant, so a typical microstate must have coarse-grained behavior which is time-symmetric.  In other words, to make a black hole in a typical member of its microcanonical ensemble of dimensionality $e^{2\pi A}$, we must slowly build it up over a time of order $M^3$ by sending in low-$\ell$ photons at the Hawking temperature in just such a way that no radiation comes out until we are done.  This is an entropy-decreasing process; it looks like the time-reverse of the usual Hawking evaporation.  Once we finish building the typical black hole, it will evaporate in the usual manner and the whole process will look time-symmetric.  

The geometry of the interior in a typical microstate is somewhat mysterious; the Penrose diagram \ref{evappen} is not time symmetric, and if we try to make it so then we invariably end up drawing a past singularity as well as a future singularity.  Do these two singularities meet in the middle?  Is there a piece of smooth geometry between them?  Or could it be the case that there is no global geometry describing the interior and exterior of a typical state?  We will see in section \ref{paradoxsec} below that there are indeed some reasons to suspect that there may not be a smooth interior for typical states, although it is too early to reach a definite conclusion.

\subsection{Scrambling and recovery of quantum information}\label{scrambsec}
Say that we throw a quantum diary into a black hole.  How long do we have to wait before its contents comes out in the Hawking radiation?  This question was studied by Hayden and Preskill in a clear and by-now-classic paper \cite{Hayden:2007cs}; I will here just give a sketch of their arguments.  The problem can be broken into two parts
\begin{itemize}
\item[(1)] How long does it take the black hole to absorb the information? 
\item[(2)] Once the information has been absorbed, how much radiation needs to come out before we can recover it? 
\end{itemize}
Before answering these questions, I will first say a little about what it means to recover quantum information.  

Say we have a quantum system in a state 
\be
|\psi\ran=\sum_i C_i |i\ran,
\ee
where $|i\ran$ is some complete basis for a Hilbert space of dimensionality $2^n$.  For example it could be a a state of a system of $n$ qubits.  Fully describing this quantum state requires specifying $2^{n-1}$ complex numbers (the $C_i$'s modulo normalization), but this is not what is usually meant by the ``quantum information'' contained in the state.  Classically an $n$-bit string is (obviously) specified by only $n$ bits of information, and this should be true for quantum information as well. The exponential comes from trying to classically describe $n$ qubits.  If I want to give a quantum state to you, it would be extremely inefficient for me to write down all of the $C_i$'s, send them to you in the mail, and then have you prepare a system in the state $|\psi\ran$.  I should really just send you the quantum state itself!  In other words quantum information is carried by qubits, not bits.  When we talk about sending or recovering quantum information, this is what we always mean: we have some set of physical operations which at the end of the day enable the transportation of an arbitrary quantum state $|\psi\ran$ of some number of qubits from one place to another, without anybody having to measure it. \footnote{An important point here is that the state does not have to be carried by the same physical qubits at the end as it was in the beginning.  For example quantum teleportation \cite{PhysRevLett.70.1895} is a famous protocol by which we can send an arbitrary state of $n$ qubits from one place to another by exchanging only $n$ classical bits of information.}

The ability to send and receive quantum information is significantly hampered by the ``no-cloning'' theorem of quantum mechanics \cite{Wootters:1982zz,Dieks:1982dj}.  This theorem says that it is impossible to find a system $C$ such that, after adjoining it to an arbitrary quantum state $|\psi\ran$ times an empty register $|0\ran$ of the same dimensionality,  we have time evolution
\be
|\psi\ran|0\ran|\phi\ran_C\to |\psi\ran|\psi\ran |\phi'\ran_C.
\ee

The proof follows immediately by contradiction when we try to use this evolution and the linearity of quantum mechanics to clone the state $\frac{1}{\sqrt{2}}(|\psi\ran+|\chi\ran)$ for some $|\chi\ran$ orthogonal to $|\psi\ran$.  This means that in some sense we can think of quantum information as being conserved: I can send a quantum state to you only if I lose it myself.\footnote{You might object that I can easily prepare multiple copies of a quantum state for which I know the $C_i$'s.  This is true, but missing the point.  In sending classical information it is not necessary to know the message to send it; for example it could be in a sealed envelope or be encrypted.  What we want is a single procedure that works for a single copy of any state of the qubits.  Measuring the state to determine the $C_i$'s and then sending them does not count, since it requires many initial copies of the same state.}  

There is a convenient method for determining whether or not a particular protocol successfully transfers quantum information from one place to another.  Say that we have a procedure which for any $|\psi\ran$ implements
\be
|\psi\ran_A |0\ran_B\to |0\ran_A|\psi\ran_B.  
\ee
This protocol transfers quantum information from system $A$ to system $B$.  Now let's introduce an additional auxiliary system $C$, of the same dimensionality as $A$ and $B$, and maximally entangle it with $A$.  By linearity we then have the evolution
\be
\frac{1}{\sqrt{|A|}}\sum_i |i\ran_A |0\ran_B |i\ran_C\to |0\ran_A \frac{1}{\sqrt{|A|}}\sum_i |i\ran_B |i\ran_C.
\ee
The evolution thus transfers the \textit{purification} of $C$ from $A$ to $B$.\footnote{See appendix \ref{BPapp} for more on the idea of purification.}  More generally we might imagine an evolution
\be\label{infoU}
|\psi\ran_A |0\ran_B\to |0\ran_AU_B|\psi\ran_B,  
\ee
where $U_B$ is some unitary transformation on B.  This is typically still counted as successfully transferring the quantum information, since after all the receiver can get back to the previous evolution by acting with $U_B^\dagger$.  The evolution once we maximally entangle $A$ with $C$ is then
\be
\frac{1}{\sqrt{|A|}}\sum_i |i\ran_A |0\ran_B |i\ran_C\to |0\ran_A \frac{1}{\sqrt{|A|}}\sum_i U_B|i\ran_B |i\ran_C.
\ee
In either case we can we can theoretically test if the transfer was successful by comparing the final states $\rho_{AC}$ and $\rho_A\otimes\rho_C$ in trace norm.  They will be close to equal if and only if the purification has been successfully switched.

Let's now apply this discussion to the quantum diary falling into a black hole.\footnote{We could try to phrase the following discussion more rigorously in terms of decomposing the states at past and future null infinity, as we did in section \ref{pcurvesec} above, but I won't attempt it.  The new subtlety that arises is that in arguing that the matrix $U$ that appears below is unitary and doesn't act on $E$ or $S$, we need to use the \textit{cluster decomposition property} of the S-matrix.  This property is a crude form of locality which says that the S-matrix approximately factorizes for widely separated systems; it is definitely true in quantum field theory and is usually expected to be true even in quantum gravity, see for example \cite{Fitzpatrick:2014vua} for a recent discussion in AdS/CFT.  To really justify the discussion given here, we'd need to study this more quantitatively to make sure it works.} Following Hayden and Preskill we first assume that the black hole absorbs the diary instantly, neglecting question (1) above.  We can model this as follows.  We begin with a diary $D$ maximally entangled with a reference system $S$.  We then throw the diary into a black hole $B$, which we model by acting on the joint $BD$ system with a random unitary $U$.  We can also include the radiation process as part of this unitary, so we re-interpret the $BD$ system after $U$ has acted as a tensor product of some Hawking radiation $R$ and a remaining black hole $B'$.  The question is then how big $R$ has to be before we can recover the quantum diary, or in other words how long we have to wait before 
\be
||\rho_{SB'}-\rho_S\otimes \rho_{B'}||_1\ll 1.  
\ee
If black hole $B$ started in a pure state, then we can just think of the diary as being an extra piece of the infalling matter that created the black hole. From Page's theorem we then already know the answer for how long we have to wait until information gets out; until the Page time the Hawking radiation is maximally mixed and carries no quantum information.  

\begin{figure}
\begin{center}
\includegraphics[height=5cm]{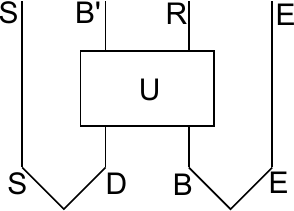}
\caption{The Hayden/Preskill experiment.  We throw a diary $D$ entangled with a reference system $S$ into a black hole $B$ that is already maximally entangled with its early Hawking radiation $E$.  The diary is absorbed by the black hole, which then partially evaporates into a remaining black hole $B'$ and some more radiation $R$.  The question is to find when the purification of $S$ is transferred to $ER$.}\label{hpfig}
\end{center}
\end{figure}

What Hayden and Preskill realized however is that the situation is more interesting if we wait until after the Page time to throw in the diary.  In this situation the black hole is already maximally entangled with its early radiation $E$; I illustrate the situation in figure \ref{hpfig}.
Using the same random unitary technology that we used above in proving Page's theorem, one can then show that \cite{Hayden:2007cs}
\be\label{hpresult}
\int dU ||\rho_{SB'}-\rho_S\otimes \rho_{B'}||_1\leq \sqrt{\frac{(|D|^2-1)(|B'|^2-1)}{|D|^2|E|^2-1}}\approx \frac{|D|}{|R|},
\ee
where in the approximation I've used that $|B'||R|=|E||D|$ and assumed all systems are large enough we can ignore the $1$'s.  Intuitively this result says the following; say that the number of bits radiated after throwing in the diary is $c$ bits more than were contained in the diary itself.  Then the righthand side is $2^{-c}$, which rapidly becomes extremely small.  So the information essentially comes out as fast as it possibly could!  This result led Hayden and Preskill to describe ``old'' black holes as ``information mirrors''.  Beware however that, as in equation \eqref{infoU}, a unitary transformation on $ER$ will in general be necessary to put the quantum state of the diary back into useable form.  This unitary transformation might be quite difficult to perform, and indeed as we will discuss later this is probably the case \cite{Harlow:2013tf}.  

Finally let's return to the question of how long it takes the black hole to absorb the diary and re-emit it, or in other words how long it takes $U$ to act.  A first guess is to just treat the diary as a massless particle sent inwards from some fixed radius $r_0$, and then ask how much Schwarzschild time goes by before it reaches the stretched horizon at $r-r_s\approx \frac{\ell_p^2}{r_s}$ (recall that the stretched horizon is the surface a Planckian proper distance outside the actual horizon.)  From the expression \eqref{nullg} for the trajectory of a radial null geodesic, we see that the time is
\be\label{prescrt}
\Delta t \approx t_{scr}\equiv r_s\log \frac{r_s}{\ell_p},
\ee
which is the same timescale we encountered previously around equation \eqref{tscrambling} in our discussion of the trans-Planckian problem.  Here we are interpreting it as the time for the diary to appear to have been completely thermalized from the point of view of an outside observer; by the same calculation as we did below equation \eqref{tscrambling} any signal sent from the diary at that point would need to have superplanckian energy to be distinguishable from the thermal atmosphere.  The timescale \eqref{prescrt} is also the time it takes for geometric perturbations of the black hole to ring down to Planckian amplitude \cite{Price:1986yy}.\footnote{The reason for this is straightforward; the scrambling time is the time it takes to function $r_s e^{-t/r_s}$ to become of order $\ell_p$.}  For these reasons the timescale $t_{scr}$ is usually called the \textit{scrambling time}.  

In fact scrambling is a technical notion in quantum information theory; we say that a piece of quantum information is scrambled into a system if it cannot be recovered from any subfactor of the system that is smaller than some order one fraction of the whole.  Hayden and Preskill's result \eqref{hpresult} shows that a random unitary accomplishes this, and indeed were the transformation $U$ not to have this property it wouldn't necessarily be the case that we could recover the diary from $RE$ since there might be some information left in $B$.  It is not immediately clear that the evolution of a black hole for a time of order $t_{scr}$ is really ``sufficiently random'' to scramble the system in this technical sense, but Hayden and Preskill provided plausible evidence for this from the theory of quantum circuits.\footnote{See appendix \ref{compsec} for an introduction to quantum circuits.}  What they pointed out is that a level of scrambling which is sufficient for equation \eqref{hpresult} to hold can be produced by much smaller quantum circuits than the exponential-sized ones which would be needed to produce Haar-typical $U$'s.  For a system of $n$ qubits there exist families of quantum circuits called $\epsilon$-approximate unitary two-designs, with the property that \eqref{hpresult} holds to within accuracy $\epsilon$ if we replace the average over all unitaries by an average just over one of these familes \cite{dankert2009exact}.  Moreover these circuits have a depth which scales like $O\left(\log n \log\frac{1}{\epsilon}\right)$. Hayden and Preskill then modeled the black hole horizon as a set of $S$ qubits positioned at the stretched horizon.  They further imagined that any two of the qubits can interact pairwise via two-qubit gates, and that each layer of the circuit requires one Planck unit of proper time to execute.  A time step of order $\ell_p$ near the horizon is redshifted to a time step of order $r_s$ in Schwarzschild time, so the total execution time for an $\epsilon$-approximate unitary two-design on these qubits will be of order $r_s\log S$, which is consistent with \eqref{prescrt}.\footnote{Actually Hayden and Preskill didn't quite get this right, since they imposed locality on the qubit interactions and then had to cancel it by having the circuit run faster.  This was corrected in \cite{Sekino:2008he}.}

The scrambling time \eqref{prescrt} is quite short compared to the evaporation time, notice that the Planck scale appears only inside of the logarithm, so even for a solar mass black hole the scrambling time is only about $10^{-4}$ seconds.  It is also quite fast compared to what is expected for more conventional systems with comparable entropy, and in the aftermath of the Hayden and Preskill paper there was quite a bit of activity aimed at understanding what sorts of physical systems might realize this ``fast scrambling'' \cite{Sekino:2008he,Lashkari:2011yi,Barbon:2011pn,Barbon:2012zv}.  More recently Shenker and Stanford have been able to independently derive the scrambling time \eqref{prescrt} for certain ``large'' black holes by using the AdS/CFT correspondence \cite{Shenker:2013pqa,Shenker:2013yza}, I will describe this in a bit more detail in section \ref{RTsec} below.

\subsection{Black hole complementarity}\label{compsec1}
So far in this section we have only studied the physics of the S-matrix; the interior region of the black hole has disappeared from view.  This was not an accident, essentially everything we really know about the quantum mechanics of black holes rests on the precise formalism of the S-matrix (or its AdS equivalent that I will introduce in the next section).  Unfortunately there does not seem to be any simple way to relate the experience of an infalling observer who crosses the horizon to properties of the S-matrix.  What then are we to do?  It seems clear to me at least that we cannot really claim to understand black holes until we understand the experience of the infalling observer.  

The black hole interior has been an extremely hot subject in the last year or two, which I will return to in the final two sections of these notes, but as a teaser I will now discuss what up until recently was the best-known attempt to grapple with the implications of unitary evaporation for the experience of an infalling observer.  This is analysis by Susskind and Thorlacius of the possibility of seeing quantum cloning by jumping into a black hole \cite{Susskind:1993mu}. 

\begin{figure}
\begin{center}
\includegraphics[height=5cm]{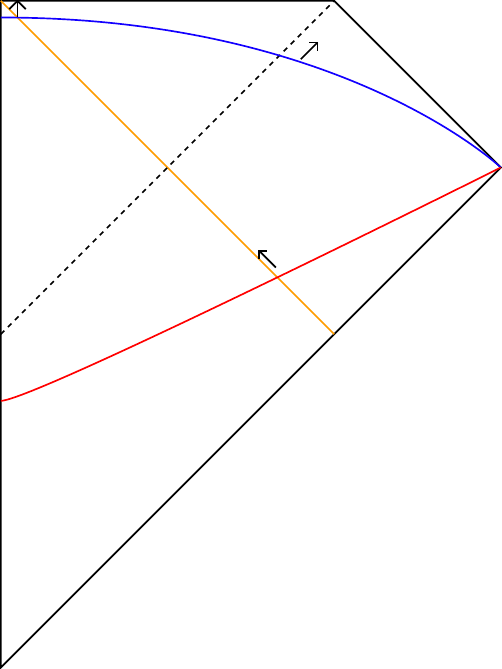}
\caption{Black hole cloning.  If we believe that Hawking evaporation is unitary, then the quantum information on the red slice about how the black hole was made is apparently present at two places on the blue slice.}\label{cloning}
\end{center}
\end{figure}
The setup is shown in figure \ref{cloning}.  In order to describe what is seen by the infalling observer in a quantum mechanical manner, we apparently need to introduce some type of degrees of freedom that live behind the horizon.  How these degrees of freedom relate to those that describe the in and out states at past and future infinity is of course an important question that any theory of the interior needs to address, but for now I will not try to do this.  The strongest assumption we could make is that there is a valid low-energy field theory description of the quantum state throughout the red and then the blue spatial slices in the figure.  These types of slices are sometimes called ``nice slices'', since they seem to be the kind of thing you would want to draw to get a complete Hamiltonian description of the spacetime including the interior.  

What Susskind and Thorlacius emphasized, following a suggestion of Preskill, is that if we believe the evaporation to be unitary, this description of the slices cannot be consistent with quantum mechanics.   The reason is that the quantum information on the red slice that is in the infalling shell apparently is cloned on the blue slice; we could throw in arbitrary quantum states and they would both stay inside in the shell and reappear in the Hawking radiation.  This type of evolution violates the quantum ``no-cloning'' theorem described in the previous section; it is inconsistent with the linearity of quantum mechanics.  Both pieces of quantum information seem to be ``real'', somebody could jump in riding the shell and find the state up in the left corner or somebody could stay outside and see the information come out in the Hawking radiation.  What Susskind and Thorlacius pointed out (and which was later refined by Hayden and Preskill in \cite{Hayden:2007cs}) is that no \textit{single} observer can see both copies.  In the diagram this is essentially obvious by causality, since anybody who waits around long enough (ie after a Page time) to get out the quantum information will then jump in far too late to be able to see the infalling shell on this slice.  We could try to avoid this by bending the blue slice back down towards the horizon, but this causes an enormous redshift of the infalling shell from the point of view of the person jumping in late, which prevents her from seeing the second copy.  This argument was significantly improved by Hayden and Preskill, who pointed out that a much more stringent test arises if the infalling observer waits until a Page time has passed and then throws in a diary as we discussed in the previous section.  We saw that in this case the information now comes out after a time which is only of order the scrambling time $M \log M$, which is much shorter.  Hayden and Preskill nonetheless showed that the observer is still just barely not able to see any quantum cloning.

You might be tempted to say that a contradiction is a contradiction, regardless of whether somebody sees it or not.  And indeed this contradiction certainly does show that imagining that there is unitary quantum mechanical evolution on these ``nice slices'' is inconsistent with having the evaporation process be unitary.  But Susskind and Thorlacius instead interpreted the inability to see the potential cloning as an argument \textit{against} using these slices in the first place.  If nobody can see the quantum state on the whole slice, why should it exist?  They argued that quantum mechanics only needs to describe the experiences of individual observers, who are appropriately restricted by causality in what they can do.  They called this idea \textit{black hole complementarity} \cite{Susskind:1993if,Susskind:1993mu,Lowe:1995ac,Kiem:1995iy}.  

If this seems like it is getting too philosophical to you, I'm sympathetic.  After all nobody ever managed to really implement black hole complementarity into an actual theory, so its consistency (or indeed its precise definition) was never clear.  But it is worth saying that this type of argument does have at least one very successful historical analogue; the Heisenberg Uncertainty principle.  Heisenberg's realization that it is operationally impossible to measure both the position and momentum of a particle could have been dismissed by a hide-bound ``classical physicist'' as follows: ``Of \textit{course} the particle has both a position and a momentum, that is part of what makes it a particle.  What do I care if you can't figure out how to measure it!''.  This would have been a profoundly wrong retort; the correct interpretation is that Heisenberg's operational limitation is an essential part of the consistency of a new type of theory where ``particle'' no longer means what it used to.  For black hole complementarity we have yet to find the new theory, and indeed recent work that I will discuss in sections \ref{paradoxsec}, \ref{altsec} suggest that more new ideas are needed before such a theory can be found, but black hole complementarity may yet play an important role in its eventual formulation.

\section{Holography and the AdS/CFT correspondence}\label{adssec}
The debate over whether or not black hole evaporation is unitary persisted for a while, but major progress in string theory in the 1990's managed to eventually push most people in the field into the pro-unitarity camp (including Hawking himself \cite{Hawking:2005kf}).  The first important part of this was the microscopic calculations of black hole entropy in string theory already mentioned in section \ref{entsec}, but the main source of the shift was the explicit realization in AdS/CFT \cite{Maldacena:1997re,Witten:1998qj,Gubser:1998bc} of the holographic principle \cite{tHooft:1993gx,Susskind:1994vu}.  In this section I give a brief overview of these ideas, focusing mostly on material relevant for understanding the black hole information problem.  It will by no means be complete; AdS/CFT is an enormous subject, only two years after its discovery the standard introduction \cite{Aharony:1999ti} already weighed in at 261 pages.  Readers are obviously encouraged to turn there for more details on AdS/CFT, although actually many of the things I will describe in this section were not known at the time of that review.  In this section I will restore the Planck scale, and instead mostly work in units where the AdS radius $r_{ads}$ is set to one.  Given the wide variety of interesting AdS/CFT examples in various spacetime dimensions, in this section I will mostly work in $d+1$ spatial dimensions instead of just $3+1$.

The precision of the AdS/CFT correspondence means that this section will necessarily have a higher density of technical results than the previous ones; in the end this is a good thing, but readers who get bogged down in the middle are encouraged to skim and skip ahead as needed.  The upshot is that the correspondence provides compelling evidence for the unitarity of black hole evaporation.

\subsection{Entropy bounds and the holographic principle}\label{holosec}
Perhaps surprisingly, the idea that black holes have microstates and are described by unitary dynamics has interesting implications for the statistical mechanics of other systems.\footnote{For a broad review of the ideas in this subsection, see \cite{Bousso:2002ju}.}  

As a first example, say that we have an object of linear size $L$ and energy $E$.  By lowering it to within a proper distance of order $L$ of the horizon of a black hole whose Schwarzschild radius $r_s$ is much greater than $L$, we can use the gravitational redshift to decrease the energy of the object as seen from infinity by a factor of order $\frac{L}{r_s}$.  If we then drop the object into the black hole, the black hole mass changes by $\Delta M=\frac{LE}{r_s}$.  The change in the Bekenstein-Hawking entropy of the black hole is
\be
\Delta S_{bh} \propto LE.
\ee
This experiment potentially is a challenge to the second law of thermodynamics; if the increase in the black hole entropy is less than the entropy of the system we threw in then the total entropy would decrease.  This led Bekenstein to conjecture that for any system we must have
\be\label{bekbound}
S<CLE,
\ee 
where $C$ is some O(1) coefficient not determined by this argument \cite{Bekenstein:1980jp}.  This conjecture is called the \textit{Bekenstein Bound}, and it has the rather surprising feature that, although it was motivated from an argument about black holes, the Planck scale does not appear on either side of the inequality.  A priori there does not seem to be an independent reason  for it to be true, so it at first appears to be a nontrivial constraint on the type of nongravitational systems that can be consistently coupled to gravity.  Indeed there has been quite a lot of controversy in the literature, both about whether or not the bound is true and whether or not it needs to be true to preserve the second law \cite{Unruh:1982ic,Marolf:2003wu,Marolf:2003sq}.  Most of the controversy stems from the issue that the precise definitions of the quantities ``S'', ``L'', and ``E'' appearing in the bound are not clear.  Recently however Casini has given a simple and elegant proof that a precise version of the bound holds in any relativistic quantum field theory \cite{Casini:2008cr}.\footnote{The proof of the Bekenstein-Casini bound is based on the positivity of \textit{relative entropy}, which for any two density matrices is defined as $S(\rho|\sigma)\equiv \mathrm{tr}\rho \log \rho-\mathrm{tr} \rho \log \sigma$.  Casini interprets the left-hand side of the bound as the renormalized Von Neumann entropy $-\mathrm{tr}\rho_V \log \rho_V+\mathrm{tr}\rho_V^0 \log \rho_V^0$ of some region $V$ in an excited state $\rho_V$, where $\rho_V^0$ is the reduced density matrix of the ground state in the same region.  The right hand side of the bound is interpreted as the renormalized expectation value $\mathrm{tr} \rho_V K-\mathrm{tr} \rho_V^0 K$ of the modular Hamiltonian $K\equiv -\log \rho^0_V$ in the excited state $\rho$; the bound then follows immediately from the positivity of $S(\rho_V|\rho_V^0)$.  I encourage the reader to read his paper for the details; the motivation and discussion are transparent, as is the explanation of why the theorem avoids various potential objections such as increasing the number of species.}  This makes it clear that the Bekenstein-Casini bound is not really a constraint on matter theories, but it is nonetheless a deep and surprising property they possess which does not seem particularly natural unless we think about black holes. 
 
To get a bound that involves the Planck scale explicitly, Susskind has suggested a different thought-experiment \cite{Susskind:1994vu}.  Consider a stationary object of entropy $S$ and energy $E$, which is contained in a sphere of area $A$.  If we assume that it isn't a black hole, then $E$ must be less than the mass $M$ of a black hole with horizon area $A$.  Now consider the process where we collapse a spherical shell of matter onto this object whose energy is $M-E$; this results in the formation of a black hole of mass $M$ and area $A$.  In order for this process to not violate the second law we apparently need
\be\label{hbound}
S\leq \frac{A}{4G},
\ee
which is called the \textit{holographic entropy bound}.  Roughly it says that the maximal amount of entropy in a spacetime region scales with the area of the boundary of the region.  From a quantum field theory point of view this is surprising; typically we are used to entropies scaling extensively with the volume of a spacetime region.  For example if we imagine a lattice of Planckian spacing with some finite number of degrees of freedom at each point on the lattice, then the logarithm of the Hilbert space dimension would scale with the volume of the lattice in Planck units.  So the holographic entropy bound is saying that the number of degrees of freedom in spacetime is much less than we might have naively thought; if you try to excite more you make a black hole!  The bound \eqref{hbound} as stated has several problems \cite{Bousso:2002ju}, but a more general covariant version of it \cite{Bousso:1999xy,Bousso:1999cb} has survived many quantitative tests and been proven to hold in a wide variety of classical and semiclassical situations \cite{Flanagan:1999jp,Wall:2010cj,Wall:2011hj,Bousso:2014sda}.  

Clearly if the holographic entropy bound is correct then there is a large amount of non-locality in whatever the correct theory of quantum gravity is.  Indeed the area scaling of the entropy led 't Hooft and Susskind \cite{tHooft:1993gx,Susskind:1994vu} to conjecture that a true theory of quantum gravity must in some sense live in one fewer dimensions than naively expected; Susskind called this idea  \textit{the holographic principle}.

\subsection{Statement of the AdS/CFT correspondence}
The holographic principle has so far had its most precise realization in the widely celebrated Anti de Sitter/ Conformal Field Theory correspondence.  AdS/CFT was originally discovered by studying the low-energy limit of brane systems in string theory \cite{Maldacena:1997re}.  In the most well-known example, one looks at a stack of $N$ D3 branes in type IIB string theory in ten dimensions.\footnote{This argument may be difficult for readers who are unfamiliar with string theory to follow, but there are only so many things I can review!  These objects are defined for example in \cite{Polchinski:1998rr}, but don't worry, they won't be on the final.}  At large $N$ the branes backreact and produce a nontrivial geometry which approaches five dimensional Anti de Sitter space times an $\mathbb{S}^5$ in the vicinity of the branes; the AdS radius in Planck units is of order $N^{1/4}$.  At any $N$ however, the region near the branes has a low-energy description given by a particular $3+1$ dimensional conformally-invariant quantum field theory called maximally supersymmetric $SU(N)$ Yang-Mills theory, with a gauge coupling constant given by $g_{YM}^2=4\pi g$, where $g$ is the string coupling constant.  In the region of overlapping validity these two theories must be the same, and since the latter one makes sense at any $N$ (and $g$) it is natural to conjecture that the equivalence holds at finite $N$ (and $g$).  For reasons we will soon see, the AdS description is often called the ``bulk'' theory while the CFT is called the ``boundary'' theory.\footnote{This terminology is convenient but it can be misleading.  The situation here is different from an often-encountered one in condensed matter physics, where there are ``edge modes'' at the boundary of some system that also has quasiparticle excitations in its interior.  In that type of system, the two types of excitations exist in the same theory.  In AdS/CFT, the two theories are equivalent, and we should use either one description or the other but not both.}  Rather than unpack the details of this argument however, with the benefit of hindsight I will instead just present AdS/CFT as a self-consistent framework.  I will state the correspondence precisely below, but I'll first briefly introduce the ingredients.  

\subsubsection{Anti de Sitter space}
\begin{figure}
\begin{center}
\includegraphics[height=5cm]{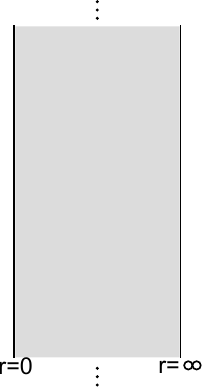}
\caption{The Penrose diagram for Anti de Sitter space.  The left side is the origin of polar coordinates at $r=0$ and the right side is the timelike spatial boundary of AdS at $r=\infty$.  The time coordinate $t$ is the proper time at $r=0$, so we see immediately that signals can be sent out to the boundary and return in a finite amount of proper time at the center.  The diagram continues infinitely far to the past and future so it isn't really compact; this can be fixed by an additional coordinate transformation, but the current form of the diagram is actually more useful so people usually don't do this.}\label{adspen}
\end{center}
\end{figure}

So far in these notes we have considered geometries which asymptote to ordinary flat Minkowski space.  In the presence of a nonzero vacuum energy however, Minkowski space is not a solution of Einstein's equations.  If the vacuum energy is negative, the simplest solution is Anti de Sitter space, which in $d+1$ spacetime dimensions has metric
\be\label{adsmetric}
ds^2=-\left(1+\left(\frac{r}{r_{ads}}\right)^2\right)dt^2+\frac{dr^2}{1+\left(\frac{r}{r_{ads}}\right)^2}+r^2d\Omega_{d-1}^2.
\ee
Here we have $t\in (-\infty,\infty)$, $r\in [0,\infty)$.  The length $r_{ads}$ is related to the vacuum energy density $\rho_0$ as
\be
\frac{1}{r_{ads}^2}=-\frac{16\pi G \rho_0}{d(d-1)};
\ee
for the remainder of this section I will work in units where $r_{ads}=1$.  This geometry manifestly has the property that for $r\ll 1$ it resembles Minkowski space in spherical coordinates.  It is not obvious in this presentation, but it has an isometry group $SO(d,2)$ which is large enough to send any point in the spacetime to any other point; the geometry is homogeneous.  As $r\to \infty$ it does \textit{not} approach Minkowski space, so it has its own interesting boundary structure.  As usual we can describe this more intuitively with a Penrose diagram, which can be derived by introducing a new coordinate $r=\tan \rho$:
\be
ds^2=\frac{1}{\cos^2 \rho}\left[-dt^2+d\rho^2+\sin^2\rho d\Omega_{d-1}^2\right].
\ee
We have $\rho\in [0,\pi/2)$, so we can conformally compactify by discarding the diverging prefactor and including the boundary at $\rho=\pi/2$.  The diagram is shown in figure \ref{adspen}.

The main lesson of the AdS Penrose diagram is that we should think of AdS as a box. Massless particles sent out from the center get all the way out to the boundary and back in a finite proper time $\pi$, as seen by somebody at the center.\footnote{Provided that we choose reflecting boundary conditions at $r=\infty$, as we probably should if we want to view AdS as a closed system.}  This is also true for massive particles; say you are floating in the center of AdS and you throw a ball away from you.  It will go out some finite distance (unlike a massless particle it won't make it all the way to the boundary), but eventually it will turn around and return to you after a time of order one in AdS units.  These observations are formalized in the statement that the boundary is \textit{timelike}; it has the topology of $\mathbb{R}\times\mathbb{S}^{d-1}$ with the $\mathbb{R}$ being temporal.   

As in flat space there is a natural notion of asymptotically AdS spacetime  \cite{Henneaux:1985tv}.  I won't define this rigorously, but roughly a spacetime is asymptotically AdS if its only boundary is timelike and in the vicinity of that boundary the geometry approaches that of AdS near $r\to \infty$.  In a theory of quantum gravity with nontrivial states there will be backreaction, so this type of geometry will need to be included to get anything interesting.  These states will lie in representations of the AdS symmetry $SO(d,2)$ in the same way that in Minkowski space excitations can be characterized by their Lorentz transformation properties. One of the generators of this symmetry is the AdS version of the ADM Hamiltonian $H$, which generates translations of the boundary coordinate $t$, so quantum gravity in asymptotically AdS space is a strong candidate for a closed Hamiltonian system.  Moreover since excitations reach the boundary and return in finite time, we can think of it as ``gravity in a box''; in particular we should expect that the spectrum of $H$ is discrete.
  
\subsubsection{Conformal field theory}
A conformal field theory is a relativistic quantum field theory which is also invariant under a larger set of spacetime transformations, the \textit{conformal group}, which is generated by the usual Poincare transformations, rescalings of the coordinates $x^{'\mu}=\lambda x^\mu$, and \textit{special coordinate transformations}
\be
x^{'\mu}=\frac{x^\mu+a^\mu x^2}{1+2x_\nu a^\nu+a^2 x^2}.
\ee
More abstractly the conformal group is defined as the set of transformations of Minkowski $\mathbb{R}^{d}$ which preserve angles but not necessarily lengths.  It is isomorphic to $SO(d,2)$, which already suggests a connection to $AdS_{d+1}$.    

The simplest example of a CFT is a free massless scalar field, which in $3+1$ Minkowski space is invariant under the dilatation transformation
\begin{align}\nonumber
x^{'\mu}&=\lambda x^\mu\\
\phi'(x')&=\lambda^{-1} \phi(x).
\end{align}

CFT's have many beautiful properties which I do not have time to go into, but two are crucial for us.  First of all in any CFT we can always find a special set of local operators, called \textit{primary operators}, which transform simply under conformal transformations.\footnote{More precisely a local operator is primary if when it is located at $x=0$ its commutators with the special conformal generators are zero.}  In particular under dilatations they transform as
\be
\mathcal{O}'(x')=\lambda^{-\Delta}\mathcal{O}(x),
\ee
where $\Delta$ is the \textit{conformal dimension} of the primary operator $\mathcal{O}$.  In a unitary conformal field theory $\Delta$ is real and positive, and if $\mathcal{O}$ is a scalar operator it obeys $\Delta\geq\frac{d-2}{2}$.  Derivatives $\partial^n\mathcal{O}$ of a primary operator are called \textit{descendants}; they have conformal dimension $\Delta+n$, meaning that they rescale as $\lambda^{-\Delta-n}$ under dilatations, but they are no longer primary.  Primary operators have simple correlation functions, for example in any CFT a scalar primary $\mathcal{O}$ of dimension $\Delta$ has a time-ordered two-point function\footnote{Remember that time-ordering means that operators at earlier times appear to the right of operators at later times.  $\epsilon$ is a positive infinitesimal quantity which fixes the interpretation of the branch cut.  For the special case of a massless free field in $3+1$ dimensions this formula follows from the $m\to 0$ limit of equation \eqref{to2pt}.}
\be\label{cft2pt}
\lan\Omega|T\mathcal{O}(x,t)\mathcal{O}(0,0)|\Omega\ran=\frac{1}{(|x|^2-t^2+i\epsilon)^\Delta}.
\ee

\begin{figure}
\begin{center}
\includegraphics[height=3cm]{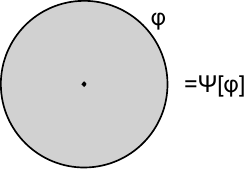}
\caption{The state-operator correspondence.  The path integral over the ball $\rho<1$ with boundary condition $\phi$ and an operator $\mathcal{O}$ at $\rho=0$ computes a wave functional $\Psi[\phi]$ which has energy $\Delta$ if $\mathcal{O}$ has dimension $\Delta$.  Moreover given such a state, we can construct the operator by evolving the state radially inwards assuming no operators are present until we are left with something at the center that must be local.  }\label{stateop}
\end{center}
\end{figure}

Secondly it is often interesting to study a CFT on the ``cylinder'' $\mathbb{R}\times \mathbb{S}^{d-1}$, with metric
\be\label{cftglobal}
ds^2=-dt^2+d\Omega_{d-1}^2.
\ee
As in our previous discussion of the Hilbert space of quantum field theory in section \ref{qftrevsec}, a natural basis of states for this system are field configurations on $\mathbb{S}^{d-1}$.  Another natural basis is the set of eigenstates of the Hamiltonian $H$ that generates $t$ translation.  In fact in CFT's there is a natural bijection between local operators of dimension $\Delta$ and these energy eigenstates.  The bijection is based on the observation that Euclidean $\mathbb{R}^{d}$
\be\label{eucplane}
ds^2=d\rho^2+\rho^2 d\Omega_{d-1}^2
\ee
is conformally equivalent to $\mathbb{R}\times \mathbb{S}^{d-1}$ in the sense of section \ref{penrosesec};  indeed the coordinate transformation $\rho=e^\tau$ puts the metric \ref{eucplane} into the form
\be
ds^2=e^{2\tau}\left(d\tau^2+d\Omega_{d-1}\right).
\ee
Note that dilatations of $\rho$ become time translations of $\tau$, which after analytic continuation becomes $t$. In a conformal field theory we can essentially ignore conformal equivalences in the metric; they do not affect angles and they should basically leave the theory invariant.\footnote{More carefully one can show that for correlation functions of scalar primary operators we have
\be\label{conftrans}
\lan \mathcal{O}_1(x_1)\ldots \mathcal{O}_n (x_n)\ran_{e^{-2\omega} g}=e^{\Delta_1\omega(x_1)+\ldots +\Delta_n\omega(x_n)}\lan \mathcal{O}_1(x_1)\ldots \mathcal{O}_n (x_n)\ran_{g} e^{A[g,\omega]},
\ee
where $A[g,\omega]$ is some known local functional of $g$ and $\omega$ that weakly depends on the theory under consideration but is independent of which operators appear in the correlation function.  It is called the \textit{conformal anomaly functional}, and is nonzero only when $d$ is even.  A similar equivalence holds for wavefunctionals computed from path integrals on conformally equivalent geometries.}  The bijection then works by using the Euclidean path integral on the ball $\rho<1$ in $\mathbb{R}^{d}$ with a local operator of dimension $\Delta$ at $\rho=0$ to construct a quantum state of energy $\Delta$ at $t=0$ on $\mathbb{R}\times \mathbb{S}^{d-1}$, as shown in figure \ref{stateop}.  This is called the \textit{state-operator correspondence}.

\subsubsection{The dictionary}
We can now state the AdS/CFT correspondence.  The modern version is as follows \cite{Heemskerk:2009pn}:
\begin{itemize}
\item Any relativistic conformal field theory on $\mathbb{R}\times \mathbb{S}^{d-1}$ with metric \eqref{cftglobal} can be interpreted as a theory of quantum gravity in an asymptotically $AdS_{d+1}\times M$ spacetime.  Here $M$ is some compact manifold that may or may not be trivial.
\end{itemize}
This statement is really something of a definition; at the moment we don't know an alternative precise theory of quantum gravity in asymptotically AdS space, so we are instead using something well-defined (the CFT) to say what it is we mean by quantum gravity in AdS space.  In order for this definition to be good, we need to address several issues:
\begin{itemize}
\item[(1)] What is the detailed map between the theories?  Given an AdS quantity we are interested in, how do we compute it in the CFT?
\item[(2)] In what cases does the CFT lead to a gravity theory with a good semiclassical description?  In other words, under what circumstances is the AdS radius large in Planck units?
\end{itemize}

Question (1) is answered by the ``dictionary'' of AdS/CFT; the list of CFT expressions for interesting bulk quantities.  I'll here describe some of the most important entries in the dictionary.  First of all the Hilbert space of physical states of the bulk is by definition identical to the CFT Hilbert space.  Moreover symmetry generators of $SO(d,2)$ in the CFT are identified with the corresponding bulk symmetry generators of asymptotically AdS space.  In particular the Hamiltonian is the same on both sides.  Quantities which only depend on the space of states and the Hamiltonian, for example the thermal partition function or the free energy at finite temperature, are thus computed by their CFT expressions by definition.  We'd also like to have something like a CFT expression for a local bulk field, but as we already argued in section \ref{holosec} local bulk fields shouldn't exist in a true theory of quantum gravity such as AdS/CFT is claiming to provide.  In section \ref{smatsec} however, we argued that states at past and future infinity in Minkowski space \textit{could} be described in terms of free fields.  Along similar lines we might guess that in AdS we could be able to make precise sense of the \textit{boundary limits} of local bulk fields.  Indeed the AdS/CFT dictionary postulates for example that if $\mathcal{O}$ is a scalar primary CFT operator then there is a bulk scalar field $\phi$ such that 
\be\label{extdict}
\lim_{r\to\infty}r^{\Delta}\phi(t,r,\Omega)\equiv \mathcal{O}(t,\Omega).
\ee
In other words if we extrapolate a bulk field to the boundary, stripping off a normalization factor, then we get a quantity which is exactly described in the CFT as a primary operator with dimension $\Delta$.\footnote{This version of the dictionary, sometimes called the ``extrapolate'' dictionary, was first proposed in \cite{Banks:1998dd}.  Its equivalence to another version \cite{Witten:1998qj,Gubser:1998bc}, sometimes called the ``differentiate'' dictionary, was shown to all orders in perturbation theory in \cite{Harlow:2011ke}.}  In an expectation value of products of bulk operators, the limit in \ref{extdict} should be taken simultaneously for all operators together.\footnote{Other choices of how we scale the operators will lead to CFT correlation functions computed on spaces other than \eqref{cftglobal}.  There are several interesting options, for example we can instead get CFT correlation functions on $\mathbb{R}^d$ (``AdS-Poincare slicing'') or on $\mathbb{R}\times \mathbb{H}^{d-1}$ (``AdS-Rindler slicing''), but I won't describe these in detail here.}

As an example of the extrapolate dictionary, consider a free massive scalar field in AdS.  Its time-ordered two point function is\footnote{See for example \cite{Burgess:1984ti}.  The most elegant way to derive this involves starting with the Euclidean Green's function on the hyperbolic disk, which depends only the geodesic distance and obeys the wave equation, and analytically continuing.  For an example see an analogous calculation for dS space in appendix B of \cite{Harlow:2011ke}.}
\be
\lan\Omega|T\phi(x)\phi(x')|\Omega\ran=(2\Delta-d)^{-1}2^{-2\Delta}\pi^{-d/2}\frac{\Gamma(\Delta)}{\Gamma(\Delta-d/2)} u^\Delta F\left(\Delta, \Delta+\frac{1-d}{2},2\Delta-d+1,u\right),
\ee 
where 
\be
u=\frac{2}{1+\cosh\ell},
\ee
where $\ell$ is the geodesic distance between the two points and $F$ is a hypergeometric function.  $\Delta$ is related to the mass $m$ as
\be\label{Deltaeq}
\Delta=\frac{d}{2}+\frac{1}{2}\sqrt{d^2+4m^2}.
\ee
In terms of the coordinates in \eqref{adsmetric}, we have
\be
u=\frac{1}{1-r r' \cos\alpha+\sqrt{(1+r^2)(1+r'^2)}\cos\left[\left(t-t'\right)(1-i \epsilon)\right]}.
\ee
Here $\alpha$ is the angle between the two points on $\mathbb{S}^{d-1}$ and $\epsilon$ is an infinitesimal positive quantity that picks out the right branch of $u^{\Delta}$.  Applying the dictionary \eqref{extdict} we then have
\be\label{sphere2pt}
\lan\Omega| T\mathcal{O}(t,\alpha)\mathcal{O}(t',0)|\Omega\ran\propto \left(\frac{1}{\cos\left[\left(t-t'\right)(1-i \epsilon)\right]-\cos\alpha}\right)^\Delta,
\ee
which is the correct two-point function in conformal field theory on $\mathbb{R}\times \mathbb{S}^{d-1}$ for a scalar operator of dimension $\Delta$.\footnote{This can be determined from \eqref{cft2pt} and \eqref{conftrans}.}

The extrapolate dictionary works not only for scalars; indeed any CFT has a unique energy momentum tensor $T_{\mu\nu}$ which is a spin two primary operator of dimension $d$.  Its bulk dual is the metric tensor, which of course should exist in any theory of gravity.  Moreover if the CFT has global symmetries then by Noether's theorem it must have conserved currents of dimension $d-1$, and these are dual to gauge fields in the bulk.  There are also other interesting items in the dictionary, among them Wilson lines \cite{Maldacena:1998im} and von Neumann entropy \cite{Ryu:2006bv,Hubeny:2007xt,Nishioka:2009un,Lewkowycz:2013nqa} in the boundary theory.  I will briefly discuss the latter in section \ref{RTsec} below.

Returning now to question (2), recall that in the original example of AdS/CFT the number $N$ of $D3$ branes is what set the AdS radius in Planck units.  It was only in the large $N$ limit that bulk gravity was approximately classical.  We'd now like to extend this to a general statement about AdS/CFT.  A general CFT has a large $N$ limit if there exists a parameter $N$, possibly discrete,  such that the set of primary operators whose dimensions do not scale with $N$ have the following properties:\footnote{This list is not necessarily exhaustive, I'm not sure if a completely sharp definition of what we mean by having a large-$N$ limit exists.  Also note that I have intentionally excluded theories with higher-spin symmetry in the bulk; although these exist \cite{Vasiliev:1990en} and sometimes have simple gravity duals  \cite{Klebanov:2002ja,Giombi:2009wh,Gaberdiel:2010pz}, they are sufficiently opaque that it is unclear to what extent they are good models for conventional gravity.}
\begin{itemize}
\item There is a finite set\footnote{Or a discrete tower with $O(1)$ spacing if there is a nontrivial compact manifold $M$.} of ``single-trace'' primary operators $\mathcal{O}_i$, each of which has spin $\leq 2$.  There is only one, the stress tensor, with spin exactly two and $\Delta=d$.  If we normalize their two-point functions so that \eqref{sphere2pt} (or its higher-spin generalization) holds with no extra prefactor, then the three-point function of any three of them is suppressed by powers of $\frac{1}{N}$.  
\item For any collection of single-trace operators $\{\mathcal{O}_{i_1},\ldots ,\mathcal{O}_{i_n}\}$, there exists a ``multi-trace'' operator $\mathcal{O}_{i_1}\ldots\mathcal{O}_{i_n}$ with dimension $\Delta=\Delta_{i_1}+\ldots+\Delta_{i_n}+O(1/N)$.
\item If we normalize the two-point functions of multi-trace operators to $O(N^0)$, then their correlation functions with each other and with other single trace operators are $O(1/N)$ \textit{unless} their components can be matched in pairs.  So for example $\lan\mathcal{O}_i(x)\mathcal{O}_j(y) \mathcal{O}_i\mathcal{O}_j(z)\ran$ is $O(N^0)$, but  $\lan\mathcal{O}_i(x)\mathcal{O}_j(y) \mathcal{O}_i\mathcal{O}_k(z)\ran$ with $k\neq j$ is $O(1/N)$.  Moreover if they can be matched in pairs, then to leading order in $1/N$ the correlation function is the sum over all such matchings of the product of the two-point functions of the matched pairs.  This property is called \textit{large-N factorization}. 
\item \textit{All} operators whose dimensions are $O(N^0)$ are either single-trace primary operators, multi-trace primary operators, or their descendants.
\end{itemize}
These properties may seem somewhat adhoc, but they can be easily remembered by considering bulk Feynman diagrams in a theory where all interactions are proportional to $1/N$.  The parameter $N$ is always proportional to some power of the AdS radius in Planck units.  The last requirement is crucial; via the state operator correspondence it says that the low-energy spectrum of the CFT is consistent with weakly-coupled low-energy effective field theory in AdS.  More explicitly the states corresponding to single-trace operators are single-particle states in the bulk, while the states corresponding to multi-trace operators are multi-particle states.  We now study this in a bit more detail.  

\subsection{Perturbations of the AdS vacuum}\label{adspertsec}   
Consider a free massive scalar field in $AdS_{d+1}$, written in coordinates where the metric is \eqref{adsmetric}.  As usual we begin by looking for positive-frequency normalizable modes for use in equation \eqref{field}, which we can expand as
\be\label{adsmoded}
f_{\omega\ell \vec{m}}(t,r,\Omega)\equiv r^{-(d-1)/2}Y_{\ell \vec{m}}(\Omega) \psi_{\omega \ell}(r).
\ee
As usual $\psi_{\omega\ell}$ obeys a Schrodinger equation
\be\label{adsschr}
-\frac{d^2}{dr_*^2}\psi_{\omega\ell}+V(r)\psi_{\omega\ell}=\omega^2 \psi_{\omega\ell},
\ee
with the ``tortoise coordinate''  now being
\be
r_*=\arctan r
\ee
and the potential being
\be
V(r)=\left(1+\frac{1}{r^2}\right)\left[\left(m^2+\frac{(d-1)(d+1)}{4}\right)r^2+\ell(\ell+d-2)+\frac{(d-1)(d-3)}{4}\right].
\ee
This potential diverges at the boundary $r_*=\pi/2$, so $\psi_{\omega \ell}$ is confined to lie in the region $r_*\in (0,\pi/2)$.  The basic features of the solutions are thus the same as for the infinite square well system.  In particular the frequency $\omega$ will be quantized with a constant high-energy density of states (at fixed $\ell$).  This is consistent with the ``gravity in a box'' intuition for physics in $AdS$; single particle states created by the creation operators for these modes have no continuous quantum numbers.  The ``ground state'' of this Schrodinger problem with $\ell=0$ corresponds to a particle localized ``at rest'' in the center of AdS, and excited states and/or states with $\ell>0$ correspond to the particle moving around. 

In fact in this case the modes can be found analytically \cite{Breitenlohner:1982jf}, and the quantization condition is\footnote{For people who are familiar with it, this is the ``standard quantization'' of the field \cite{Breitenlohner:1982jf,Klebanov:1999tb}, where the boundary conditions require the modes to behave like $r^{-\Delta}$ at infinity.  Generically normalizability requires this choice, but if $\frac{d}{2}<\Delta<\frac{d+2}{2}$ then we can instead choose modes which behave like $r^{\Delta-d}$ at infinity.  The operator $\mO$ will then have dimension $\Delta_-\equiv d-\Delta$, and formulas for that case can be obtained from those here by replacing $\Delta \to\Delta_-$.}
\be\label{adsquant}
\omega_{n\ell}=\Delta+\ell+2n \qquad \qquad n=0,1,2,\ldots,
\ee
with $\Delta$ given by \eqref{Deltaeq}.  As expected, the lowest energy state for the particle has $\ell=n=0$.  

Although we won't really need it, for posterity a properly normalized (in the KG norm) expression for the mode is
\begin{align}\nonumber
r^{-\frac{d-1}{2}}\psi_{n\ell}(r)=&\frac{1}{\Gamma(\Delta-d/2+1)}\sqrt{\frac{\Gamma(n+\Delta-d/2+1)\Gamma(n+\Delta+\ell)}{\Gamma(n+1)\Gamma(n+\ell+d/2)}}\\
&r^{-\Delta}\left(1+\frac{1}{r^2}\right)^{-\frac{\ell+\Delta+2n}{2}}F\left(-n,-n+1-\ell-d/2,\Delta-d/2+1,-\frac{1}{r^2}\right),\label{adsmode}
\end{align}
where $F$ is a hypergeometric function that goes to $1$ as $r\to \infty$ and is actually equivalent to some polynomial for $n= 0,1,2,\ldots$  This quantization condition is necessary for this mode to also be normalizable near $r=0$.

The formula \eqref{adsquant}  has a nice CFT interpretation via the state-operator correspondence.  The state created by the $\ell=n=0$ creation operator is the CFT state produced by inserting the single-trace primary $\mathcal{O}$ dual to $\phi$ into center of the Euclidean path integral as in figure \ref{stateop}, and the various excited states come from inserting its descendants.  As a simple check consider the states with energy $\omega=\Delta+2$ in $AdS_4$.  There are two types: a single state with $\ell=0, n=1$ and five states with $\ell=2, n=0$.  In the CFT we get descendants of this dimension by acting on $\mathcal{O}$ with two derivatives.  There is an angular momentum singlet where we contract the two derivatives and a traceless symmetric tensor where we don't; the latter has five linearly independent components so the degeneracies match.  It isn't hard to generalize this counting to arbitrary $d$, $n$, and $\ell$, and needless to say it works.

We can also understand multiparticle states along these lines.  The ground state with no particles is just the CFT ground state, produced by inserting the identity in the Euclidean path integral.  The rest of the multiparticle Fock space can be built in the CFT by inserting multitrace operators.  In fact this discussion can be condensed into the single statement that to leading order in $1/N$ the operator Fourier transform $\mathcal{O}_{n\ell\vec{m}}$ of the single trace primary operator $\mathcal{O}$ has an algebra consistent with it being proportional to the lowering operator $a_{n\ell\vec{m}}$ for the mode $f_{n\ell\vec{m}}$ \cite{Banks:1998dd}.  We can determine the constant of proportionality by comparing equations \eqref{field}, \eqref{extdict}, and \eqref{adsmode} in the limit $r\to \infty$, to find
\be
\mathcal{O}_{n\ell\vec{m}}=\frac{1}{\Gamma(\Delta-d/2+1)}\sqrt{\frac{\Gamma(n+\Delta-d/2+1)\Gamma(n+\Delta+\ell)}{\Gamma(n+1)\Gamma(n+\ell+d/2)}} \,a_{n\ell\vec{m}}.
\ee
Since the left hand side of this equation is a CFT operator, together with equation \eqref{field} this then allows us to write a CFT expression for a local bulk field at \textit{any} bulk point!  This may seem disturbing, given our general arguments that this should not be possible, but remember that this construction is only valid to leading order in $1/N$, and only in states close to the vacuum.  It can be ``fixed up'' perturbatively in $1/N$ \cite{Kabat:2011rz,Heemskerk:2012mn}, but there is no reason to expect a generalization that holds non-perturbatively and good reasons not to.\footnote{You  might also worry about the gauge invariance of ``local'' bulk operators, but this can be dealt with by first fixing a gauge and then defining these operators \cite{Kabat:2012av,Heemskerk:2012np}.  This is somewhat similar to the situation with computing primordial density perturbations produced during inflation in cosmology \cite{Maldacena:2002vr}.  In both cases however at higher orders in perturbation theory it is not completely clear whether or not the gauges used are ``physical'' in the sense that the quantities which appear simple are actually quantities that we observe.}  The regime of validity of this construction of local bulk fields has recently been reinterpreted in the language of quantum error correction, a beautiful subject that is unfortunately beyond the scope of these notes \cite{Almheiri:2014lwa,Pastawski:2015qua}.

\subsection{One-sided AdS black holes at fixed energy}\label{microadsbhsec}
The main new feature of black holes in AdS is that their Hawking radiation is reflected back by the boundary in finite time.  For small enough black holes this is not really important, since after all the entire black hole could evaporate before the radiation gets to the boundary, but as the Schwarzschild radius of the black hole approaches the AdS radius we eventually reach a point where the radiation is being reflected back into the black hole as fast as it is being emitted.  At this point the black hole never evaporates, so large enough black holes in AdS are eternal.  One thus often hears discussion of ``big'' and ``small'' black holes in AdS, with the distinction almost always meaning that the big ones are stable and the small ones aren't.\footnote{This distinction is a bit subtle, since as we will see the transition happens at different values of the energy in the microcanonical and canonical ensembles.  People usually seem to have the canonical transition in mind, and since as we will see it happens at higher energy it is always safe to assume this when a big black hole is being discussed.}  

The crossover point between stability and instability can be estimated by a simple statistical argument \cite{Horowitz:1999uv}.  A typical state of energy $E$ in the CFT will have some fraction $x$ of its energy in a black hole and the rest in the radiation field.  We are interested in finding the $x$ which maximizes the total entropy, which (in $AdS_4$, ignoring $O(1)$ factors, and temporarily restoring $r_{ads}$) is approximately
\be
S\approx(E \ell_p)^2 x^2+(E r_{ads})^{3/4}(1-x)^{3/4}.
\ee
Here the first term is the black hole entropy and the second term is the entropy of the radiation field, which we can think of being in a box of linear size $r_{ads}$.\footnote{I've assumed here that the Schwarzschild radius of the black hole is small enough compared to $r_{ads}$ that we can use the usual Minkowski formula for the entropy, we will see momentarily that this is self-consistent.}  At low energies the second term dominates and the function decreases monotonically; we maximize the entropy by taking $x=0$ so typically there is no black hole.  This matches onto our Minkowski intuition that black holes should evaporate.  At sufficiently large $E$ however the first term dominates so we win by taking $x$ to be very close to one; in fact there is a local maximum that is near but not quite at $x=1$.  Thus almost all of the energy (and entropy) are contained in a single black hole that never evaporates.  The crossover apparently happens when the two terms are of comparable size, which happens when
\be\label{MCcross}
Er_{ads}=\left(\frac{r_{ads}}{\ell_p}\right)^2\left(\frac{r_{ads}}{\ell_p}\right)^{-\frac{2}{5}}.
\ee
I have written it this way because the first term is the energy in AdS units of a black hole whose Schwarzschild radius is of order $r_{ads}$, so we see that the crossover happens when the black hole is parametrically smaller in $\frac{r_{ads}}{\ell_p}$ than the AdS radius \cite{Horowitz:1999uv}.\footnote{This holds up in $AdS_{d+1}$, with the suppression being $\left(\frac{r_{ads}}{\ell_p}\right)^{-\frac{(d-1)(d-2)}{2d-1}}$.  If there is a large compact manifold, such as for the $AdS_5\times \mathbb{S}^5$ example, then $d+1$ is the total number of large dimensions.}  

This crude argument can't really tell us what happens when $r_s$ of order the size of the AdS radius or larger, so to proceed further we need to address this.  The $AdS_{d+1}$ version of the Schwarzschild geometry, which is the unique spherically symmetric solution of Einstein's equation with negative vacuum energy, has (for $d\geq 3$) metric (again setting $r_{ads}=1$)
\be\label{adsschg}
ds^2=-f(r)dt^2+\frac{dr^2}{f(r)}+r^2 d\Omega_{d-1}^2.
\ee
Here
\be\label{fdef}
f(r)=\left(r^2+1-\frac{\alpha}{r^{d-2}}\right),
\ee
and $\alpha$ is related to the $AdS$ version of the ADM mass $M$ as
\be
\alpha=\frac{16\pi G M}{(d-1)\Omega_{d-1}}.
\ee
The Schwarzschild radius $r_s$ is the unique positive root of $f$.  By demanding that the Euclidean version of this geometry be smooth at $r=r_s$ as we did for the Minkowski black hole in section \ref{eucblacksec}, one finds the temperature is
\be\label{adsT}
T=\frac{d-2+dr_s^2}{4\pi r_s}.
\ee
For $r_s\ll 1$ and $d=3$ you can check that this agrees with \eqref{Thawk} above. 
Combining this with the expression
\be\label{adsM}
M=\frac{(d-1)\Omega_{d-1}}{16\pi G} r_s^{d-2}\left[1+r_s^2\right]
\ee
for the energy we can integrate to find the entropy
\be\label{adsS}
S=\frac{\Omega_{d-1} r_s^{d-1}}{4 G}=\frac{A}{4G}.
\ee
Thus as we keep increasing the energy, $r_s$ and the entropy both continue to grow.  There is less and less room for the radiation gas, so the black hole continues to win entropically.  Through the AdS/CFT dictionary we thus arrive at the following statement: \textit{at sufficiently large energy, almost all states in the CFT have a bulk description as a single gigantic black hole}.  More carefully, black hole states dominate the microcanonical ensemble of the CFT at sufficiently large energy.  This provides a very concrete realization of Bekenstein's proposal that black hole entropy should actually count microstates.

In the special case of $d=2$ we can actually quantitatively confirm this result.  For $d=2$ we have to replace the AdS-Schwarzschild geometry \eqref{adsschg} by the BTZ black hole \cite{Banados:1992wn}
\be
ds^2=-(r^2-r_s^2)dt^2+\frac{dr^2}{r^2-r_s^2}+r^2d\theta^2,
\ee
but the entropy and temperature are still obtained by the $d\to 2$ limits of \eqref{adsT} and \eqref{adsS}.  The energy differs from the $d\to 2$ limit of \eqref{adsM} by an $r_s$-independent shift, such that we still have $M\to 0$ as $r_s\to 0$.\footnote{Note that with this convention the energy of empty AdS, which we obtain from taking $r_s^2\to -1$, is actually negative.  Solutions with $-1<r_s^2<0$ have a naked singularity and are usually considered to be unphysical, so there is thus a nontrivial energy gap between the vacuum and the ``lightest BTZ black hole''.  It might seem more natural to define the energy so that the vacuum energy is zero, but for various reasons in $1+1$ dimensional CFT's the convention I use here is standard.}  The thermal partition function of a general unitary $1+1$ CFT with a discrete spectrum of primary operator dimensions quantized on a circle of radius $L$ has been shown by Cardy \cite{Cardy:1986ie} to scale at high temperature as
\be\label{cardyform}
Z[\beta]= e^{\frac{\pi^2 c L}{3\beta}}\left[1+O\left(e^{-4\pi^2 \Delta L/\beta}\right)\right],
\ee
where $\Delta$ is the dimension of the lowest-dimension nontrivial primary and $c$ is the ``Virasoro central charge'' of the CFT.  This central charge can also be computed in the bulk \cite{Brown:1986nw}, giving
\be
c=\frac{3 r_{ads}}{2G}.
\ee
$c$ thus plays the role of the parameter we have been calling $N$ in higher dimensions, so it should be large in a CFT with a semiclassical bulk dual.  The energy and entropy we compute from \eqref{cardyform} using \eqref{statmech} are 
\begin{align}\nonumber
E&=\frac{\pi^2 L c}{3\beta^2}\\
S&=\frac{2\pi^2 c L}{3\beta}=2\pi \sqrt{\frac{c LE}{3}}.
\end{align}
Replacing $L\to r_{ads}=1$ (as the $r\to\infty$ behavior of the metric requires) and comparing with \eqref{adsT} and \eqref{adsS}, we find precise agreement between the entropy of the CFT and the entropy of the black hole.  The energy also agrees with \eqref{adsM} up to the above-mentioned shift.  Moreover from \eqref{cardyform} this calculation stops being correct when $\beta\sim L$, which we will now see is exactly the order of the temperature where the black hole stops dominating the canonical ensemble.\footnote{An important subtlety here is that the asymptotics of \eqref{cardyform} are only really rigorous in the ``high temperature limit'' where we keep $c$ fixed and take $\beta \ll L$, whereas for the Bekenstein-Hawking entropy formula to be valid we also want the ``semiclassical limit'' $c\gg 1$ but only need $\beta \lesssim L$ to be above the Hawking Page transition.  Taking $c$ to be large may in principle interfere with \eqref{cardyform} if the density of states grows too rapidly with $c$.  For a nice recent analysis of this issue see \cite{Hartman:2014oaa}, who confirm the validity of the Cardy formula in the semiclassical limit.}  

\subsection{One-sided AdS black holes at fixed temperature and the Hawking-Page transition}
The transition from unstable to stable black holes can also be studied at finite temperature instead of finite energy, where it is possible to be more rigorous along the lines of section \eqref{eucblacksec} \cite{Hawking:1982dh,Witten:1998zw}.  We should expect the transition to happen at a temperature corresponding to a higher energy than \eqref{MCcross}, since at finite temperature it is possible for energy to be absorbed by the heat bath instead of being reflected back into the black hole, which makes it harder for the black hole to win in the canonical ensemble.  As in section \eqref{eucblacksec}, one proceeds by evaluating the Euclidean gravitational action 
\be\label{eucactads}
I_E=-\frac{1}{16\pi G}\int_{\mathcal{M}} d^{d+1}x \sqrt{-g}\left(R+d(d-1)\right)-\frac{1}{8\pi G}\int_{\partial \mathcal{M}} d^{d}x \sqrt{\gamma}K,
\ee
of the AdS-Schwarzschild geometry \eqref{adsschg} and comparing it to the Euclidean gravitational action of pure $AdS$ space with compactified Euclidean time.  In doing this one needs to cutoff the geometry at some large $r_c$ and ensure that the physical radius of the temporal $\mathbb{S}^1$ at this cutoff matches for the two geometries.  I will not work out the details, but the result is that 
\begin{align}\nonumber
I_E[g_{ads}]&=-\frac{(d-1)\beta \Omega_{d-1}}{8\pi G}\frac{1}{\sqrt{1+1/r_c^2}}r_c^{d-1}(1+r_c^2)\\
I_E[g_{sch}]&=I_E[g_{ads}]+\frac{\beta\Omega_{d-1}}{16\pi G}r_s^{d-2}\left(1-r_s^2\right)+O(1/r_c^2).
\end{align}
Here $r_s$ is related to $\beta$ by \eqref{adsT}.  The divergent parts of $I_E[g_{ads}]$ can all be canceled by adding a series of boundary terms to \eqref{eucactads} that depend only on the induced metric $\gamma$ at the boundary.  These terms correspond to possible counterterms in the CFT, and we are always free to add them without destroying the variational interpretation of \eqref{eucactads}.  In fact it is necessary to add them if we wish the ground state energy to be zero.

In any event the main point is that, unlike what we found in asymptotically flat space, $I_E[g_{sch}]-I_E[g_{ads}]$ changes sign at $r_s=1$ \cite{Hawking:1982dh}.  The two saddle points exchange dominance in the expression \eqref{saddlept} for the partition function.  This is the transition we are interested in; at sufficiently large temperatures the black hole wins, while at lower temperatures the thermal gas in AdS wins.  As expected, the transition happens at a higher temperature than what we found in the previous section for the microcanonical ensemble.  

This discussion has the great advantage over our discussion in section \eqref{eucblacksec} that we now actually know what we are computing; the thermal partition function in the CFT.  We saw already in the previous section that in $1+1$ dimensions this can be checked explicitly, and in higher dimensions the CFT interpretation of this transition is still fairly well understood \cite{Witten:1998zw,Aharony:2003sx}.  The basic idea is that for a CFT quantized on a spatial $\mathbb{S}^{d-1}$, when the temperature is considerably larger than the inverse sphere radius the system basically behaves like a gas of $\sim N^\alpha$ free particles in $d-1$ spatial dimensions, where $\alpha$ is some $O(1)$ constant.  This is consistent with our expressions \eqref{adsM}, \eqref{adsS} for the black hole energy and entropy, which at high temperature scale as $T^{d}$ and $T^{d-1}$, as they must if these quantities are to be extensive.  This will no longer be true when the temperature is less than the inverse sphere radius, which had better be the case since we know that we now have a gas of free particles in $d$ spatial dimensions in the bulk. On the CFT side the thermodynamics are now dominated by the constant modes of the fields, which in the special case of large $N$ gauge theories like the $\mathcal{N}=4$ Super Yang Mills theory in $3+1$ dimensions are dominated by the holonomies of the gauge fields about the Euclidean thermal circle.  It is reassuring to see the bulk and boundary descriptions of the physics line up in this manner.

\subsection{Fields in the AdS-Schwarzschild background}\label{fieldsonadss}
We now briefly study fields propagating in the exterior of the AdS-Schwarzschild background.  We can decompose the modes as \eqref{adsmoded}, as we did for pure AdS, but the tortoise coordinate is now defined by
\be
\frac{dr_*}{dr}=\frac{1}{f(r)},
\ee
with $f(r)$ defined by \eqref{fdef}.  For simplicity I'll take $r=\infty$ to be $r_*=0$, in which case we have
\be\label{adstort}
r_*=-\int_r^\infty\frac{dr'}{f(r')}.
\ee
The potential appearing in the effective Schrodinger equation \eqref{adsschr} is now 
\begin{align}\nonumber
V(r)=\frac{f(r)}{r^2}\Bigg[&\left(m^2+\frac{(d+1)(d-1)}{4}\right)r^2\\
&+\left(\ell(\ell+d-2)+\frac{(d-1)(d-3)}{4}\right)+\frac{(d-1)^2}{4}\cdot\frac{\alpha}{r^{d-2}}\Bigg].
\end{align}
The details here aren't too important, the main point is that, since $f(r)$ has a simple root at $r_*$, the tortoise coordinate $r_*$ now runs from $-\infty$ to $0$ as $r$ runs from $r_s$ to $\infty$.  Moreover the effective potential now vanishes as $r_*\to -\infty$, so the effective Schrodinger problem is no longer confined to a box; the modes will now have a continuous frequency spectrum.  As in our discussion of the brick wall model however, this continuum is presumably discretized by Planckian physics near the horizon.\footnote{We will see CFT evidence for this in section \ref{bigUsec} below.}    

It is worthwhile to note that the decomposition into ``modes in the atmosphere'' and ``modes in the radiation'' that we found for Minkowski black holes is not really valid once $r_s \gtrsim 1$.  Once this is the case then the black hole potential barrier and the AdS barrier at $r_*=0$ essentially merge, so ``the zone'' just fills the whole AdS space.\footnote{A shallow barrier will still exist for $\ell\gg r_s^{\frac{2d}{d-2}}$, but the valley outside of it will be at very high energy so any mode localized there would be highly Boltzmann suppressed.}

It is obviously interesting to understand to what extent we can extend our discussion of constructing local bulk field operators from section \ref{adspertsec} to the AdS-Schwarzschild geometry.  The situation is more subtle than it was around the vacuum, since there we had a very clear picture of the structure of the Hilbert space from the state-operator correspondence.  Acting on the vacuum with low-dimension primaries we were able to reproduce the detailed Fock space of low-energy field theory in AdS.  A big black hole in a pure state by contrast is dual to a high-energy pure state in the CFT, which we can realize via the state-operator correspondence as the insertion of a very high dimension operator in radial quantization.  This state is part of a densely spaced ensemble in the CFT; the subspace of states in some energy width of order the temperature has a dimensionality of order $e^S$.  Understanding the details of this set of states is a hopeless task, so it is not immediately clear to what extent we can ``find'' the effective field theory Hilbert space buried within.  The continuum spacing of the classical modes is certainly consistent with this dense spacing, and introducing something like the brick wall will produce some set of states which have about the right density of states, but we'd really like to have a prescription which does not require introduction of an arbitrary cutoff.  To proceed further it so far seems to be necessary to make some sort of \textit{typicality} assumption about the state of the black hole.  At least to leading order in $1/N$ this allows us to replace the detailed choice of pure state with a thermal density matrix, about which much more is known.  In particular one can argue that the Fourier modes of CFT primary operators continue to behave in thermal expectation values as if they were creation and annihilation operators, but now for the Schwarzschild modes we've been discussing \cite{Papadodimas:2012aq}.  This then seems to allow an expression \eqref{field} for the bulk field, at least for fields located \textit{outside} of the horizon.  One can then attempt to extend the definition behind the horizon \cite{Fidkowski:2003nf,Papadodimas:2012aq,Kabat:2014kfa}, at least for the two-sided Schwarzschild geometry I'll discuss in more detail momentarily, but it seems to me that there is considerably more to be said about this than what is currently known, especially about the regime of validity of the perturbative expansion in $1/N$.\footnote{It has been pointed out that, even if we stay outside the horizon, this construction cannot immediately be written in position space with the bulk field realized as an integral of some kernel times the position space CFT operator, contrary to the situation near the vacuum where it can \cite{Leichenauer:2013kaa}.  This does not seem to me to be an insurmountable obstruction, for two reasons: first of all we can just work in terms of the modes and not ask for such a formula.  Secondly, if we allow ourselves to smear the bulk operator over a small region, then there \textit{is} an expression of the desired form.  In this sense the non-convergence discussed in \cite{Leichenauer:2013kaa} is similar to the observation that formally the standard expression $\int \frac{d^4k}{(2\pi)^4} \frac{e^{ikx}}{k^2+m^2-i\epsilon}$ for a free field propagator is divergent at large $k$, which can also be resolved by smearing (see \cite{Morrison:2014jha} for a similar perspective).}  

\subsection{Collapsing shells and the two-sided AdS wormhole}\label{ads2sidesec}
\begin{figure}
\begin{center}
\includegraphics[height=5cm]{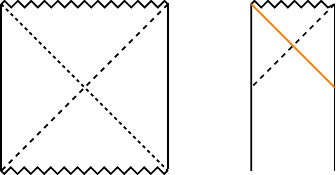}
\caption{The two-sided AdS-Schwarzschild wormhole, and a one-sided big AdS black hole formed from collapse.}\label{ads2}
\end{center}
\end{figure}

So far I have not been particularly specific about the global structure of the geometry of AdS black holes.  As in the asymptotically flat case, the full AdS-Schwarzschild geometry describes a wormhole connecting two asymptotic regions.  Here each ``exterior'' region is asymptotically AdS.  We can see this explicitly by introducing an AdS version of Kruskal coordinates:\footnote{The nontrivial factors of $f'(r_s)$ are needed to ensure that $U$ and $V$ stay real under analytic continuation.  In section \ref{krusksec} we had $f'(r_s)=1/r_s$, so in the $r_s=1$ convention we were using we didn't need them.}
\begin{align}\nonumber
U&\equiv -e^{\frac{r_*-t}{2}f'(r_s)}\\
V&\equiv e^{\frac{r_*+t}{2}f'(r_s)}.
\end{align}
As in section \ref{krusksec} there are two singularities at some positive value of $UV$, but now there are also two AdS boundaries at $UV=-1$.  Defining $U=T-X$ and $V=T+X$ we can write the metric as
\be
ds^2=4\frac{f(r)}{f'(r_s)^2}e^{-r_*f'(r_s)}\left(-dT^2+dX^2\right)+r^2 d\Omega_{d-1}^2.
\ee
Although not completely obvious, using \eqref{adstort} one can see that this geometry is smooth at the horizon at $UV=0$.  As usual the Penrose diagram is more illuminating however, it is shown on the left part of figure \ref{ads2}.  I also show the one-sided geometry for a stable AdS black hole, created by some sort of collapsing shell.  Note that in order to avoid the shell reflecting off of the boundary and going back in as we go back in time, we need to have it enter the system from the outside at $t=0$.  Such a one-sided state will \textit{not} be typical, after all its time reverse would be a black hole which spontaneously spits out all of its mass into a single narrow shell.  A typical big black hole would be one that we assemble over a time that is exponentially long in $N$.  

\begin{figure}
\begin{center}
\includegraphics[height=4.5cm]{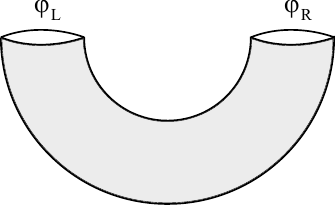}
\caption{The CFT construction of the Hartle-Hawking state.  In the CFT lives on the boundary of the tube, so the field configurations at either end describe CFT field configurations on $\mathbb{S}^{d-1}$.}\label{cfthh}
\end{center}
\end{figure}

As with the two-sided asymptotically Minkowski black hole we discussed earlier, there is a natural choice of bulk ground state for the AdS wormhole: the Hartle-Hawking \eqref{hhs}.  As the geometry has two asymptotically AdS boundaries, the ``extrapolate'' dictionary strongly suggests that it should be realized as a state in the Hilbert space of \textit{two copies} of the CFT \cite{Maldacena:2001kr}.  Indeed by comparison with equation \eqref{hhs} there is a natural candidate for the CFT version of the Hartle-Hawking state:
\be\label{CFTHH}
|\psi_{HH}\ran\equiv \frac{1}{Z}\sum_i e^{-\beta H/2}|i^*\ran_L|i\ran_R,
\ee
where the states $|i\ran_R$ are now understood as being energy eigenstates of a single copy of the CFT, and $|i^*\ran_L=\Theta|i\ran_R$, where $\Theta$ is an antiunitary operator that exchanges the two CFT's and reverses the direction of time in each.\footnote{$\Theta$ should not be confused with the natural $CPT$ operation on a single copy of the CFT on a sphere, which reverses time but also reverses a longitudinal direction within the sphere.}  This state is generated by the CFT Euclidean path integral on an interval times a sphere, as shown in figure \ref{cfthh}.  As in the asymptotically Minkowski case we discussed earlier, the Hartle-Hawking (or Hartle-Hawking-Israel) state is also often called the thermofield double state. 

This proposal for the CFT description of the two-sided wormhole may seem obvious, but it has a rather surprising consequence.  The Hamiltonian of the joint system is just the sum of the two CFT Hamiltonians; there are no interactions between the two CFTs.  This is consistent with the bulk picture, where the ADM Hamiltonian is a sum of two boundary terms, one localized at each of the two boundaries, but it leads to the rather striking conclusion that two completely non-interacting systems can nonetheless have an alternate description where there is a single connected geometry where observers from the right and the left can jump in and meet each other in the middle.  This is made possible by the diffeomorphism invariance of gravity; the bulk interactions that enable such a meeting are buried in the Hamiltonian constraint of canonical gravity and are invisible in the gauge-invariant CFT description of the system.  This has led to a more general proposal that ``entanglement generates geometry'' \cite{VanRaamsdonk:2010pw}, which recently has been given the rather amusing name of ``ER=EPR'' \cite{Maldacena:2013xja}.\footnote{If you live in a cave then this refers to the classic work of Einstein and Rosen on wormholes and Einstein, Podolsky, and Rosen on entanglement \cite{Einstein:1935tc,Einstein:1935rr}.}

The two-sided AdS Schwarzschild wormhole is perhaps the best understood of all black hole-type systems, and it has justly taken a central role in many recent analyses of black hole physics.  We will meet it again frequently in the remainder of these notes.  

\subsection{The information problem in AdS/CFT}
We now have all the pieces on the table, so we can return to the black hole information problem.  I will discuss it for small black holes in this subsection and big black holes in the following one.  The obvious thing to do is embed Hawking's original thought-experiment of forming a black hole and watching it evaporate into AdS/CFT and see what happens.  We will need the black hole to be large enough to be semiclassical:
\be\label{adssc}
E r_{ads}\gg \left(\frac{r_{ads}}{\ell_p}\right),
\ee
but small enough to evaporate:
\be\label{adsevap}
E r_{ads}\ll \left(\frac{r_{ads}}{\ell_p}\right)^{\frac{d^2-1}{2d-1}}.
\ee
Here \eqref{adsevap} is the $d+1$-dimensional version of \eqref{MCcross}; remember that it is stability in the microcanonical ensemble that decides whether or not a black hole of fixed energy evaporates.  To create the black hole, we can act with the CFT creation operators we defined in section \ref{adspertsec} to create an infalling spherical shell of matter that from the bulk point of view is expected to collapse into a black hole.  We can then evolve this state forward in the CFT and see what it looks like after a time which is greater than the bulk evaporation time.  This evolution is unitary, so to the extent that AdS/CFT is a definition of the bulk theory, this resolves the information problem in the sense of telling us the answer; information is preserved.  

Because of the strongly-coupled nature of the CFT it is difficult to actually compute the result of this evolution, but after we have evolved long enough for the CFT to thermalize there is a fairly simple argument that the state we get should typically have a bulk interpretation as a cloud of radiation in a pure quantum state with no significant projection onto any state whose semiclassical interpretation is not clear.  Recall from section \ref{adspertsec} that in the CFT we can produce multi-particle states by acting repeatedly on the vacuum with the Fourier modes of single-trace operators.  Once we act with enough operators to get to energies of order \eqref{adssc} there will occasionally be states where the bulk wave packets are so close together that they collapse to form small black holes, but as long as we stay below the energy \eqref{adsevap} we expect from the bulk discussion around equation \eqref{MCcross} that states where the bulk radiation cloud is weakly coupled should be the most entropic.\footnote{For $E\ell_p\ll 1$ this follows from our large $N$ assumptions about the spectrum of low dimension operators in the theory, in the intermediate energy regime we are considering here there is perhaps an additional assumption that the states we produce this way continue to dominate the CFT spectrum as suggested by the bulk.  In some special cases we can confirm it directly, see for example \cite{Shenker:2011zf}, but for this argument to really be airtight we would need to be able to show this in general in the strongly-coupled CFT.  For the conclusion to fail however we would need to be totally wrong in this estimate, missing a few states wouldn't change anything.}  I will refer to such states as ``radiation states'';  classically it is basically obvious that if the set of radiation states has a larger entropy than its complement we should expect to find ourselves in such a state after thermalization, but quantum mechanically it is a little less straightforward.  The set of radiation states is closed under quantum superposition,  so we can then write the CFT Hilbert space within some energy band as a direct sum of a ``radiation'' subspace whose states can be produced by acting with CFT creation operators from section \ref{adspertsec} that are separated enough in the bulk to avoid a breakdown of bulk effective field theory, and its complement, which I'll call the ``black hole'' subspace.  At risk of hitting a mosquito with a sledgehammer, we can then use our unitary integration technology to show that a typical state in this energy band has almost no projection onto the black hole subspace, or more carefully that a typical state is exponentially close in the trace norm to its projection onto the radiation subspace.  As in our discussion of Page's theorem, we can write the typical state as
\be
|\psi(U)\ran=U|\psi_0\ran,
\ee
where $|\psi_0\ran$ is some reference state and $U$ is chosen randomly from the Haar measure.  The average trace norm distance (see section \ref{pagesec}) between $|\psi(U)\ran$ and its projection onto the radiation subspace then obeys
\begin{align}\nonumber
\int dU&\Big|\Big||\psi(U)\ran \lan \psi(U)|-\frac{1}{\lan\psi(U)|\Pi_{rad}|\psi(U)\ran}\Pi_{rad}|\psi(U)\ran \lan \psi(U)|\Pi_{rad}\Big|\Big|_1\\\nonumber
&=2\int dU \sqrt{\lan \psi(U)|\Pi_{bh}|\psi(U)\ran}\\\nonumber
&\leq 2 \sqrt{\int dU \lan \psi(U)|\Pi_{bh}|\psi(U)\ran}\\
&=\frac{2e^{-(S_{rad}-S_{bh})/2}}{\sqrt{1+e^{-(S_{rad}-S_{bh})}}}.
\end{align}
Here $\Pi_{rad}$ is the projection onto the radiation subspace, $\Pi_{bh}$ is the projection onto the black hole subspace, and as in the proof of Page's theorem I have used Jensen's inequality and the unitary matrix technology of appendix \ref{Uintsec}.  Since we have $S_{rad}>S_{bh}$ by assumption, and since if we aren't right up against the energy scale \eqref{adsevap} their difference will be fairly large, we see that for all practical purposes the pure quantum state in the CFT that results from collapsing a shell and then waiting for the system to equilibrate will almost always be ``all radiation''.  

What this ``before'' and ``after'' analysis leaves unclear is what happened in the middle; we have not yet achieved Strominger's success criterion of actually computing the Page curve.  Nonetheless up to some mild assumptions about the structure of the CFT Hilbert space, we have ruled out remnants and information loss.\footnote{The argument against remnants can be strengthened considerably if we allow ourselves to enable large black holes in AdS to evaporate by locally coupling the CFT to an external system \cite{Rocha:2008fe}.  In this case to preserve unitarity using remnants we would need arbitrarily entropic low-energy states in the CFT, which certainly do not exist.}  This is most assuredly progress.

\subsection{Unitarity for big AdS black holes}\label{bigUsec}
The discussion of the information loss problem in the previous section was somewhat messy in that it required a treatment of the CFT realization of the bulk effective field theory Hilbert space of the radiation gas; in particular we needed an assumption about the CFT spectrum for $E\ell_p\gg 1$ which, although well-motivated, we would prefer to do without.  In fact Maldacena has pointed out that in the context of big non-evaporating AdS black holes there is still an issue analogous to the information loss problem, but which has a cleaner resolution in AdS/CFT \cite{Maldacena:2001kr} (see also \cite{Barbon:2003aq,Barbon:2014rma}).

Consider two CFT's entangled in the Hartle-Hawking state \eqref{CFTHH}.  The essence of the information problem is that according to bulk quantum field theory the correlation between an object that we throw in early and the radiation that comes out late vanishes as the radiation comes out later and later; what Maldacena pointed out is that this assertion can also be tested simply by considering the two-point function of two primary operators in one of the CFT's in the limit that we take their time separation to be large.  The idea is that in the naive bulk theory this two point function, interpreted via the ``extrapolate'' dictionary \eqref{extdict} as the boundary limit of a two-point function of bulk fields, will decay exponentially for arbitrarily long times.  By contrast in the CFT it will decay only until it is of order $e^{-S}$, after which it will undergo chaotic quasi-periodic behavior.  The latter is characteristic of unitary evolution in a system with finite entropy, so the former is inconsistent with unitarity.  Thus the bulk will reproduce the ``coarse-grained'' behavior of this CFT two-point function, but will not get the detailed late-time structure right; this lends considerable support to option (3) of section \ref{infpsec}.  The goal of this section is to explain these statements in some more detail.  

Let's first get some idea of what sort of time-dependence is expected for quantum field theory correlation functions in the thermal ensemble.  A full treatment of this subject requires some careful analytic continuations along the lines of section \ref{eucblacksec}, but we can quickly get to the main point from our existing results on the Rindler decomposition.  Recall from sections \ref{rindsec1}-\ref{rindsec2} that the reduced density matrix in the right Rindler wedge is thermal, so if we study the correlation functions in the Minkowski vacuum restricted to the wedge then we can reinterpret them as thermal QFT correlators with respect to the Rindler time $\tau_R$.  In particular consider the free massless scalar field in $3+1$ dimensions; the time-ordered two-point function is given by the $m\to 0$ limit of equation \eqref{to2pt}:
\be
\lan \Omega |T \phi(t,x)\phi(t',x')|\Omega\ran=\frac{1}{4\pi^2}\frac{1}{|x-x'|^2-(t-t')^2}.
\ee
If we take both points to lie in the right Rindler wedge, with $\tau_R=\tau$ and $\tau_R'=\xi_R=\xi_R'=\vec{y}=\vec{y}'=0$, then using \eqref{rindcoord} we can rewrite this as
\be\label{expcorr}
\lan \Omega | T \phi(\tau,0)\phi(0,0)|\Omega\ran=\frac{1}{8\pi^2}\frac{1}{1-\cosh\tau}.
\ee
Thus we see that at large time separation in the thermal ensemble the correlation decays exponentially in time.  Recall that in \eqref{rindcoord} we suppressed a length scale that sets the effective temperature for an observer at $\xi_R=0$, so it is the temperature that sets the exponential decay constant.  This is quite intuitive; whatever perturbation we put in at $\tau=0$ will be rapidly thermalized, so it will be very hard to detect the perturbation with a single local operator at much later times.\footnote{Note that conceptually this is a different exponential decay than the exponential decay with distance that we found for \textit{massive} fields in the vacuum expectation value \eqref{free2pt}.}  Moreover we saw in section \ref{fieldsonadss} that fields on the AdS-Schwarzschild background effectively all live in the thermal atmosphere, where as discussed in section \ref{2siderindsec} the Rindler wedge is a good model, so we should expect that correlation functions of boundary operators in the right exterior should have the same qualitative behavior of exponential decay at large timelike separation. This can be confirmed explicitly for the special case of $AdS_3$ and with somewhat more difficulty in higher dimensions \cite{Maldacena:2001kr}. 

Now let's consider what sort of behavior is expected for unitary systems with finite entropy. We will be interested in ``thermodynamic'' systems, where for the temperature ranges we are interested in the canonical entropy $S=-\tr \rho(\beta) \log \rho(\beta)$ is very large.  Moreover we'll assume that the energy spectrum is densely spaced in the vicinity of the energy that dominates the canonical ensemble, with typical energy spacing of order $e^{-S}$ times the temperature.  The energy levels are distributed chaotically, with the only degeneracies arising from a small number of compact symmetries such as rotations, parities, etc.  In such a system we would like to understand the long-time behavior of things like
\begin{align}\label{thermal2pt}
G(t)&\equiv \frac{1}{Z}\tr \left(e^{-\beta H}\mO(t)\mO(0)\right)\\
&=\frac{1}{Z}\sum_{ij}e^{-\beta E_i +i(E_i-E_j)t}|\mO_{ij}|^2,
\end{align} 
where $\mO(t)$ is a Heisenberg picture operator and $\mO_{ij}\equiv \lan i|\mO(0)|j\ran$, with $|i\ran$ a complete basis of energy eigenstates.  One simple way to do this is to compute the time average of $|G(t)|^2$ over some very long time $T$:
\be\label{averageG}
\frac{1}{T}\int_0^T dt |G(t)|^2=\frac{1}{Z^2}\sum_{ij,i'j'}e^{-\beta(E_i+E_{i'})}|\mO_{ij}|^2|\mO_{i'j'}|^2 \left[\frac{1}{T}\int_0^T dt e^{i(E_i-E_j+E_{j'}-E_{i'})t}\right].
\ee
As $T\to \infty$, the quantity in square brackets here is equal to one when $E_i-E_j+E_{j'}-E_{i'}=0$ and is zero otherwise; given our assumptions about the spectrum, it will be nonzero only if $E_i-E_j=E_{i'}-E_{j'}=0$ or $E_{i}-E_{i'}=E_j-E_{j'}=0$.  Thus we see that the average is finite in the limit $T\to\infty$, which means that, unlike for the black hole correlation function, $G(t)$ cannot decrease monotonically to zero at late times.  

If we don't know anything about $\mO$ we can't say much more about the late time behavior of $G(t)$, but under quite reasonable assumptions we can estimate its typical size.  First of all I will assume that there is a symmetry which acts on $\mO$ as $\mO \to-\mO$.  This is more of a convenience than a necessity, but it is consistent with $\mO$ being a CFT operator dual to a bulk field $\phi$ which has such a symmetry.  What this buys us is that, perhaps after some reshuffling of the $|i\ran$'s within degenerate eigenspaces of $H$, we have $\mO_{ii}=0$; this implies that the thermal one-point function of $\mO$ is zero.  We also would like to ensure that $G(t)$ is ``stable'' under small changes, either of $\beta$ or of the high-energy structure of the theory.  Concretely let's consider
\be
G(0)=\frac{1}{Z}\sum_{ij}e^{-\beta E_i}|\mO_{ij}|^2.
\ee
We'd like this quantity to be order one, and to be a reasonably continuous function of $\beta$.  For the sum over $j$ to even converge we need $\mO_{ij}$ fall off sufficiently fast at fixed $i$ and large $E_j$, and since we'd like to think of the operator as being a ``probe'' we can further demand that it falls off fast enough that the sum is dominated by the region where $E_j-E_i\ll \lan H\ran=\frac{1}{Z}\tr H e^{-\beta H}$.\footnote{In fact these statements will not be true for local field operators, which \text{do} actually have significant matrix elements between low and high energy states; this is the origin of the short-distance singularities in their correlation functions, which here would say that $G(t)\to\infty$ as $t\to 0$.  We can fix this by ``smearing'' the operators against wave packets whose size in space and time is of order $\beta$, which will not affect the late-time behavior that we are interested in.}  Since even in this region of $j$ there are of order $e^{S}$ states, we apparently need the individual $\mO_{ij}$'s to be $O(e^{-S/2})$.  Moreover to get a reasonable function of $\beta$ we need the dependence of $|\mO_{ij}|$ on $i,j$ to be a reasonably smooth function of $E_i$ and $E_j$; we make no similar restriction on the $i,j$ dependence of its phase.\footnote{These properties of $\mO_{ij}$ are sometimes referred to as the ``eigenstate thermalization hypothesis'', with the hypothesis being that for local Hamiltonians $H$ the local operators $\mO$ that we are interested in obey them \cite{deutsch1991quantum,Srednicki:1995pt}.  More generally if we do not assume there is an $\mO\to-\mO$ symmetry the eigenstate thermalization hypothesis says that
\be
\mO_{ij}=\mO(E_i)\delta_{ij}+e^{-S\left(\frac{E_{i}+E_{j}}{2}\right)/2}f(E_i,E_j)R_{ij},
\ee
where $\mO(E)$ and $f(E,E')$ are real smooth functions of their arguments, but $R_{ij}$ is a complex (both in the sense of ``not real'' and the sense of ``complicated'') $O(1)$ function of $i$ and $j$.}  With these assumptions we then see that the time average \eqref{averageG} is of order $e^{-2S}$, and thus that the typical late-time behavior of $G(t)$ is of order $e^{-S}$.  Over long time scales the correlator will fluctuate erratically, and over even longer timescales it will even sometimes come back up to an $O(1)$ value.

How then are we to reconcile this with the endless exponential decay of \eqref{expcorr}? The resolution of course is that, as we found in the brick wall model of section \ref{bricksec}, the entropy of the Rindler wedge is infinite due to the infinite collection of modes near the horizon in tortoise coordinates.  By studying the correlation at very large $\tau$ separation, we are directly taking advantage of these degrees of freedom near the horizon.  If the black hole entropy is actually finite, as it certainly is in AdS/CFT, then the exponential decay must eventually stop.  That the CFT agrees with the black hole result for this correlation function at short times (this is shown in more detail in \cite{Papadodimas:2012aq}), but disagrees at long times as required by unitarity, is compelling evidence for the unitarity of black hole evaporation.\footnote{In \cite{Maldacena:2001kr} Maldacena also pointed out that subleading saddles in the path integral can give rise to long-time corrections which are of order the $e^{-S}$ required by unitarity.  This is sometimes misunderstood as an argument that including such saddle points is \textit{sufficient} to resolve the information problem and reproduce the chaotic late-time CFT behavior, but this is most likely not the case \cite{Barbon:2003aq}. There are presumably other non-perturbative effects in quantum gravity beyond those suggested by Euclidean gravity, and it would be unreasonable to expect a semiclassical interpretation for all of them.}  

\subsection{Von Neumann entropy and the Ryu-Takayanagi Formula}\label{RTsec} 
We've now completed the background in AdS/CFT we need for the rest of the notes, but there is one more aspect of the correspondence that is too important to avoid mentioning entirely.  This is the proposal of Ryu and Takayanagi (RT), later generalized by Hubeny, Rangamani, and Takayanagi (HRT), for a holographic expression for the von Neumann entropy of a subregion in the boundary theory \cite{Ryu:2006bv,Hubeny:2007xt}.  A vast literature understanding and using this conjecture has appeared in the last few years, here I will only state the conjecture and mention two applications of interest to black hole physics.  

\begin{figure}
\begin{center}
\includegraphics[height=4cm]{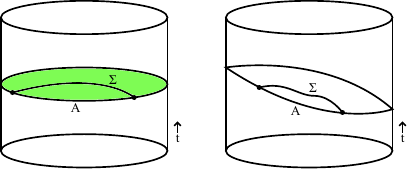}
\caption{Two examples of the RT/HRT proposal, shown in $AdS_3/CFT_2$ for simplicity.  On the left there is a time-translation symmetry (the geometry is ``stationary''), and I have chosen the spatial slice in the boundary to respect this.  The extremal surface $\Sigma$ is a geodesic of minimal area that lies in the green bulk Cauchy slice that also respects the symmetry.  This was the setup originally studied by Ryu and Takayanagi.  On the right I show a more arbitrary slice in a possibly non-stationary geometry; the extremal surface no longer lies in any preferred bulk slice.}\label{hrt}
\end{center}
\end{figure}

The basic idea is as follows; we have taken the CFT to live on $\mathbb{S}^{d-1}\times \mathbb{R}$, and in the Schrodinger picture we can pick out any particular Cauchy slice with topology $\mathbb{S}^{d-1}$ and study a quantum state of the CFT on that slice.  A natural thing we can do with the state is decompose the slice into a region $A$ and its complement $B$ and then compute the von Neumann entropy
\be
S_A=-\tr \rho_A \log \rho_A.
\ee 
The RT/HRT proposal says that to compute $S_A$ using the bulk gravity theory, we should look for the codimension two extremal-area surface $\Sigma$ in the bulk with the property that $\partial\Sigma=\partial A$.\footnote{Technically we also require that the surface $\Sigma$ is homologous to the region $A$ on the boundary \cite{Headrick:2007km}.}  If there is more than one such $\Sigma$, we take the one of smallest area.  The proposal is then that to leading order in $1/N$ we have
\be
S_A=\frac{A(\Sigma)}{4G},
\ee
where $A(\Sigma)$ is the area of $\Sigma$ in the bulk geometry.  The basic idea is illustrated in figure \ref{hrt}.  

The RT/HRT proposal has passed many nontrivial checks, especially for $AdS_{3}/CFT_2$, where both sides can be computed explicitly in many cases \cite{Hartman:2013mia,Faulkner:2013yia}.  It also obeys nontrivial properties of entropy like strong-subadditivity \cite{Headrick:2007km,Wall:2012uf}.  Indeed last year a fairly general ``heuristic proof'' was given by Lewkowycz and Maldacena \cite{Lewkowycz:2013nqa}, although the range of validity of this argument is still being actively explored by the community.  The RT/HRT conjecture has also recently been extended to bulk theories more general than Einstein gravity \cite{Dong:2013qoa}, and it also has been used to derive formulae for the areas of more general non-extremal bulk surfaces \cite{Balasubramanian:2013lsa,Czech:2014wka}.  

One word of warning, the RT/HRT proposal is often described as computing entanglement entropy, but this is misleading since it is supposed to work even if the total state is not pure.

From the point of view of these notes however, the most important applications of the RT/HRT proposal involve its use in spacetimes with black holes.  Two especially interesting examples of this are its use to study the two-sided AdS Schwarzschild geometry, by Hartman and Maldacena \cite{Hartman:2013qma}, and perturbations thereof, by Shenker and Stanford \cite{Shenker:2013pqa,Shenker:2013yza}.  The calculation of Hartman and Maldacena builds on an observation of \cite{Morrison:2012iz}, who emphasized that at $t=0$ in the Hartle-Hawking state \eqref{CFTHH} of two CFT's there is considerable ``local entanglement''.  What this means is that if I take a suitable region $A_L$ in the left CFT and its mirror region $A_R$ in the right CFT, their mutual information $I(A_L,A_R)\equiv S_{A_L}+S_{A_R}-S_{A_LA_R}$ is nonvanishing, and in fact is proportional to the black hole entropy.  This statement can be confirmed in the bulk by using the RT/HRT proposal \cite{Morrison:2012iz}.  What Hartman and Maldacena studied is how this statement evolves as we evolve time forward simultaneously on the two sides.\footnote{Note that evolving time forward on one side and backward on the other is a symmetry of the HH state \eqref{CFTHH}, but evolving both sides forward is nontrivial.}  Let's say that the linear size of the region is $L$, and that the temperature is large enough compared to the Hawking-Page transition that we can have $\beta\ll L\ll 1$ (this is what ``suitable'' means).  Hartman and Maldacena then argued that the mutual information $I(A_L,A_R)$ starts out positive and decreases linearly with time, changing nontrivially over a time of order $\beta$, until at a time of order $L$ it drops to zero.  In $1+1$ dimensions they were able to also confirm this behavior directly in the CFT.  Moreover they found that the extremal surface used in computing $S_{A_L A_R}$ over this time scale extends through the wormhole, with the details of the result depending on the metric in the interior.  The successful matching with the CFT calculation provides good evidence that, at least in this special case, the CFT knows about the interior geometry, which should thus be taken seriously despite the paradoxes I describe in the following section.  

The decrease of the mutual information $I(A_L,A_R)$ found by Hartman and Maldacena is reminiscent of thermalization, since it represents the ``dilution'' of a special property of the state at $t=0$, the local entanglement, into the rest of the system as time evolves.  Soon after Shenker and Stanford introduced a modification of this setup in which the connection to thermalization in the CFT can be made even more explicit.  Their idea was to instead study how this special property of the state is affected by introducing a small perturbation to the system at an earlier time $t=-t_w$ (with $t_w>0$).  They modeled this in the bulk theory by sending in a spherical shell of matter on one side of the wormhole at this earlier time, whose energy was of order the temperature $\beta^{-1}$.  They found that the local effect of this perturbation on the extremal surface at $t=0$ grows exponentially in $t_w$, for essentially the same reason that the center of mass collision energy grew exponentially in our discussion of the trans-Planckian problem in section \ref{infpsec}, and thus that the backreaction of the shell becomes significant for the RT/HRT calculation at a time of order 
\be
t_w\approx \beta \log S.
\ee
This is nothing other than the scrambling time \eqref{prescrt}, which we argued in a rather different way in section \ref{scrambsec} is the relevant time scale for a black hole to absorb information.  The effect of this backreaction is to make the wormhole ``longer'', which causes the mutual information to vanish since the minimal-area extremal surface used in computing $S_{A_L A_R}$ switches to a pair of surfaces that do not extend through the wormhole (this is also what causes it to drop to zero in the Hartman/Maldacena calculation).  This is essentially a bulk illustration of the ``butterfly effect'' in the CFT evolution; we perturb a state at $t=-t_w$ which was carefully tuned to produce local entanglement at $t=0$, but our perturbation prevents it from doing so.  The black hole entropy $S$ appears in the calculation since, unlike in the Hartman/Maldacena setup, we need to wait for the perturbation to be mixed throughout the system.  It is gratifying to see such a chaotic effect in the CFT's emerging from a straightforward bulk calculation.  

\section{Paradoxes for the infalling observer}\label{paradoxsec}
Having at least provisionally settled the information paradox, it is now time to return to the description of the black hole interior.  In section \ref{compsec1} we saw that unitarity of the evaporation process leads to a possible inconsistency in the description of the interior; a violation of the no-cloning theorem \cite{Susskind:1993mu}.  We saw however that this cloning seemed to be unobservable \cite{Susskind:1993mu,Hayden:2007cs}, and thus considered the possibility that the principle of ``black hole complementarity'' could allow us to formulate a theory in which no observer sees a violation of quantum mechanics, avoiding both non-unitarity outside and cloning inside.  

Many people were reasonably satisfied with this state of affairs, but there were always some lingering doubts and in particular there was a sizeable contingent of people who were never convinced \cite{Unruh:1995gn,Mathur:2009hf,Giddings:2011ks}.  If black hole complementarity is consistent and correct, shouldn't we be able to find a real theory of the interior that realizes it?  As we will now see, it seems to be the case that black hole complementarity as originally formulated cannot be consistent.  We will also see in the following subsections that there are a number of problems with any naive attempt to ``reconstruct'' the interior in AdS/CFT using the same machinery as we did for perturbations of the vacuum in \ref{adspertsec}.  As of now the status of these arguments is somewhat controversial; the reader will notice a definite decrease in the precision of the arguments as we move in from the boundary, but in my view they raise serious obstructions that any self-respecting physicist would need to address before claiming to possess a satisfactory theory of black holes.

\subsection{The entanglement-monogamy problem}\label{ampssec}
The argument against the consistency of complementarity I will present is due to Almheiri, Marolf, Polchinski, and Sully \cite{Almheiri:2012rt}, who I'll refer to as AMPS, but its basic building blocks have a long history.  Throughout the discussion of the information problem, there was some concern that changing the state of the evaporating modes from Hawking's result would be dangerous to the infalling observer \cite{Giddings:1994pj,Polchinski:1995ta,Mathur:2009hf,Giddings:2011ks,Avery:2011nb}.  In particular the main quantitative piece of the AMPS argument, based on the strong subadditivity of Von Neumann entropy, is due to Mathur \cite{Mathur:2009hf}.  Many aspects of the argument were also independently realized by Braunstein \cite{Braunstein:2009my}, who suggested the term ``energetic curtain'' for what AMPS later called a ``firewall''.  The contribution of AMPS was to assemble these pieces and use them to attack complementarity in a concrete way.  The argument I present in this section differs in detail from their argument, in particular neither strong subadditivity nor a discussion of black hole mining \cite{Brown:2012un} is needed, but the basic idea is the same.  Readers who have made it this far without reading appendix \ref{BPapp} should do so now unless they are already familiar with its contents.

\begin{figure}
\begin{center}
\includegraphics[height=8cm]{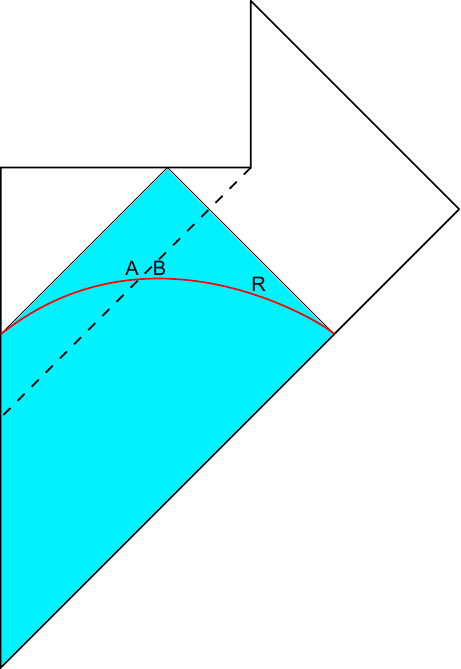}
\caption{A spatial slice in the diamond of an infalling observer.}\label{alice}
\end{center}
\end{figure}
The goal of the AMPS argument is to put all of the ``moving parts'' of the black hole information problem into the past lightcone of a single observer, preventing any use of complementarity to avoid an observable violation of effective field theory or quantum mechanics.  The basic setup is shown in figure \ref{alice}.  Rather than assuming that there is some well-defined state of the quantum fields on an entire nice slice such as the blue one in figure \ref{cloning}, we make the weaker assumption that there exists such a state only on the red slice that stays within the past lightcone of an infalling observer Alice in figure \ref{alice}. This is still a strong assumption, one that goes well beyond the asymptotic S-matrix and boundary correlator assumptions we've used so far in these notes, and it almost certainly will not survive in the correct theory of the interior.  Nonetheless it has proven quite difficult to come up with consistent alternatives that are able to reproduce quantum field theory in the expected limits, so for now we will make this assumption; it might be called ``taking the quantum mechanics of effective field theory seriously''.  We will not need to assume anything about the dynamics of how different such time slices are related, we can get into trouble just by trying to find \textit{any} quantum field theory state for this slice that is consistent both with the horizon being smooth for an infalling observer and with unitarity.

To proceed, in figure \ref{alice} I've indicated three important sets of modes.  The mode $B$ is one of the thermally-occupied Schwarzschild modes \eqref{schmode} in the atmosphere, confined between the horizon and the barrier of the effective potential \eqref{schpot} (I've assumed here that we have waited long enough after the black hole formation that the geometry in the vicinity of $B$ is well-approximated by the upper corner of the Schwarzschild geometry).  The mode $A$ is the ``mirror'' of $B$ behind the horizon; it is analytically continued up from right-moving modes near the horizon of the left exterior region of the two-sided Schwarzschild geometry in figure \ref{kruskfig}.  If the horizon is smooth, we expect $A$ and $B$ to be highly entangled as in equation \eqref{rindvacstate}.  If not then at least in quantum field theory we would expect something singular to happen at the horizon, as explained in section \ref{firewallsec}.\footnote{You may worry that we have evolved the modes up from the bifurcate horizon in figure \ref{kruskfig}, which is not part of the diagram in figure \ref{alice}, but the Hamiltonian is conserved by this evolution so we can recast the discussion of section \ref{firewallsec} entirely as a calculation at some later time where the two relevant modes \textit{are} given by $A$ and $B$ in figure \ref{alice}.}

You might think that unitarity does not impose any significant constraint on the red slice of figure \ref{alice}; after all this observer never gets to measure the S-matrix, so what would unitarity mean operationally?  One answer is provided by Page's theorem; once the black hole is sufficiently ``old'', we saw in section \ref{pagesec} that we should expect a considerable degree of entanglement between the black hole $BH$ and its early radiation $R$.  In fact we saw that, ignoring energy conservation, they should be maximally entangled.  Once we include energy conservation what should be true instead is that $BH$ will typically be \textit{thermally} entangled with $R$, in the sense that $\rho_{BH}=\frac{1}{Z_{BH}} e^{-\beta H_{BH}}$.  Moreover as discussed in section \ref{pursec} this means that, provided that $B$ is sufficiently weakly interacting with the rest of the black hole, it can be purified by some tensor factor $R_B$ in the early radiation in the sense that $\rho_{BR_B}$ is close to a pure quantum state.  This is a statement that lies within the infalling observer's diamond; it must then be a property of the state on the red slice. This is what we need to form a contradiction; if $B$ is strongly entangled with $R_B$, then it cannot also be significantly entangled with $A$ without violating a principle; the amusingly-titled monogamy of entanglement \cite{koashi2004monogamy}.  

We can illustrate the contradiction more precisely using some entropy inequalities from appendix \ref{BPapp}.  Let's begin by assuming that $B$ and $R_B$ are thermally entangled, so that we can take $S_{BR_B}=0$.  From the non-negativity of the mutual information $I_{A,BR_B}\equiv S_A+S_{BR_B}-S_{ABR_B}$ and the triangle inequality \eqref{triangle}, we see that
\be
S_A=S_{ABR_B},
\ee
and thus that
\be
I_{A,BR_B}\equiv S_A+S_{BR_B}-S_{ABR_B}=0.
\ee
We saw in appendix \ref{BPapp} that the mutual information between two tensor factors vanishes only if the state is a product state, so we thus have
\be
\rho_{ABR_B}=\rho_A\otimes \rho_{BR_B},
\ee
which is incompatible with any entanglement, or indeed correlation of any kind, between $A$ and $B$.  Conversely we could assume that $A$ and $B$ are thermally entangled, in which case we would conclude there is no correlation between $B$ and $R$.  We thus seem to be led to the apparently absurd conclusion that either unitarity is violated or sufficiently old black holes have singular horizons; this is the ``firewall'' paradox.  

One obvious issue with this argument is whether or not $B$ really is ``sufficiently weakly interacting'' with the rest of the black hole to allow us to approximate the thermal density matrix $\rho_{BH}$ as a product between a thermal density matrix for $B$ and a thermal density matrix for the rest of the black hole.  This is not a trivial point; we saw in section \ref{firewallsec} that gradient interaction between the two Rindler wedges leads to a high ``energy cost'' for firewalls, which causes them to be Boltzmann suppressed  in a thermal distribution defined with respect to the Minkowski Hamiltonian $H$.  The difference here however is that the asymptotic time translation symmetry of Minkowski space, which is conjugate to the conserved energy, behaves in the vicinity of the Schwarzschild horizon as a boost $K$ rather than $H$.  From our expression \eqref{qftboost} for the quantum field theory boost operator, we see that the gradient interactions between the two sides are suppressed by an extra factor of $x$ at the origin, which prevents them from being as restrictive.  Alternatively we saw in section \ref{bricksec} that modes with Schwarzschild energy of order the temperature can be thought of as living in a box of size $\log \frac{r_s}{\ell \ell_p}$ in the tortoise coordinate $r_*$.  This allows us to form wave packets of narrow frequency that are nonetheless well-localized away from the horizon, so any contribution from the interactions near the horizon is suppressed appropriately.  Nonetheless I think that given the stakes it would definitely be worthwhile to understand these energetics better, especially in the context of a careful treatment of the Hamiltonian formulation of canonical gravity, but I leave this to future work.\footnote{Even if there is some subtlety in general with the energetics of $B$, it seems unlikely that it would be so radical as to impose entanglement across the horizon for the lowest $\ell$ modes; these are the ones which directly carry out the information, and it is difficult to see how the evaporation could be unitary if they do not come out entangled with $R$.  For this reason people sometimes contemplate an ``s-wave firewall', which corrupts only those $B$ modes with $O(1)$ angular momentum.  This has some resonance with the second energetic argument just given, since for sufficiently large $\ell$ we cannot think of $\log \frac{r_s}{\ell \ell_p}$ as being large; perhaps energetics prevent a full firewall but allow an s-wave one?}

\subsection{Firewall typicality}\label{ftypsec}
The AMPS argument of the previous section does not obviously apply to big asymptotically AdS black holes; they do not evaporate.  One way to deal with this is to enable them to evaporate, either by coupling the CFT to an external system or by ``mining'' them, but it would be nice to have a version of the paradox that does not require evaporation.  Indeed the only thing we really needed the evaporation for in the previous section was to argue that the black hole was in a thermally mixed state; at least if we are playing by the usual rules of quantum mechanics, and assuming some crude form of locality, then any local experiment in the vicinity of the black hole should not care whether this mixed state is purified by the early Hawking radiation.\footnote{We will see some tension with this statement in section \ref{mwsec} below, as well as in section \ref{ereprsec}.}  If we instead interpret the thermal density matrix as a classical ensemble of pure states, then the energetic argument of the previous section suggests that firewalls are ``typical'' in this ensemble; unlike in ordinary Minkowski space they are not energetically suppressed \cite{Bousso:2013wia}. 

One can make this ``typicality'' argument more precisely for big AdS black holes \cite{Marolf:2013dba}.  Consider again our Schwarzschild mode $B$, which has a frequency of order the temperature and is localized away from the horizon.  Since we are now considering big AdS black holes, meaning for simplicity black holes that are stable in the canonical ensemble and thus have a Schwarzschild radius at least of order the AdS radius, the mode $B$ will automatically be in the thermal atmosphere (``the zone''), since as we saw in section \ref{fieldsonadss} this fills the entire space.  The AdS/CFT description of such black holes is as excited states of a conformal field theory quantized on a spatial sphere cross time.  We also saw in section \ref{fieldsonadss} that to leading order in $1/N$ we can plausibly interpret the operator Fourier transform $\mO_{\omega \ell m}$ of a single-trace primary $\mO$ dual to a bulk field $\phi$ as the annihilation operator for a bulk Schwarzschild mode such as $B$, and in particular we can define a number operator
\be
N_B\equiv \mO_{\omega \ell m}^\dagger\mO_{\omega \ell m}+O(1/N).
\ee  
The operator $\mO_{\omega \ell m}^\dagger\mO_{\omega \ell m}$ exactly commutes with the CFT Hamiltonian, so to leading order in $1/N$ it seems we can simultaneously diagonalize $N_B$ and $H$.  This is rather problematic, since it implies that we can find a complete basis for the microcanonical ensemble (the subspace of all states built from energy eigenstates within a narrow energy width centered at some large energy $E_0$) with the property that each basis element is an eigenstate of the occupation number for $B$.  Such states are very far from the expected entanglement of equation \eqref{rindvacstate}, so they cannot be expected to locally resemble the Minkowski vacuum.  

We'd now like to argue that the existence of a complete basis of ``bad'' states implies that almost all pure states sampled from the microcanonical ensemble are ``bad'', but to really make this argument we need to make one final assumption based on the linearity of quantum mechanics.  Namely we assume that in addition to the existence of a CFT operator $\mO_{\omega \ell m}$ which annihilates $B$, there is \textit{also} a CFT operator $\tO_{\omega \ell m}$ which annihilates $A$.  Unlike $\mO_{\omega \ell m}$ we do not have an explicit expression for this operator, but it is quite natural to assume it exists; after all the occupation number for $A$ is an observable and thus it should correspond to a self-adjoint operator according to the general principles of quantum mechanics.  Given this operator we can then define the annihilation operator
\be
c_{1\omega \ell m}=\frac{1}{\sqrt{1-e^{-\beta\omega}}}\left(\mO_{\omega \ell m}-e^{-\beta\omega/2}\tO_{\omega\ell (-m)}^\dagger\right),
\ee
which we saw in sections \ref{rindsec2}, \ref{firewallsec} can be used as ``diagnostic'' for firewalls.  Indeed we saw that its number operator $c^\dagger c$ will annihilate any state with a smooth horizon, whereas it will have $O(1)$ expectation value in a disentangled state.  If we now compute the expectation value of $c^\dagger c$ in the microcanonical ensemble however, we are free to use the basis of $N_B$ eigenstates we just described.  In this case the average will clearly be $O(1)$.  Moreover it is a general fact (see section 2.4 of \cite{Harlow:2014yoa} for a review) that the microcanonical expectation value of some operator is exponentially close to the expectation value of the same operator in a randomly chosen pure state from the ensemble.  Thus $c^\dagger c$ will have an $O(1)$ expectation value in almost all pure states; since this argument can be applied to any mode, firewalls that corrupt all $B$ modes are typical \cite{Marolf:2013dba}.  

This argument has several technical points which have not been completely explicated in the literature, in particular I'm not really sure that the $1/N$ corrections are under control, and I am also not totally sure about the bulk interpretation of $\mO_\omega$.  Nonetheless so far no decisive objection has been raised, and even if this argument does not end up being correct it will be illuminating to understand why it fails.  

One aspect of the argument of this subsection which is less satisfying than the original AMPS argument of the previous subsection is that it is less operational; there is no obvious low-energy experiment which illustrates the paradox.  I'll return to this in section \ref{savecomp} below.

\subsection{The creation operator problem}
The paradoxes of the previous two subsections arose from taking the quantum mechanics of bulk effective field theory seriously.  In this subsection and the next, I will describe two additional arguments that suggest that one might not want to do this.  

I'll first focus on the mode $A$, which lives just behind the horizon.  One way to think about our construction of the CFT representation of creation and annihilation operators for the mode $B$ is that it proceeds by solving the bulk equations of motion in from the boundary, with the boundary conditions given by the extrapolate dictionary \eqref{extdict} \cite{Heemskerk:2012mn}.  A similar evolution for the mode $A$ however would involve either propagating forward to the singularity or backwards through the trans-planckian bulk collision with the infalling shell, as described in section \ref{infpsec}.  This means that any attempt to produce a CFT expression for creation/annihilation operators for $A$ via the same method will require an understanding of trans-planckian bulk physics; a tall order \cite{Almheiri:2013hfa}.  

In fact there is a more basic problem with finding a CFT expression for this mode.  Consider a candidate raising operator $\tO^\dagger$ for $A$.  Within effective field theory we expect this operator to \textit{lower} the energy of the quantum state; remember that the Schwarzschild isometry acts within the vicinity of the horizon as a boost, and the $A$ mode has negative boost energy.  Moreover within effective field theory there are no states that this operator can annihilate.  If we assume that these two properties are true for the CFT operator $\tO^\dagger$ we reach a contradiction; the density of states of the CFT decreases as we go to lower energy, so it is impossible for an operator that lowers the energy on all states to not have any states it annihilates.  Thus we seem to have an obstruction to a naive representation of creation operators for the $A$ modes within the CFT \cite{Almheiri:2013hfa}.

It is not totally clear to me that constructing the interior really requires an operator with the assumed properties, as mentioned above I would like to see a more careful treatment of energetics and bulk diffeomorphism invariance, but it is interesting to note that this argument seems to \textit{oppose} the firewall typicality argument of the previous subsection; it says that one of the assumptions of that argument, the existence of a $\tO$ operator in the CFT with the expected properties from effective field theory, can't be true.  On some level this is encouraging; it means that rather than accepting firewalls we should look for a less naive description of the interior.\footnote{This objection does \textit{not} invalidate the argument that there is a complete basis of $N_B$ eigenstates spanning the microcanonical ensemble; this itself is already uncomfortable.}  

\subsection{The Marolf-Wall paradox}\label{mwsec}
Another interesting obstruction to a naive ``inclusion'' of effective field theory into the CFT was pointed out by Marolf and Wall \cite{Marolf:2012xe} (see also \cite{Avery:2013exa} for more discussion).  The idea is as follows; consider a single large-N CFT in a thermal density matrix
\be
\rho_{CFT}=\frac{1}{Z}e^{-\beta H},
\ee
with $\beta>1$ so that black holes dominate the ensemble.  In ordinary quantum mechanics, we are free to interpret this state either as a classical probability distribution for energy eigenstates or as being purified by an auxiliary system.  But for the black hole interior it seems like there is a difference between the two cases.  

Say that we view this density matrix as a probability distribution for one-CFT states.  According to the rules of quantum mechanics there should be linear operators on the single CFT whose expectation values we can compute to see what is going on in the interior.  But alternatively say we view this density matrix as being purified by a second copy of the same CFT, and moreover say that we choose the joint system to be in the thermofield double state \eqref{CFTHH}.  As described in section \ref{ads2sidesec}, this system is usually interpreted as describing the two-sided AdS/Schwarzschild geometry, with the two sides connected by a wormhole.  For definiteness we can interpret the left CFT as the ``auxiliary'' one and the right CFT as the one we started with.  There is now something of a paradox; in the two sided system there is nothing to stop us from acting on the left CFT with a unitary operator that sends a signal into the wormhole.  An observer on the right side could jump into the wormhole and receive this signal.  But if he/she uses the right-CFT operators we just motivated in the single-CFT case, their expectation values will be identical whether or not we send a signal from the left!  Any dictionary for the interior which could detect this signal would clearly need to involve operators from both CFT's.   

What are we to make of this?  Marolf and Wall suggested that we need additional degrees of freedom beyond the two CFT's in describing this setup.  They wanted to use these extra degrees of freedom to distinguish between two black holes which just happen to be entangled with each other but don't share a common interior (interpretation one of the previous paragraph) and two entangled black holes connected by a wormhole (interpretation two).  I find it more natural to say that there are just two different interpretations of the two-CFT system, one which assumes there is a bridge and the other which doesn't.  The Hamiltonian and Hilbert space are the same in both cases but the dictionary for observables is different.  It is interesting that we seem to have this choice in defining the dictionary, there does not seem to be an analogous ambiguity for observables that are not behind horizons, since they can always be unambiguously evolved back to the boundary and matched onto the ``extrapolate'' dictionary.  

\section{Proposals for the Interior}\label{altsec}
We have now seen that there are several interesting obstructions to a quantum description of the black hole interior. In this section I will describe what I consider to be some of the more promising ideas that have been proposed to resolve the paradoxes of the previous section.  Since none of these ideas are unambiguously successful, I will be brief.  The reader should beware that my own biases will be at most on display in this section.  

\subsection{Complementarity from computational complexity?}\label{savecomp}
The AMPS argument of section \ref{ampssec} presented a thought-experiment testing the unitarity of black hole evaporation that appears to be doable by a single observer without violating causality.  Could it be however that there is some principle other than causality that prevents the experiment from being done?  If so, then it may be that some version of black hole complementarity as described in section \ref{compsec1} may yet provide an escape from the apparent inconsistency of unitary evaporation and smooth infall without requiring an observable breakdown of some principle of physics.  Several reasons why the AMPS experiment might be impossible have been proposed \cite{harlow2,Hui:2013jfa,Ilgin:2013iba,Freivogel:2014fqa}, but in my view the most robust is my own observation with Patrick Hayden that the computational complexity of ``distilling'' the purification $R_B$ from the Hawking radiation is so severe that it almost certainly requires an exponential amount of time in the entropy of the black hole \cite{Harlow:2013tf}.  This far exceeds the evaporation time, which is only of order $S^{3/2}$ in Planck units, so the infalling observer of figure \ref{alice} will not succeed in extracting $R_B$ until long after the black hole has evaporated and she can no longer jump in and see a contradiction.  In the remainder of this subsection I explain why this is probably the case, and then briefly comment on to what extent this is a satisfactory resolution of the paradoxes of the previous section.

To be concrete we can model the old black hole of section \ref{ampssec} as a qubit system, with the mode $B$ taken to be a single qubit, its complement $H$ in the black hole taken to be $m$ qubits, and the radiation $R$ being a further $n$ qubits.  The black hole will be old in the Page sense if $n\gg m$.  We can take the state of the system to be a random pure state of $BHR$, which by Page's theorem will be maximally mixed on $BH$.  By the Schmidt decomposition of appendix \ref{BPapp}, we can represent the state as 
\be\label{psi}
|\psi\ran=\frac{1}{\sqrt{2|H|}}\sum_{bh}|b\ran_B |h\ran_H U_R |bh0\ran_R,
\ee
where $b$ and $h$ label convenient bases for $B$ and $H$, and we have chosen a ``local'' basis for the radiation in the sense that a state like $|10110\ldots \ran_R$ is analogous to a state where all the radiation modes have definite occupation number.  $U_R$ is some unitary transformation on the radiation that relates this basis to the natural basis of the Schmidt decomposition where the entanglement of the state $|\psi\ran$ is manifest.  $U_R$ is defined only up to an arbitrary unitary on the orthogonal complement of the subspace spanned by $|bh0\ran$.  In order to really observe a violation of entanglement monogamy along the lines of the AMPS experiment, our infalling observer must first use a quantum computer to act with $U_R^\dagger$ to ``distill'' the purification of $B$ in the radiation into an easily useable form.  We would like to assess the ``difficulty'' of doing this, as defined using the quantum circuit model described in appendix \ref{compsec} (the reader who is unfamiliar with quantum complexity theory should now read this appendix).  

The first obvious objection to any claim that implementing $U_R^\dagger$ requires a time that is exponential in $n$ is that the black hole certainly does not require exponential time to \textit{produce} the state \eqref{psi}.  Indeed we should probably accept that there exists a polynomial-sized circuit $U_{dyn}$ with the property that acting on the state $|0\ran_B|0\ran_H |000\ran_R$ it produces the state $|\psi\ran$; this amounts to the quite plausible assumption that quantum gravity can be ``efficiently simulated''.\footnote{In the cases where we really understand it, such as the BFSS matrix model or AdS/CFT, this seems likely to be the case \cite{feynman1982simulating,lloyd1996universal,Jordan:2011ne}.}   The problem however is that even if we have available a polynomial-sized circuit for $U_{dyn}$, we cannot use its inverse to decode the Hawking radiation, since that would only work if we are also able to act on the degrees of freedom $H$ that remain inside the black hole (this is explained in more detail in \cite{Harlow:2013tf}).  The argument above equation \eqref{psi} does ensure the existence of a distilling $U_R^\dagger$ that acts only on the radiation, but it is non-constructive and gives no information about the complexity of this distillation.  Without a construction, we are left with the basic fact reviewed in appendix \ref{compsec} that almost all unitaries on $n$ qubits require a circuit whose size is of order $2^{2n}$ to implement.

Of course it could still be the case that we can somehow use the simplicity of $U_{dyn}$ to argue in a more complicated way that there must be a polynomial sized circuit for $U_R$, but Patrick and I were able to give a complexity theoretic argument that this is probably not the case.  We can phrase this as a general question about quantum circuits:
\begin{itemize}
\item \textbf{HAWKING DISTILLATION PROBLEM:} Say we are given a product Hilbert space of three qubit systems $B$, $H$, and $R$, along with a polynomial-sized circuit $U_{dyn}$ that prepares a quantum state $|\psi\ran=U_{dyn}|0\ran$, where $|0\ran$ is the product state that is all $0$'s for all three factors, and moreover say that $|\psi\ran$ has the property that it is maximally mixed on $BH$.  Also say that the dimensionality $|B|$ is some $O(1)$ number.  Does there exist a small quantum circuit $U_R^\dagger$, meaning a circuit whose size is at most polynomial in $m=\log_2|H|$ and $n=\log_2|R|$, that distills the purification of $B$, in the sense that
\be
U_R^{\dagger}|\psi\ran\approx\frac{1}{\sqrt{|B||H|}} \left( \sum_b |b\ran_B |b\ran_{R_B}\right)\left(\sum_{h} |h\ran_H U_R'|h0\ran_{R_{\bar{B}}}\right).
\ee
I have here allowed for some remaining scrambling $U_R'$ of the purification of $H$, although this can be removed by a basis redefinition of $H$.  The approximation should be understood as saying the two states are close in the trace norm.  
\end{itemize}
Patrick and I argued based on the assumed hardness of a quantum complexity class called QSZK (Quantum Statistical Zero Knowledge) that the answer to this question is typically no, but I will instead give a more elegant pair of arguments due to Scott Aaronson that lead to the same conclusion.\footnote{The first argument is also explained here: http://www.scottaaronson.com/talks/hawking.ppt, the second came out of a recent discussion between Scott and me.}  As with most complexity theoretic arguments, it is too difficult to directly prove that no efficient distillation $U_R^\dagger$ exists.  So what one does instead is show that if it \textit{did} exist, this would enable us to do something which is widely expected to be difficult.

The difficult task Scott uses is the inversion of ``one-way functions'', meaning functions which are easy to evaluate but difficult to invert.  It is not a priori clear that such functions should exist, but they are widely expected to; indeed almost all of modern cryptography is based on their assumed existence \cite{arora2009computational}.  In quantum language we can think of a one-way function as a map $f$ from $m$-bit strings to $n$-bit strings, with $n\geq  m$, such that there exists a polynomial-sized circuit $U_{f}$ where
\be\label{Uf}
U_f|x,0\ran=|x,f(x)\ran,
\ee
but no polynomial-sized circuit $U_{f^{-1}}$ such that
\be\label{Ufinv}
U_{f^{-1}}|0,f(x)\ran=|x,0\ran.
\ee
Note that although $U_f$ is invertible, we cannot use its inverse to invert the function; if we give it $|0,f(x)\ran$ it will not necessarily do anything useful.\footnote{In fact the standard definition of one-way functions is slightly weaker than this, it demands only that no efficient algorithm exist which can invert the function on a fraction of inputs which at most polynomially small in $m$.  Since we will see that solving Hawking Distillation cracks the stronger version this distinction is not relevant for our purposes, although it could be if we tried to define a version of Hawking Distillation that allowed for ``imperfect distillation''.}  What Scott was able to do was show that if our Hawking Distillation Problem has a positive answer, then one can use that answer to efficiently invert \textit{any} candidate injective one-way function; in other words no injective one-way function could exist.\footnote{Injectivity is not a very strong restriction; if $n>2m$ then we avoid the birthday paradox and the function $f$ will typically be injective.}  This may not sound so bad to physics readers, but in computer science and cryptography it would be a genuine catastrophe (or revelation!), of almost a similar order of magnitude as a proof that $P=NP$ \cite{impagliazzo1995personal}.  

Indeed say we are given a candidate injective one-way function $f$ on $m$ bit strings and an efficient circuit $U_f$ that implements it, as in equation \eqref{Uf}.   It is not difficult (homework!) using the Hadamard and CNOT gates of appendix \ref{compsec} to come up with a polynomial-sized quantum circuit which, given the all zero state of $BHR$, prepares the state
\be\label{psiscott}
|\psi\ran=\frac{1}{\sqrt{2|H|}}\sum_h|h\ran_H\Big(|0\ran_B|0,h,0\ran_R+|1\ran_B|1,f(h)\ran_R\Big).
\ee
Now let's assume that the Hawking Distillation Problem has a positive answer: there then must exist an efficient distilling unitary transformation $U_R^\dagger$ which acts as
\begin{align}
U_R^\dagger |0,h,0\ran_R&=|0,g(h)\ran_R\\
U_R^\dagger |1,f(h)\ran_R&=|1,g(h)\ran_R,
\end{align}
where $|g(h)\ran$ is some given set of $2^m$ states of the radiation minus a qubit.
But now by combining $U_R$ and $U_R^\dagger$ with the $X$ operation that flips the first qubit we can directly construct an efficient circuit $U_{f^{-1}}$ which implements \eqref{Ufinv}, contradicting the one-way nature of $f$.\footnote{More carefully it implements it with the assistance of a single extra ancillary qubit, which does not increase the complexity in any practical sense.}  Thus the Hawking Distillation problem is tractable in general only if there do not exist injective one-way functions.  Since there is strong evidence that they do exist, Hawking Distillation is most likely a hard problem.

A subtlety with this argument however is that, although we need to do something hard to distill manifest entanglement, the state \eqref{psiscott} already exhibits \textit{classical correlation} with the radiation in the sense that its mutual information with the first qubit of $R$ is nonzero.  This is in fact already sufficient to prevent maximal entanglement between $B$ and something else, so a slight modification of the AMPS experiment is sufficient to argue that the state \eqref{psiscott} can't have a smooth horizon even if we can't distill manifest entanglement.  What we would really like to argue is that in the setup of of the Hawking Distillation problem, even distilling classical correlation between $B$ and $R$ is usually exponentially difficult.  In fact a small modification of Scott's argument, again due to Scott, is able to resolve this.  Namely we consider instead the state
\be\label{psiscott2}
\frac{1}{\sqrt{2|H|}}\sum_{h_1,h_2}|h_1,h_2\ran_H \Big( |h_1\cdot h_2 \ran_B |f(h_1),h_2,0\ran_R+|h_1\cdot h_2+1\ran_B |f(h_1),h_2,1\ran_R\Big),
\ee
where we have split $H$ into two equal size pieces $H_1$ and $H_2$.  Here $f$ is a candidate injective one-way function on $m/2$ bits, and by $h_1 \cdot h_2$ I mean the inner product of the two strings, computed mod $2$.  It is not difficult to see that this state satisfies the conditions of the Hawking Distillation problem; it can be simply generated using a circuit $U_f$ that computes $f$ as in equation \eqref{Uf}, and it is maximally mixed on $BH$.  Now let's imagine that there exists an efficient circuit $U_R^\dagger$ that is able to distill a bit of $R$ that is classically correlated with $B$ in the sense that their mutual information is close to $\log 2$ (or close to one if we define the entropy with a base two logarithm).  We must then have
\begin{align}\nonumber
U_R^\dagger|f(h_1),h_2,0\ran_R&=|h_1\cdot h_2,g(h_1,h_2)\ran_R\\
U_R^\dagger|f(h_1),h_2,1\ran_R&=|h_1\cdot h_2+1,g'(h_1,h_2)\ran_R,
\end{align}
which means that using $U_R^\dagger$ to determine $h_1\cdot h_2$ given $(f(h_1),h_2)$.  This however again allows us to invert the function \cite{goldreich1989hard}; for example given $f(h_1)$ we can determine the first bit of the input by choosing $h_2$ to be $1$ for the first qubit and $0$ for the rest.   Thus the assumed existence of one-way functions also ensures the exponential difficulty of distilling classical correlation.  

Given these arguments, it seems highly likely that computational complexity prevents an AMPS experiment from being done.  What then are we to conclude?  One option would be to declare victory for black hole complementarity and move on; by the standards of the mid 1990's this would be what we would do.  What the past few years have taught us however is that those standards were too low; without an actual theory of the interior we cannot be sure that complementarity actually provides the mechanism whereby unitarity is made consistent with a smooth experience for the infalling observer.  One problem is that somebody might later come up with another thought-experiment that cannot be resolved in the same way, and then we would be back in the soup.\footnote{Indeed there have been several attempts to come up with thought-experiments that evade our computational constraints.  Some of these involve direct manipulation of the microscopic degrees of freedom from the outside of the system, and thus cannot be performed by observers who are actually part of the system and are restricted to local operations allowed by perturbative semiclassical physics \cite{Almheiri:2013hfa}, while others require the ability to precisely manipulate the degrees of freedom in the atmosphere without introducing any decoherence \cite{Oppenheim:2014gfa}.  At the moment I do not find either proposal convincing; it is not clear what we should expect to come out of direct manipulation of UV degrees of freedom, indeed because of the fundamental non-locality in holography I expect that large scale action-at-a-distance is possible in the bulk by such operations, so asking for a semiclassical description of such an experiment may well be futile.  The second type of proposal requires large amounts of machinery in the vicinity of the black hole, and it is hard to see how this could be possible without introducing at least some decoherence.  In \cite{Oppenheim:2014gfa} a mechanism was suggested to ``correct'' the decoherence, but implementing it again seems to require an exponentially long quantum computation.}  Even if they do not however, there are basic physical questions such as ``what happens if we form and evaporate a little black hole inside of a big one?'' that seem to really require a theory of the black hole interior that goes beyond effective field theory.  It is only after we find this real theory of the interior that we will be able to see whether or not limitations from computational complexity play an important role in its consistency; at best they currently can be thought of as a tool to use in looking for that theory.  In the historical analogy suggested in section \ref{compsec1} above, it could be that our computational complexity restrictions on experiments are analogous to the uncertainty principle, but if so then we have not yet discovered the theory analogous to quantum mechanics.  Until we do, there will always be stodgy classical physicists saying ``of course you can measure the position and momentum, you just need to think more about how to do it.''\footnote{One direction for a more ``positive'' approach to complementarity is called ``strong complementarity'', and operates under the basic assumption that different observers have different Hilbert spaces and observables, with the only consistency condition being that they must agree on the results of experiments that they can both do \cite{Bousso:2012as,harlow2,Banks:2012nn,Harlow:2013tf,Almheiri:2013hfa}.  So far this idea is somewhat ambiguous and no precise version of it has appeared.}   In particular the firewall typicality arguments of section \ref{ftypsec} are not expressed as operational contradictions accessible to a low-energy thought-experimenter in the bulk, and thus are not immediately addressed by complexity theoretic concerns. Perhaps these arguments are resolved by noting that they are based on assumptions that are analogous to asserting that a particle has both a position and a momentum, despite the unobservability of this notion.  To confirm (or refute) this however, we need a theory.

\subsection{Nonlinearity?}\label{ereprsec}
I'll now turn to proposals that, unlike complementarity, give some sort of positive prescription for what the theory of the interior might be.  

One obvious idea for evading the entanglement monogamy paradox is to just declare that in the theory of quantum gravity, the interior mode $A$ is simply defined to be ``whatever is entangled with $B$''.  In the language of section \ref{ampssec} this is sometimes called $A=R_B$.  Versions of this idea have been proposed in many places \cite{Bousso:2012as,Verlinde:2012cy,Papadodimas:2012aq,Harlow:2013tf,Maldacena:2013xja}; I have recently written a general discussion of what I consider to be the best-defined of the options, that of Raju and Papadodimas \cite{Papadodimas:2013wnh,Papadodimas:2013jku,Verlinde:2013qya}, and I refer the reader there for many more details \cite{Harlow:2014yoa}.\footnote{Recently \cite{Marolf:2015dia} re-emphasized the central points of \cite{Harlow:2014yoa} in a more general but less precise way.}  In this section I will just make a few general points about the features of these proposals, in the context of a simple qubit model.

Before beginning however, it may already seem crazy that we should think of the interior of an old black hole as having anything to do with the distant Hawking radiation that has already been emitted.  We know that there is some type of non-locality in holography, but this may seem to ``just be too much''.  An illuminating argument in this regard was emphasized by Maldacena and Susskind, and by van Raamsdonk, who suggested taking the Hawking radiation cloud of a black hole that has just evaporated halfway and collapsing it into a second black hole that is thermally entangled with the first in the thermofield double state \cite{Maldacena:2013xja,VanRaamsdonk:2013sza}.  From our discussion of section \ref{ads2sidesec} it then seems at least somewhat plausible that the two black holes will be connected by a wormhole.  Assuming that this is the case, from the Penrose diagram in figure \ref{schpenrose} (or figure \ref{ads2}) we can clearly affect the interior of the original black hole by throwing things into the other black hole.  It thus seems that at least there is one unitary operation we can do on the Hawking radiation which directly affects the interior.\footnote{We can make this more precise by starting with a large AdS black hole in a CFT and allowing it to evaporate into a second copy of the CFT by coupling the two through a simple local coupling at the boundary as in \cite{Rocha:2008fe}.  By then manipulating the second copy we can literally prepare the thermofield double state of the two CFT's, which we have already decided is quite plausibly described in the bulk by a wormhole connecting the two asymptotically AdS regions.}  Given that this is the case, it doesn't seem so unreasonable to try to look for a general theory of the interior that allows us to use whatever the black hole is entangled with as part of the theory.  

In trying to turn this observation into a theory of the interior however one runs into a set of fairly serious problems that so far have resisted a clear resolution.  For simplicity I will illustrate them in the simple qubit model of the previous section, where we think of the factors $B$, $H$, and $R$ as qubit systems.  I will again model $B$ as a single qubit, so by the Schmidt decomposition I can represent a typical state of the system as
\be\label{sdstate}
|\psi_+\ran=\frac{1}{\sqrt{2}}\left(|0\ran_B|\wt{0}\ran_{HR}+|1\ran_B|\wt{1}\ran_{HR}\right).
\ee
I will also imagine that the ``smooth'' horizon state for $A$ and $B$ is $|00\ran_{AB}+|11\ran_{AB}$, and I have labeled the states of HR appearing in the Schmidt decomposition appropriately.  The problems arise from the interpretation of the three other states
\begin{align}\nonumber
|\psi_-\ran&\equiv\frac{1}{\sqrt{2}}\left(|0\ran_B|\wt{0}\ran_{HR}-|1\ran_B|\wt{1}\ran_{HR}\right)\\
|\chi_\pm\ran&\equiv\frac{1}{\sqrt{2}}\left(|0\ran_B|\wt{1}\ran_{HR}\pm|1\ran_B|\wt{0}\ran_{HR}\right).
\end{align}
If we are trying to define the interior modes using whatever $B$ is entangled with, then since all of these states have maximal entanglement it is tempting to look for a definition where they all ``look smooth''; after all they are just as typical as the state$|\psi_+\ran$ that we started with.  We then find ourselves confronted by the following issues \cite{Almheiri:2013hfa,Bousso:2013ifa,Harlow:2014yoa} (see also \cite{Bousso:2012as,Chowdhury:2013mka}):
\begin{itemize}
\item \textbf{Nonlinearity:} By taking simple superpositions of these four states we can produce states where $B$ is pure and unentangled with anything; in fact there is a complete basis of such states.  We saw in section \ref{ftypsec} that it is impossible for such states to have unexcited horizons.  This means that we cannot view ``unexcitedness'' as a conventional observable realized by a self-adjoint operator on the Hilbert space; if it were then since all four of these states are unexcited the ``unexcitedness operator'' is the identity on this subspace and any superposition must also be unexcited, contradicting the fact that there \textit{are} excited states.   There is thus a basic tension between wanting all four states to have a smooth horizon and the linearity of quantum mechanics.  This situation is sometimes described as ``state-dependence'': for a given observable, say the excitation number of the $A$ qubit behind the horizon, one tries to use a different self-adjoint operator depending on which of the four states the system is in \cite{Almheiri:2013hfa}.
\item \textbf{Frozen Vacuum:} Say that we wish to excite the horizon by acting with the $X$ operator on $B$.  From a semiclassical point of view this should produce an excitation at the horizon, but here this operator just permutes us amongst the four states which are all taken to be smooth.  We then do not seem to be able to realize these excited states, even though they exist semiclassically \cite{Almheiri:2013hfa,Bousso:2013ifa,Harlow:2014yoa}.  
\item \textbf{Non-Unitary Measurement:} Now say we want to design an apparatus which measures the $Z$ operator on the interior qubit $A$ which is entangled with $B$.  A straightforward argument \cite{Harlow:2014yoa} shows that, unlike in conventional quantum mechanics, the measurement process cannot be described as unitary evolution on the system together with the apparatus.  This is essentially a consequence of the non-linearity; it is a general problem for attempts to formulate ``state-dependent'' quantum mechanics.  It is also a good concrete criterion for distinguishing ``illegal'' state-dependence of the type needed here from more conventional experiments that naively seem to involve state-dependent observables \cite{Harlow:2014yoa}.  
\item \textbf{Non-uniqueness:} We saw in appendix \eqref{BPapp} that the purification $R_B$ is not uniquely defined; we are free to conjugate it by any unitary transformation which fixes the subspace of states appearing in the Schmidt decomposition - here the space spanned by $\{|\wt{0}\ran_{BH},|\wt{1}\ran_{BH}\}$.  This means that any attempt to define interior operators acting directly on the subfactor $R_B$ will not be unique.  How then are we to choose which operators to use?
\item \textbf{Commutator Problem:}  Once we want to define interior operators out of operators that act nontrivially on $H$ and $R$, it is no longer clear that these operators will commute with ``simple'' operators on $H$ and $R$ \cite{Almheiri:2013hfa}.  This raises the possibility of acausal signaling; we can in principle use this commutator either to create a firewall by doing something simple on the distant radiation or communicate from inside the black hole to the outside.  
\end{itemize}
So far no version of $A=R_B$ has appeared which can give completely satisfying resolutions to these objections; the proposal that comes closest is that of Raju and Papadodimas \cite{Papadodimas:2013wnh,Papadodimas:2013jku}, which among other things is able to ``postpone'' the commutator problem to operations that involve some order one fraction of the radiation, but in my view it still is ultimately vulnerable to (somewhat more precise) versions of these problems \cite{Harlow:2014yoa}.\footnote{In particular the operations for which the proposal fails are vastly simpler than the exponentially complex quantum computations considered in the previous subsection; there does not seem to be any serious obstruction to doing them.}  In the end if some version of $A=R_B$ is to work, it will need to come with a completely developed measurement theory that generalizes and replaces that of quantum mechanics, and it will need to make clear predictions for any thought-experiment we can reasonably imagine doing.     

\subsection{Postselection?}
Another interesting proposal for modifying quantum mechanics to obtain a consistent description of the black hole interior is the quantum postselection formalism of Horowitz and Maldacena \cite{Horowitz:2003he}.  Quantum postselection is a generalization of quantum mechanics to allow probabilities to depend not only on the initial state of the system, as is usual, but also on some additional ``final state'' that in general has no relation to the initial state \cite{PhysRev.134.B1410}.  To concisely state the proposal, it is convenient to first introduce a condensed formalism for describing the predictions of quantum mechanics for successive experiments.  Say that we have a quantum system that begins life in a (possibly mixed) quantum state $\rho_i$.  If we then measure a series of observables $A_1, A_2, \ldots, A_n$, it is not too hard to show that the joint probability distribution for the outcomes $a_1, a_2,\ldots, a_n$ for all the experiments is given by
\be\label{qmP}
P(a_1,a_2,\ldots,a_n)=\tr \left(\Pi_{a_n} \ldots \Pi_{a_1}\rho_i \Pi_{a_1}\ldots \Pi_{a_n}\right),
\ee
where $\Pi_{a_j}$ is the projection operator onto the outcome $a_j$ from measuring $A_j$.  This remarkable formula combines the Born Rule and the collapse of the wave function into a single equation.  It elegantly captures some of the more surprising features of quantum measurement theory:
\begin{itemize}
\item If two sequential observables do not commute, then the probability distribution depends on which we measure first.  If they commute then it does not.
\item Regardless of commutators, the probability distribution for the outcomes of the first $m$ measurements is independent of any measurements that come later.  For example
\be\label{qmcausal}
P(a_1)\equiv\sum_{a_2}P(a_1,a_2)=\tr \left (\Pi_{a_1}\rho_i \Pi_{a_1}\right).
\ee
This is essential for quantum mechanics to respect causality; our results for measurements today should not depend on what we decide to do tomorrow.  
\item Averaging over results of a measurement in the \textit{middle} of the chain does affect the probability distribution for the results of the other measurements.  For example if $[A_1,A_2]\neq 0$ we usually have
\be
P(a_2)\equiv \sum_{a_1} P(a_1,a_2) \neq \tr \left (\Pi_{a_2}\rho_i \Pi_{a_2}\right);
\ee
``averaging out'' a measurement is not in general equivalent to not doing the measurement, unlike in classical physics where we allow measurements that acquire any information we like about the system without disturbing it.\footnote{A sufficient condition for a sequence of measurements to allow such averaging is for the \textit{decoherence functional} $D(a_1,\ldots a_n,a'_1\ldots a'_n)\equiv \tr \left(\Pi_{a_n} \ldots \Pi_{a_1}\rho_i \Pi_{a'_1}\ldots \Pi_{a'_n}\right)$ to be diagonal in the sense of vanishing unless $a_j=a'_j\,\,\, \forall j$.  Such sequences of projection operators are sometimes called ``consistent histories'', although this term is rather misleading since there is nothing inconsistent about more general sequences of measurements.  What ``consistent'' really means here is that when $D(a,a')$ is diagonal, we are allowed to think of the system as having a definite ``classical history'' in the sense that we can imagine that each observable has a well-defined value at each point in time whether or not we look at it.  Of course this ``classical history'' is just us lying to ourselves about the fundamental indeterminacy of quantum mechanics, but it can be useful in understanding under what circumstances classical mechanics emerges from quantum mechanics \cite{GellMann:1992kh}.}  
\end{itemize}
Now to generalize to including a ``final state'', the idea is simply to replace \eqref{qmP} by 
\be\label{PqmP}
P(a_1,\ldots a_n)=\frac{1}{\mathcal{N}}\tr \left(\rho_f\Pi_{a_n} \ldots \Pi_{a_1}\rho_i \Pi_{a_1}\ldots \Pi_{a_n}\right),
\ee
where the final state is $\rho_f$ \cite{PhysRev.134.B1410} and
\be
\mathcal{N}=\sum_{a'_1\ldots a'_n}\tr \left(\rho_f\Pi_{a'_n} \ldots \Pi_{a'_1}\rho_i \Pi_{a'_1}\ldots \Pi_{a'_n}\right).
\ee

This defines a normalized set of probabilities for the results of any experiment we can imagine doing.  The first thing to notice however is that we now lose our ability to average out later measurements as in equation \eqref{qmcausal}; postselected quantum mechanics violates causality!  It also immediately grants us the ability to solve NP-complete problems in polynomial time \cite{Aaronson:2005qu}.  These may already be sufficient grounds to reject it, but as we will now see it may provide a surprisingly simple way of getting information out of a black hole without disrupting the experience of an infalling observer.  

To see what postselection has to do with black holes, following Horowitz and Maldacena we can again take seriously the effective field theory Hilbert space on a ``nice slice'' such as the blue one shown in figure \ref{cloning}.  In section \ref{compsec1} we saw that together with unitarity this led to quantum cloning, but now we are violating quantum mechanics anyway so let's press on.  We can model the Hilbert space as a tensor product
\be
\mathcal{H}=\mathcal{H}_M \otimes \HA \otimes \HB,
\ee
where now we are interpreting $M$ as the full set of ``left-moving'' modes behind the horizon, $A$ as the full set of ``right-moving'' modes behind the horizon, and $B$ as the full set of ``outgoing'' modes in the atmosphere.  We can think of them as all having dimensionality $|M|=e^{S_{BH}}$, since $M$ describes ``all things that we throw in'', $B$ describes ``all radiation that comes out'', and $A$ are ``all Hawking partners of the radiation''.  We can take the initial state to be
\be\label{initial}
|\psi_i\ran_{MAB}=|\psi\ran_M \otimes |\phi\ran_{AB}, 
\ee
with 
\be
|\phi\ran_{AB}\equiv \frac{1}{\sqrt{|M|}}\sum_a |a\ran_A |a\ran_B.
\ee
Here I have denoted the initial quantum state of the infalling matter as $|\psi\ran_M$ and taken $A$ and $B$ to be maximally entangled as required for a smooth horizon.  Were we to proceed as usual we would conclude that this state has information loss, since the state of $B$ is mixed and has no memory of $|\psi\ran_M$, but the insight of Horowitz and Maldacena was that if we now introduce a final state
\be
\rho_f=|\chi\ran \lan\chi|_{MA} \otimes \frac{I_B}{|M|},
\ee
with 
\be
|\chi\ran_{MA}=\frac{1}{\sqrt{|M|}}\sum_a S^\dagger|a\ran_M |a\ran_A,
\ee
the information actually gets out.  More explicitly, say that we want to compute the probability distribution $P(b)$ for some set of measurements on the radiation $B$.  We can represent the sequence of projectors for some set of outcomes as $C_b\equiv \Pi_{b_1}\ldots \Pi_{b_n}$, in which case it isn't too hard to see (homework!) that
\be\label{finalstateresult}
\frac{\tr\left(\rho_f C_b^\dagger \rho_i C_b\right)}{\sum_{b'}\tr \left(\rho_f C_{b'}^\dagger\rho_i C_{b'}\right)}=\lan \psi|S^\dagger C C^\dagger S \psi\ran.
\ee
In other words the outcomes of all measurements on the outgoing radiation are consistent with just taking it to be in the pure state $S|\psi\ran_B$ and using ordinary quantum mechanics!  

Of course to really address the AMPS paradox, we need to understand the implications of this proposal for experiments that involve the infalling observer \cite{Lloyd:2013bza,Bousso:2013uka}.  Consider an experiment where an infalling observer attempts to confirm that $A$ and $B$ are indeed entangled as in the state \eqref{initial}.  She can do this by measuring the projection operator $\Pi_\phi=I_M\otimes|\phi\ran\lan\phi|_{AB}$.  Using equation \eqref{PqmP}, a straightforward computation shows that the probability for observing $|\phi\ran_{AB}$ is one; she will always see the desired entanglement!  Thus we see that postselected quantum mechanics allows us to both have our cake and eat it too; the Hawking radiation is unitary, but the horizon is smooth.

Unfortunately the simple story presented so far in this section becomes more complicated once we try to include more details.  In the formalism I have presented here I have not taken into account interactions between $M$ and $A$, and doing so requires a more delicate choice of final state \cite{Gottesman:2003up}.  Moreover once we consider experiments like the AMPS experiment that involve both attempting to verify the entanglement between $A$ and $B$ and confirm the purity of the radiation, the acausality intrinsic to the proposal rears its ugly head \cite{Bousso:2013uka,Lloyd:2013bza}.  One point that is especially confusing is that the apparatus of the infalling observer must be included as part of the infalling system $M$ and must interact with $A$; since this is eventually post-selected on it may be necessary to ``undo'' any measurement that happens by a redefinition of the final state.  At the moment there do not seem to be completely satisfactory resolutions of these issues; one possibility that is perhaps worth exploring more is that the computational complexity restrictions we discussed in section \ref{savecomp} may relieve some or all of the pressure on the Horowitz/Maldacena proposal \cite{Bousso:2013uka,Lloyd:2013bza}.

\subsection{Firewalls?}
The final possibility I'll consider is that some version of a firewall actually exists.  I will take this term to mean any observable violation of low-energy effective field theory for simple experiments in the vicinity of a black hole horizon.  Six options in particular have been discussed in some detail:
\begin{itemize}
\item \textbf{Full-strength Firewall:} AMPS originally argued that the most conservative resolution of the tension in the previous section is to simply imagine that the horizon becomes singular and the interior no longer exists, either for typical big AdS black holes or for old asymptotically flat black holes \cite{Almheiri:2012rt,Marolf:2013dba}.  Given the craziness of the proposals in the previous sections, the reader may be sympathetic to considering this option more seriously.  It still has substantial drawbacks however: there is currently no dynamical explanation for why a singularity should form at the horizon ``out of nothing'', and if one does it seems rather unlikely that we should still take Hawking's calculation of the temperature and entropy seriously.  Since this calculation is what justified black hole thermodynamics in the first place, the firewall proposal is somewhat self-defeating.\footnote{Lenny Susskind has emphasized to me however that one can attempt a ``strictly exterior'' calculation of the entropy and temperature by arguing that quantum fields outside the horizon have a large backreaction in the Schwarzschild geometry if we put them at a temperature other than $T_{hawking}$.
This is true, but if we are willing to allow large backreaction right at the horizon in the form of a firewall, why shouldn't we also allow it further out in the atmosphere?  Keeping the bad behavior quarantined at the horizon may be the ``least objectionable'' thing to do, but without an explanation for how the firewall forms we cannot be sure it does not extend further.  We know from AdS/CFT that at least in some cases it does not, for example in $AdS_3/CFT_2$ where we can compare the BH entropy and the Cardy formula as in section \ref{microadsbhsec}.  Logically we can only interpret this as constraining firewalls to lie strictly at the horizon and not outside, but it seems natural (to me at least) to interpret it as validating the whole coarse-grained semiclassical picture of the horizon, including a smooth experience for an infalling observer.  Of course we still need to understand how to resolve the paradoxes of section \ref{paradoxsec} before really dismissing firewalls.}  
\item \textbf{Typical-states-only Firewall:} Another possibility is to accept that black holes formed in typical states have firewalls, but then attempt to argue that the ones that form in nature never do, even if they are old evaporating black holes.  This could be possible because, as argued in section \ref{typicalsec}, the black holes we make in short collapses are only a vanishingly small fraction of the total ensemble whose dimensionality is $e^{S_{BH}}$.  As the black hole evaporates the state of the remaining black hole becomes more typical by Page's theorem, but the state of the joint black hole/radiation system stays atypical; if we are willing to allow the radiation to be used in constructing the interior a la section \ref{ereprsec}, then we can use this to avoid the genericity argument of section \ref{ftypsec}.  Recently Susskind has been exploring arguments based on computational complexity that may support this possibility \cite{Susskind:2014rva} (see also \cite{Stanford:2014jda,Susskind:2014jwa}), although this proposal still seems to be subject to the criticism of invalidating Hawking's calculation of the temperature and entropy, with the same caveats as before.  
\item \textbf{S-Wave Firewall:} We saw in section \ref{ampssec} that at least some of the paradoxes can be satisfied by a firewall that only affects low angular momentum modes.  So far this proposal has not received too much attention in the literature, perhaps mostly because, to quote Raphael Bousso, one would like to ``get rid of firewalls entirely or don't bother''.  Nonetheless this idea does have some things going for it, perhaps chief among them that the backreaction becomes small and thus the basic structure of Hawking's calculation should go through.  One would still need a mechanism for how the S-wave firewall develops however.
\item \textbf{Non-violent Nonlocality:}  Another possibility that Giddings has been exploring is that there is a more diffuse violation of effective field theory that is spread out non-locally through the atmosphere rather than concentrated at the horizon \cite{Giddings:2012gc,Giddings:2013noa}.\footnote{Giddings objects to this proposal being included as a type of firewall, since for him a firewall is necessarily violent.  For me however the real question is whether or not there is a simple experiment an infalling observer can do that detects a violation of effective field theory; I am less concerned with how traumatic the experience is, after all the s-wave firewall is also ``non-violent''.}  The models he has studied so far indeed have the property that black hole thermodynamics tends to be modified; preventing this requires large modifications of the Schwarzschild geometry outside the horizon that may even be detectable in upcoming experiments \cite{Almheiri:2013hfa,Giddings:2013vda,Giddings:2014ova}.\footnote{At least somebody in this business is making experimental predictions...}  Once we allow such large modifications of effective field theory outside the horizon however, it is difficult to see how they will not arise in other situations as well; ``small'' violations of causality tend to have a way of not staying small.  
\item \textbf{Fuzzballs} For certain higher-dimensional black holes with large charges under various gauge form fields such as one finds in string theory, there exist so-called ``fuzzball'' solutions, which resemble the black hole solution near infinity but near the horizon cap off into higher dimensions in various ways, effectively excising the interior.  There is a vast literature discussing these solutions, see \cite{Gibbons:2013tqa} for a gentle introduction and further references.  It is sometimes argued that there might be enough of these solutions to account for the full Bekenstein-Hawking entropy of these black holes.  Despite the capping off of fuzzball geometries at the horizon, there have been some attempts to argue that an infalling observer nonetheless experiences a smooth horizon  \cite{Mathur:2012jk,Mathur:2013gua}, but this certainly is not what the fuzzball geometries naively tell us \cite{Bena:2012zi} so if it is true it will require some new idea.  In any case fuzzball solutions exist only in these special cases, so far there are no analogous solutions for uncharged black holes and in fact there are theorems forbidding their existence, which makes the relevance of these solutions for a general solution of the information problem unclear.
\item \textbf{Shut Up and Calculate in the UV Theory:} In string theory we already have available at least one candidate theory of quantum gravity; shouldn't we just see what it predicts at the horizon?  If it predicts a firewall shouldn't we then just accept it? Of course the problem with this obviously correct philosophy is that we don't understand the theory well enough to decide what it predicts; the only case where it really seems well-defined so far is in the AdS/CFT correspondence, which as we have been discussing has not yet given us a clear answer.  Nonetheless some attempts to study the black hole information problem have been made using \textit{perturbative} string theory in the bulk, for some preliminary results that go in the direction of non-locality see \cite{Lowe:1995ac,Giddings:2007bw,Amati:2007ak,Silverstein:2014yza,Dodelson:2015uoa,Dodelson:2015toa}, although at this point it is not so clear whether these results actually lead us to expect observable violations of effective field theory.  See also \cite{Horowitz:2009wm} for another stringy attempt at behind the horizon physics.  
\end{itemize}

Thus we find ourselves in the enviable situation of having an interesting problem with no really satisfying answer; if we are lucky this means that we will learn something deep.  It is my hope that what we learn can then be applied to the other profound problem of quantum gravity - making sense of cosmology.  Observers behind black hole horizons have many common features with observers in expanding universes, and what is in my view the best proposal for the global structure of spacetime, the landscape of string theory populated by eternal inflation, has perplexing difficulties which seem to be grown-up versions of the problems we are already confronting in black hole physics.  If you have made it this far in these notes I hope it is clear to you that at least one major new idea is needed to understand the black hole interior, and it is exciting to imagine where it might lead us in the future.  

\paragraph{Acknowledgments}  These notes would not have been possible without many people; this is a difficult subject, and I have benefited from many conversations with some of the world's experts.  In particular Raphael Bousso, Patrick Hayden, Juan Maldacena, Joe Polchinski, Douglas Stanford, Herman Verlinde, and my co-advisers Steve Shenker and Lenny Susskind have taught me many things over the years.  It is impossible to list everyone else I am indebted to, but certainly the set includes Scott Aaronson, Ahmed Almheiri, Nima Arkani-Hamed, Tom Banks, Charlene Borsack, Borun Chowdhury, Bartek Czech, Xi Dong, Willy Fischler, Ben Freivogel, Steve Giddings, Dan Kabat, Igor Klebanov, Tom Hartman, Matt Headrick, Idse Heemskerk, Simeon Hellerman, Veronica Hubeny, Nima Lashkari, Albion Lawrence, Stefan Liechenauer, Samir Mathur, Don Marolf, Jonathan Oppenheim, Don Page, Kyriakos Papadodimas, John Preskill, Andrea Puhm, Suvrat Raju, Mukund Rangamani, Vladimir Rosenhaus, Julie Shih, Jamie Sully, Eva Silverstein, Eliezer Rabinovici, Mark Van Raamsdonk, Erik Verlinde, Aron Wall, and Edward Witten.  You all have had to put up with me to varying degrees these past years, I hope I have given something back.  I'd like to thank two anonymous referees, and also Steve Giddings, for very detailed feedback on an earlier version of these notes, as well as Arash Ardehali for pointing out an impressive number of typos.  I'd also like to thank the participants of the 2014 Jerusalem Winter school for providing a stimulating audience for the lectures, the Israel Institute for Advanced Studies for the invitation to give them and hospitality during them, the Weizmann Institute and the Aspen Center for Physics for hospitality (twice) while the notes were being completed, as well as the Pacific Institute for Theoretical Physics at UBC, and Eliezer Rabinovici for a warm welcome to Israel during my first trip there several years ago. I am supported by the Princeton Center for Theoretical Science.  Finally I would like to dedicate this review to the memory of Professor Jacob Bekenstein, who was an active member of the audience when these lectures were delivered.  His seminal realization that the entropy of black holes should be taken seriously lies at the foundation of everything discussed in these notes.

\appendix

\section{General relativity}\label{GRapp}
In this appendix I very briefly review Einstein's theory of gravitation.  More details are available in \cite{Carroll:2004st,Wald:1984rg}. The basic idea is that the gravitational force is mediated by the geometry of spacetime, described mathematically using the \textit{metric tensor} 
$g_{\mu\nu}(x)$.  Here $\mu$ and $\nu$ are spacetime indices, running from $0$ to $3$ for $3+1$ dimensional spacetime.   The metric is a symmetric tensor in these two indices, and on physical grounds it is taken to be invertible and to have exactly one negative eigenvalue. $x$ is a spacetime coordinate, so $g_{\mu\nu}(x)$ is a tensor field and should be thought of as serving in gravity a similar function as the electric and magnetic fields do in electromagnetism.\footnote{In fact a better analogy would be between the electric and magnetic fields and the Riemann curvature tensor $R^\alpha_{\phantom{\alpha}\beta\mu\nu}$ introduced below, but I will not explore this further.}   Intuitively one can think of it as defining a ``local inner product'' at each point in spacetime; for any two vectors $V^\mu$, $U^\nu$ at a point $x$, we can define their inner product as
\be
V\cdot U\equiv g_{\mu\nu} (x)V^\mu U^\nu.
\ee
This gives a local definition of the lengths of vectors and the angles between them, which is how the metric encodes geometry.  To emphasize this the metric is often expressed as the ``invariant line element''\footnote{Although I will not discuss it in any detail, the meaning of ``invariant'' here is that this expression is unchanged under arbitrary coordinate transformations.  The change of the metric components $g_{\mu\nu}$ under such transformations is precisely cancelled by the change of the coordinate displacements (more precisely the differential forms) $dx^\mu$, $dx^\nu$.}
\be
ds^2=g_{\mu\nu}dx^\mu dx^\nu.
\ee
The simplest example of a metric is that of ordinary Minkowski space in Cartesian coordinates:
\be\label{minkmet}
ds^2=-dt^2+dx^2+dy^2+dz^2\equiv \eta_{\mu\nu} dx^\mu dx^\nu.
\ee
We see here that the condition that there is exactly one negative eigenvalue is what ensures that there is only one direction of time; in general this property is called having Lorentzian signature.\footnote{For inscrutable reasons some people use a convention where the metric has \textit{three} negative eigenvalues, switching the overall sign.  One tends to find these ``mostly-minus''-ers in the vicinity of particle physics conferences.}  For comparison with the Schwarzschild metric in the main text (and to illustrate how coordinate transformations work) it is convenient to also present the metric in spherical coordinates.  These are defined by
\begin{align}\nonumber
x&=r\sin\theta\cos\phi\\\nonumber
y&=r\sin\theta \sin \phi\\
z&=r \cos\theta,
\end{align}
and by taking differentials and substituting into \eqref{minkmet} one easily sees that
\be
ds^2=-dt^2+dr^2+r^2\left(d\theta^2+\sin^2\theta d\phi^2\right)\equiv -dt^2+dr^2+r^2 d\Omega_2^2.
\ee

Given a spacetime metric, we can determine the spacetime trajectories of test particles $x^\mu(\tau)$ by solving the \textit{geodesic equation}
\be\label{geodesiceq}
\frac{d^2 x^\mu}{d\tau^2}=-\Gamma^{\mu}_{\alpha\beta}\frac{dx^\alpha}{d\tau} \frac{dx^\beta}{d\tau}, 
\ee
where $\Gamma^\mu_{\phantom{\mu}\alpha\beta}$ is the \textit{Christoffel Connection}
\be
\Gamma^\mu_{\phantom{\mu}\alpha\beta}\equiv \frac{1}{2}g^{\mu \kappa}\left(\partial_\alpha g_{\beta \kappa}+\partial_{\beta}g_{\alpha \kappa}-\partial_\kappa g_{\alpha\beta}\right).
\ee
Here $g^{\mu\nu}$ is the \textit{inverse metric}, obeying
\be
g^{\mu\alpha}g_{\alpha \beta}=\delta^\mu_\beta,
\ee 
with $\delta^\mu_\beta$ the identity matrix.  For massive particles it is conventional to rescale $\tau$ such that we have $g_{\mu\nu}\frac{d x^\mu}{d\tau}\frac{d x^\nu}{d\tau}=-1$, in which case we can think of $\tau$ as the proper time along the trajectory.  For a massless particle we have $g_{\mu\nu}\frac{d x^\mu}{d\tau}\frac{d x^\nu}{d\tau}=0$, in which case the geodesic is said to be \textit{null}.  There is no such thing as proper time for a massless particle, so people instead conventionally rescale $\tau$ such that $\frac{dx^\mu}{d\tau}$ is the the four-momentum of the particle.\footnote{The geodesic equation \ref{geodesiceq} is invariant under linear transformations of $\tau$, but not under general relabelings.  A parametrization of a geodesic where \eqref{geodesiceq} holds is called an \textit{affine} parametrization.}  One should think of the geodesic equation as being analogous to the Lorentz force law in electromagnetism.  

Finally to determine the metric tensor we must solve Einstein's equation
\be
R_{\mu\nu}-\frac{1}{2}R g_{\mu\nu}=8\pi G T_{\mu\nu},
\ee
where $R_{\mu\nu}$ and $R$ are the \textit{Ricci Tensor} and \textit{Ricci scalar}, obtained from the \textit{Riemann curvature tensor}
\be
R^\alpha_{\phantom{\alpha}\beta\mu\nu}\equiv \partial_\mu \Gamma^\alpha_{\phantom{\alpha}\beta\nu}-\partial_\nu \Gamma^\alpha_{\phantom{\alpha}\beta\mu}+\Gamma^\alpha_{\phantom{\alpha}\mu \kappa}\Gamma^\kappa_{\phantom{\kappa}\beta\nu}-\Gamma^\alpha_{\phantom{\alpha} \nu \kappa}\Gamma^\kappa_{\phantom{\kappa}\beta\mu}
\ee
as 
\begin{align}\nonumber
R_{\mu\nu}&\equiv R^\alpha_{\phantom{\alpha}\mu\alpha\nu}\\
R&\equiv g^{\mu\nu}R_{\mu\nu}.
\end{align}
$T_{\mu\nu}$ is the \textit{energy-momentum tensor} of all matter present, and $G$ is Newton's gravitational constant.  The Riemann tensor quantifies the ``curvature'' of spacetime by encoding the extent to which nearby geodesics which start out parallel remain so.  Einstein's equation is analogous to Maxwell's equations in electromagnetism; $T_{\mu\nu}$ is the ``source'' for the gravitational field $g_{\mu\nu}$.  Mathematically it can be viewed as ten coupled nonlinear partial differential equations for the metric $g_{\mu\nu}$, together with whatever matter fields are present.  

Einstein's equation can be derived from the \textit{Einstein-Hilbert Action}
\be
S_{EH}=\frac{1}{16\pi G}\int d^4 x \sqrt{-g}R+S_{matter},
\ee
where $g$ is the determinant of $g_{\mu\nu}$ and $S_{matter}$ is the action for any matter present.  The energy momentum tensor is defined as
\be
T_{\mu\nu}(x)\equiv -2\frac{1}{\sqrt{-g}}\frac{\delta S_{matter}}{\delta g^{\mu\nu}(x)}.
\ee
A simple example is a scalar field, which has
\be
S_{matter}=\int d^4 x \sqrt{-g}\left(-\frac{1}{2} g^{\mu\nu}\partial_\mu \phi \partial_\nu \phi-V(\phi)\right)
\ee
and
\be\label{scalarT}
T_{\mu\nu}=\partial_\mu\phi\partial_\nu \phi-g_{\mu\nu}\left(\frac{1}{2}g^{\alpha\beta}\partial_\alpha \phi \partial_\beta \phi+V(\phi)\right).
\ee
One is also often interested in constant vacuum energy density $\rho_0$, which is implemented in the stress tensor as
\be
T_{\mu\nu}=-\rho_0 g_{\mu\nu}.
\ee

\section{Qubits}\label{qubitapp}
The simplest nontrivial quantum mechanical system is the spin $1/2$ particle, which in information-theoretic lingo is called a qubit.  It has a two-dimensional Hilbert space, which physicists usually think of as being spanned by an ``up'' and a ``down'' state.  In quantum information theory it is very convenient to instead describe the standard basis as $\{|0\ran,|1\ran\}$.  Moreover the Pauli operators $\sigma_x$, $\sigma_y$, and $\sigma_z$ are usually instead denoted $X$, $Y$, and $Z$.  Explicitly they act on the qubit as
\begin{align}\nonumber
X|a\ran=|a+1\ran\\\nonumber
Y|a\ran=i(-1)^{a+1}|a+1\ran\\
Z|a\ran=(-1)^a |a\ran.
\end{align}
Here all addition is done mod 2, so $1+1=0$.  Indeed this ability to use modular arithmetic is one good reason to prefer the $|0\ran,|1\ran$ notation.  Note however that, somewhat confusingly, 
\be
Z|1\ran=-|1\ran.
\ee

One is often interested in strings of qubits, that is in tensor products of some number $n$ qubits.  The $n$-qubit string has a $2^n$ dimensional Hilbert space, spanned by a basis labelled by all classical $n$-bit strings.  One then defines the Pauli operators $X_i$, $Y_i$, $Z_i$ acting on the $i$-th qubit in the obvious way; they act as the identity on all the other qubits.  

\section{Pure states, mixed states, and bipartite systems}\label{BPapp}
In this appendix I review some basic definitions and results in quantum information theory.  I will always assume all Hilbert spaces are finite-dimensional to avoid irritating technical distractions.  For many more details see for example \cite{preskillnotes,nielsen2010quantum}.  
\subsection{Definitions}
First of all a quantum state $\rho$ is a non-negative hermitian operator of trace $1$ acting on a Hilbert space $\mathcal{H}$.  The components $\rho$ in some basis are called the \textit{density matrix}, and I will sometimes lazily use this term for $\rho$ itself as well.  A quantum state is called \textit{pure} if it can be written as an outer product
\be
\rho=|\psi\ran\lan \psi|,
\ee 
where $|\psi\ran$ is some element of $\mathcal{H}$ with norm one.  In this case $|\psi\ran$ itself is also often referred to as the quantum state of the system.  A state which is not pure is \textit{mixed}.  The outcome $i$ of any measurement is always associated with some projection operator $\Pi_i$, and the probability for measuring that outcome is
\be
P(i)= \mathrm{tr} \left(\rho \Pi_i\right).
\ee

Any quantum state $\rho$ can always be interpreted (not necessarily uniquely) as a classical probability distribution over a set of mutually orthogonal pure states.  For example if we had a qubit which we knew was either in the state $|0\ran$ \textit{or} the state $|1\ran$, with $50\%$ probability either way, then we could compute the probability for measuring some outcome $i$ as
\be
P(i)=\frac{1}{2}\left(\lan 0 |\Pi_i |0\ran+\lan 1 |\Pi_i |1\ran\right)=\mathrm{tr} \Pi_i \rho_{MM},
\ee
where $\rho_{MM}$ is the \textit{maximally mixed state}, proportional to the identity operator $I$.  More generally there is always some basis where $\rho$ is diagonal, and in that basis we can interpret the diagonal elements as the classical probabilities for the system being in the various pure state eigenvectors of the density matrix.

It is not always apparent whether a given state $\rho$ is pure or mixed, so it is convenient to define a function $S$ on the space of states with the property that it is zero if and only if its argument is pure.  A function that fits the bill is the \textit{Von Neumann Entropy}
\be
S(\rho)\equiv -\mathrm{tr}\left(\rho \log \rho\right).  
\ee
The Von Neumann entropy has a long list of remarkable properties, among them:
\begin{itemize}
\item $S(U^\dagger \rho U)=S(\rho)$, for $U$ any unitary operator.
\item $S(\rho)\geq 0$, with equality if and only if $\rho$ is pure.
\item $S(\rho)\leq \log d$, where $d$ is the dimensionality of $\mathcal{H}$, with equality if only if $\rho$ is maximally mixed.
\item For any set of non-negative numbers $\lambda_i$ obeying $\sum_i\lambda_i=1$, 
\be
S\left(\sum_i \lambda_i \rho_i\right)\geq \sum_i \lambda_i S\left(\rho_i\right), 
\ee
which is called \textit{concavity}.  This roughly says that the entropy of the average over a set of states is at least equal to the average of their individual entropies and is usually larger, which is a sensible property for an entropy to have.  
\end{itemize}

\subsection{Bipartite systems}
A \textit{bipartite system} is one whose Hilbert space can be written as a tensor product\footnote{Note that the tensor product $\otimes$ is \textit{not} the direct sum $\oplus$.  It is remarkable how often people become confused by not properly distinguishing these...}
\be
\mathcal{H}=\HA\otimes \HB.
\ee
Without loss of generality I will take $|A|\leq |B|$, where $|A|$ is the dimensionality of $\HA$ and similarly $|B|$ for $\HB$.  The Hilbert space is spanned by states of the form $|a\ran\otimes |b\ran\equiv |a\ran|b\ran\equiv |ab\ran$, where $|a\ran$ and $|b\ran$ are complete bases for $\HA$ and $\HB$ respectively.  Bipartite systems are interesting because they describe the \textit{composition} of two independent physical systems. 

\subsubsection{Product operators and the partial trace}
On a bipartite Hilbert space we can define product operators $M_A\otimes N_B$, which have the nice properties that
\begin{align}\nonumber
\left(M_A\otimes N_B\right)\left(M_A'\otimes N_B'\right)&=M_AM_B'\otimes N_B N_B'\\\nonumber
\mathrm{tr}\left(M_A\otimes N_B\right)&=\mathrm{tr}\left(M_A\right) \mathrm{tr}\left(N_B\right)\\
\left(M_A\otimes N_B\right)^\dagger&=M_A^\dagger\otimes N_B^\dagger.
\end{align}
We are often interested in measuring product operators which act nontrivially only on one subfactor.  For example for an observable which acts only on system $A$, the projection operators have the form $\Pi_i\otimes I_B$, where $I_B$ is the identity operator on $\HB$.  Given some quantum state $\rho_{AB}$ of a bipartite system we can then compute the probability of outcome $i$ as
\begin{align}\nonumber
P(i)&=\mathrm{tr}_{AB}\left(\rho_{AB}\left(\Pi_i\otimes I_B\right)\right)\\
&=\tr_A\left(\rho_A \Pi_i\right),
\end{align}
where
\be\label{rhoA}
\lan a'|\rho_A|a\ran\equiv \sum_b \lan a'b|\rho_{AB}|ab\ran.
\ee
Thus given any state $\rho_{AB}$ we can define a reduced state $\rho_A$ (or $\rho_B$), which can be thought of as the quantum state of the subfactor $\HA$ (or $\HB$) for the purposes of measuring any observable that acts nontrivially only on $\HA$.  The operation \eqref{rhoA} is called the \textit{partial trace}.\footnote{The partial trace is absolutely essential to doing quantum mechanics, without it we would have to know the wave function of the entire universe just to get started thinking about the hydrogen atom!}

\subsubsection{Entanglement and the Schmidt decomposition}
Bipartite systems in quantum mechanics have the remarkable feature that they can be \textit{entangled}; the reduced density matrices $\rho_A$ and $\rho_B$ can be mixed even if the joint state $\rho_{AB}$ is pure.  Said differently, ``complete'' knowledge of the state of $AB$ does not necessarily imply ``complete'' knowledge of the states of $A$ and $B$. This situation is illuminated by introducing what is called the \textit{Schmidt decomposition}.  

Consider a pure state $|\psi\ran$ of the bipartite system.  We can always find a basis $|a,\psi\ran$ for $\HA$ in which the reduced state $\rho_A=\tr_B|\psi\ran\lan\psi|$ is diagonal, and in this basis we can expand\footnote{Here I will include a subscript $A$ or $B$ on states to be clear about which tensor factor they are in.}
\be
|\psi\ran=\sum_{ab}C_{ab}|a,\psi\ran_A|b\ran_B,
\ee
where $|b\ran$ is some arbitrary basis for $\HB$.  Note that $\rho_A=C C^\dagger$, so $C C^\dagger$ is diagonal, with diagonal elements $\lan a|C C^\dagger |a\ran\equiv p_a$ in the closed interval $[0,1]$ and obeying $\sum_a p_a=1$.  We can write the state as
\be
|\psi\ran\equiv \sum_{ab}|a,\psi\ran_A|\tilde{a},\psi\ran_B,
\ee
where $|\tilde{a},\psi\ran_B\equiv \sum_b C_{ab}|b\ran_B$.  We then have
\be
_{B}\lan \tilde{a}',\psi|\tilde{a},\psi\ran_B=\, _{A}\lan a'|C C^\dagger|a\ran_A=p_a \delta_{aa'},
\ee
so the states $|\tilde{a},\psi\ran_B$ are orthogonal.  We can make the subset with $p_a\neq 0$ orthonormal by defining $|a,\psi\ran_B\equiv \frac{1}{\sqrt{p_a}}|\tilde{a},\psi\ran_B$, which then form an orthonormal basis for a subspace of $\HB$ whose dimensionality equals the rank of $\rho_A$.  We thus arrive at the statement of the Schmidt decomposition:
\begin{itemize}
\item Given a pure state $|\psi\ran$ in a bipartite Hilbert space $\HA\otimes \HB$ with $|A|\leq |B|$, there exist constants $p_a\in [0,1]$, obeying $\sum_a p_a=1$, an orthonormal basis $|a,\psi\ran_A$ for $\HA$, and an orthonormal basis $|a,\psi\ran_B$ for a subspace of $\HB$ whose dimensionality equals the rank of $\rho_A$, such that we have
\be
|\psi\ran=\sum_a \sqrt{p_a} |a,\psi\ran_A|a,\psi\ran_B.
\ee
\end{itemize}

The Schmidt decomposition has several immediate consequences about bipartite systems when $\rho_{AB}$ is pure:
\begin{itemize}
\item $\rho_A$ and $\rho_B$ have the same nonzero eigenvalues.
\item The entropies $S_A\equiv -\tr \rho_A \log \rho_A$ and $S_B\equiv -\tr \rho_B \log \rho_B$ are equal.  In this context $S_A=S_B$ is often referred to as the \textit{entanglement entropy}.
\item The state $|\psi\ran\lan\psi|$ is entangled if and only if more than one of the $p_a$ are nonzero. If not then the state is a tensor product of two pure states on $\HA$ and $\HB$.  
\end{itemize}

\subsubsection{Purifications from the Schmidt Decomposition}\label{pursec}
If it happens that $|A|$ divides $|B|$, as for example would be the case in any qubit system, then we can give a somewhat more intuitive statement of the Schmidt decomposition.\footnote{This restriction may seem adhoc, but we can always achieve it by adding a few more ``extra'' states to $B$ whose tensor products with any state of $A$ have zero projection on $|\psi\ran$.}  Namely we can reinterpret the subspace of $\HB$ in terms of a tensor factorization
\be
\HB=\mathcal{H}_{B_A} \otimes \mathcal{H}_C, 
\ee 
where $|B_A|=|A|$ and the subspace spanned by the $|a,\psi\ran_B$'s is also the span of the states $|a\ran_{B_A}|0\ran_C$.  Here $|a\ran_{B_A}$ is an orthonormal basis for $B_A$ and $|0\ran_C$ is some particular state of $C$.  The Schmidt decomposition then says that
\be\label{purify}
|\psi\ran=\left(\sum_a \sqrt{p_a}|a,\psi\ran_A |a\ran_{B_A}\right)\otimes |0\ran_C,
\ee
so $B_A$ is ``the part of $B$ entangled with $A$'' and $C$ is its tensor complement.  In this situation people sometimes say that $B_A$ is a \textit{purification} of $A$.\footnote{We will see at the end of this section that is a slight misuse of the word; more correctly the state in parentheses in equation \eqref{purify} is a purification of the state $\rho_A$.  In these notes I will sometimes abuse the term in this manner since there doesn't seem to be an alternative widely accepted name for the tensor factor $B_A$.}  It is important to note that although $B_A$ is useful for making various arguments, it is not unique. Any unitary transformation on $B$ that fixes the subspace spanned by the $|a,\psi\ran_B$'s will leave $|\psi\ran$ invariant, but will typically act nontrivially on $B_A$, mapping it to a different subfactor $B_A'$ that still can be used to write the state $|\psi\ran$ in the form \eqref{purify}.

For our discussion of the firewall paradox it will be useful to further discuss the special case where $A$ itself factorizes
\be
\HA=\mathcal{H}_{A_1}\otimes \mathcal{H}_{A_2}.
\ee
For general states there isn't too much to say about this, but in the special case where $\rho_A=\rho_{A_1}\otimes \rho_{A_2}$ we have
\be
p_{a_1a_2}=p_{a_1}p_{a_2},
\ee
so again assuming that $|A|$ divides $|B|$ we can again introduce a tensor factor $B_A$, which we can now factorize into $B_{A_1}\otimes B_{A_2}$, and write the state as
\be\label{prodschmidt}
|\psi\ran=\left(\sum_{a_1}p_{a_1}|a_1,\psi\ran_{A_1}|a_1\ran_{B_{A_1}}\right)\otimes\left(\sum_{a_2}p_{a_2}|a_2,\psi\ran_{A_2}|a_2\ran_{B_{A_2}}\right)\otimes |0\ran_C.
\ee
Thus in this case $A_1$ and $A_2$ each get their own purification in $B$.  There are two particularly interesting cases where $\rho_A$ factorizes; the first is maximal entanglement, where $\rho_A=\frac{I_A}{|A|}$, which obviously factorizes for any tensor decomposition of $A$.  We saw from Page's theorem in section \ref{pagesec} that maximal entanglement is typical for generic states in $\HA\otimes \HB$, provided that $\frac{|A|}{|B|}\ll 1$.  The second interesting situation is \textit{thermal} entanglement, where $\rho_A=\frac{1}{Z_A}e^{-\beta H_A}$.  As I briefly remarked in section \ref{pagesec}, this arises generically in states in a subspace of fixed total energy within $\HA\otimes \HB$.  If $A_1$ and $A_2$ are sufficiently weakly interacting then $\rho_A$ again factorizes and we can write the total state as \eqref{prodschmidt}.

The idea of purification is useful more generally, but as mentioned above the standard usage of the term is slightly different than how it has come to be used in the black hole information literature; I here explain the difference.  The Schmidt decomposition makes it manifest that any quantum state $\rho_A$ of some system $A$ can always be obtained by partial trace of a pure state $|\psi\ran$ in a larger Hilbert space obtained by composition of $A$ with an auxiliary system $B$, provided that $|A|\leq |B|$.  In this situation the joint state $\rho_{AB}=|\psi\ran\lan \psi|$ is what is really called a purification of $\rho_A$.  The basic relationship between the two notions is that $B_A$ is a ``smallest piece'' of $B$ such that $\rho_{AB_A}$ is still a purification of $\rho_A$ in the proper sense.  It is clear that there is no unique purification of a state $\rho_A$; we can act on $B$ with any unitary transformation without changing $\rho_A$.\footnote{This non-uniqueness is different from the non-uniqueness in the choice of the tensor factor $B_A$. That non-uniqueness was present for a given state $|\psi\ran$ that purifies $\rho_A$, while here the non-uniqueness in general involves changing $|\psi\ran$ in a way that preserves $\rho_A$.}  The Schmidt decomposition shows however that this is all that we can do; any two purifications of $\rho_A$  are always related by a unitary on $B$.  

\subsection{Entropy inequalities in multipartite systems}
In multipartite systems the Von Neumann entropy obeys some interesting inequalities.  For bipartite systems the \textit{mutual information}
\be
I_{AB}\equiv S_A+S_B-S_{AB}
\ee
obeys
\be\label{Ipos}
I_{AB}\geq 0,
\ee
with equality if and only if $\rho_{AB}=\rho_A\otimes \rho_B$.  This object is a quantification of ``how much information  $A$ has about $B$'', or equivalently ``how much information $B$ has about $A$'', since it is symmetric between $A$ and $B$.  When $\rho_{AB}$ is pure any nonzero mutual information comes entirely from entanglement, but when $\rho_{AB}$ is mixed it can also come from classical correlation.  For example the two-qubit state
\be
\frac{1}{2}\left(|00\ran\lan 00 |+|11\ran \lan 11|\right),
\ee
which represents a state which is either $|00\ran$ or $|11\ran$ with $50-50$ probability, has mutual information $\log 2$ even though there is no ``quantum funny business'' going on.  
Bipartite systems also obey an interesting ``triangle inequality''
\be\label{triangle}
|S_A-S_B|\leq S_{AB}.
\ee
Both of these inequalities follow quite easily (homework!) from the properties of another interesting quantity, the \textit{relative entropy} of two states $\rho$ and $\sigma$ on a single Hilbert space, which is defined as
\be
S(\rho|\sigma)\equiv \tr \rho \log \rho-\tr \rho \log \sigma.
\ee
It is not hard to show that the relative entropy is non-negative, and is zero if and only if $\rho=\sigma$ (more homework!).  It has an interpretation in terms of hypothesis testing \cite{Vedral:2002zz}, but its main use so far has been in proving other statements that are more immediately interesting.

For tripartite systems we have the \textit{strong subadditivity inequality}
\be
S_{ABC}+S_B\leq S_{AB}+S_{BC},
\ee
which can be rewritten as
\be
I_{A,B}\leq I_{A,BC}.
\ee
Strong subadditivity thus rather reasonably says that $B$ and $C$ together have more mutual information with $A$ than just $B$ by itself does.  Despite the ``obviousness'' of this statement, the known proofs of strong subadditivity are all rather tedious.\cite{nielsen2010quantum}.  

\section{Unitary integrals}\label{Uintsec}
In studying random states we often want to integrate polynomials of unitary matrices over the group invariant Haar measure on $U(N)$.  The first few results are:
\begin{align}\nonumber
\int dU=&1\\\nonumber
\int dU \,U_{ij} U^\dagger_{kl}=&\frac{1}{N}\delta_{il}\delta_{jk}\\\nonumber
\int dU\, U_{ij}U_{kl}U^\dagger_{mn}U^\dagger_{op}=&\frac{1}{N^2-1}\left(\delta_{in}\delta_{kp}\delta_{jm}\delta_{lo}+\delta_{ip}\delta_{kn}\delta_{jo}\delta_{lm}\right)\\\nonumber
&-\frac{1}{N(N^2-1)}\left(\delta_{in}\delta_{kp}\delta_{jo}\delta_{lm}+\delta_{ip}\delta_{kn}\delta_{jm}\delta_{lo}\right)\\
&\ldots\label{Uints}
\end{align}
The integral vanishes for any polynomial where the number of $U$'s does not equal the number of $U^\dagger$'s.

These results are derived by using the group invariance of the measure to express the integral as a sum of invariant tensors with arbitrary coefficients, and then judiciously contracting indices using $U^\dagger U=1$ to determine the coefficients.  

\section{Quantum computation and complexity theory}\label{compsec}
In this section I review some of the barest essentials of quantum computation; for more details see  \cite{nielsen2010quantum,preskillnotes}.  A \textit{quantum computer} is a machine that enables us to act with any unitary transformation we wish on some finite dimensional quantum system, which is usually referred to as a \textit{quantum memory}. The quantum memory is usually taken to be a string of $n$ qubits, although most of the theoretical results are robust enough to apply to other models of quantum computation with little or no modification.  Quantum complexity theory is the study of how ``hard'' it is to apply whichever unitary transformation we happen to be interested in.  The preferred factorization of our $2^n$-dimensional Hilbert space into $n$ qubits is essential for the notion of ``hard'' to have nontrivial content; what one typically assumes is that it is ``easy'' to apply unitary transformations that act nontrivially only on an $O(1)$ number of the qubits.  This assumption is based on locality; it captures the intuition that however we manipulate our quantum computer, it will most likely involve some sort of device that moves around from qubit to qubit, acting on them a few at a time.  The hardness will then roughly be defined as the number of these easy transformations which are needed to produce the unitary we are interested in.

\begin{figure}
\begin{center}
\includegraphics[height=5cm]{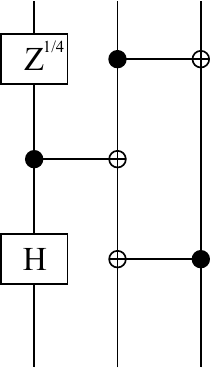}
\caption{An example of a quantum circuit on three qubits.  The vertical lines are the qubits, which are acted on by various quantum gates as we go upwards in time.  The horizontal lines are CNOT transformations, where the addition happens on the qubit with the plus.  This circuit has size five and depth three.}\label{circuit1}
\end{center}
\end{figure}

This notion of hardness is formalized in the \textit{quantum circuit model} of quantum computation.  In the circuit model one picks a small finite number of one- and two-qubit unitaries, which are referred to as \textit{gates}.  These gates can then be sequentially applied to any pair of qubits in our preferred factorization; in this manner by acting with enough gates we can generate arbitrary unitary transformations to within some tolerance $\epsilon$.\footnote{The tolerance $\epsilon$ is defined for example with respect to the trace norm distance $||U-V||_1$ on the unitary group, analogous to the definition for states above equation \eqref{norms}.}  A set of gates which is able to generate arbitrary unitaries in this way is called \textit{universal}, and it turns out that almost any set of gates will be universal.  A simple set of three gates which is universal is the Hadamard transformation
\begin{align}\nonumber
H|0\ran&=\frac{1}{\sqrt{2}}\left(|0\ran+|1\ran\right)\\
H|1\ran&=\frac{1}{\sqrt{2}}\left(|0\ran-|1\ran\right),
\end{align}
the $Z^{1/4}$ gate
\begin{align}\nonumber
Z^{1/4}|0\ran&=|0\ran\\
Z^{1/4}|1\ran&=e^{i\pi/4}|1\ran,
\end{align}
and the CNOT gate
\be
CNOT|i,j\ran=|i,i+j\ran.
\ee
As usual the addition in the CNOT gate is mod $2$.  An ordered list of these gates, along with which of the $n$ qubits they should be applied to, is called a \textit{quantum circuit}.  It is a ``digital'' representation of a unitary transformation.  The number of gates is called the \textit{size} of the circuit, and is a reasonable quantification of the ``hardness'' of the circuit.  For example if our quantum computer runs in serial then the running time will be directly proportional to the size of the circuit.  If our computer is able to run in parallel, meaning that it can simultaneously apply any number of gates which act on mutually disjoint qubits, then the running time will instead be proportional to what is called the \textit{depth} of the circuit.  This is the number such simultaneous actions that are needed to implement the circuit; the distinction is illustrated in figure \ref{circuit1}, which also shows the basic idea of a circuit diagram.
The \textit{circuit complexity} $\mathcal{C}$ of a unitary transformation is then defined as the size (or sometimes depth) of the smallest circuit that implements it to within accuracy $\epsilon$.  

Of course for a particular unitary the circuit complexity will just be some number, but what we are usually interested in is actually \textit{families} of unitaries that are defined for arbitrary values of $n$; we want the set of unitaries $U_n$ which solve some problem of interest for inputs of size $n$  ``using the same algorithm''.  It then makes sense to ask about the asymptotic scaling of the circuit complexity as we take $n$ to be large and $\epsilon$ to be small.  The exact value of the circuit complexity for some particular unitary will depend on things like our detailed choice of gates, but the asymptotic scaling with $n$ and $\epsilon$ usually will not.  Typically a family of unitaries $U_n$ is called \textit{efficient} if the circuit complexity is asymptotically bounded by some low-order power of $n$ and $1/\epsilon$.  The dependence on $\epsilon$ is usually logarithmic, so the primary source of hardness is the scaling with $n$.  Indeed a simple counting argument, reviewed for example in \cite{Harlow:2013tf}, says that almost all (in the sense of the Haar measure on $U(2^n)$) unitary transformations have circuit complexity at least
\be
\mathcal{C}\gtrsim 2^{2n} \log \frac{1}{\epsilon}.
\ee
An upper bound is a bit harder to obtain, and involves what is called the Solovay-Kitaev theorem, but one eventually finds that \cite{nielsen2010quantum}
\be
\mathcal{C}\lesssim 2^{2n} \left(\log\frac{1}{\epsilon}\right)^c,
\ee
where $c$ is some unknown number that must be greater than $1$ and is less than $2$.  

Thus the vast majority of unitary transformations require circuits of exponential size in $n$: contrary to what you may have read in the mainstream media, quantum computations are hard!  The main goal of quantum complexity theory is to find those happy cases where unitary transformations that we care about can be implemented efficiently, or failing that to explain to us why they \textit{can't} be implemented efficiently.  The former is easy to do in principle, one just needs to give us the circuit.  The latter is usually quite difficult, for example in classical complexity theory you may have heard about the continued lack of progress in proving the ``obviously true'' statement that P$\neq$ NP, so usually complexity theorists settle for arguments along the lines of the following: ``If you were able to efficiently solve the problem that you care about, then here is a way that you could then solve all sorts of other problems that nobody has been able to solve.  This means that you probably won't be able to efficiently solve your problem either.''  Physicists usually don't find this type of argument very convincing the first time they hear it, but in fact this kind of reasoning has turned out to be very powerful in computer science in the last few decades.

\section{Homework problems}
\begin{itemize}
\item Problem 1: Confirm the algebra leading to equation \eqref{penrosemet}, including the coordinate ranges for $T$ and $R$, and convince yourself that this leads to figure \ref{flatpenrose}.
\item Problem 2: Derive equation \eqref{freevac} by expressing the lowering operators $a_{\vec{k}}$ in terms of $\phi$ and $\pi=-i\frac{\delta}{\delta\phi}$, and then demanding that they annihilate the vacuum wave functional $\lan \phi|\Omega\ran$. (Hint: you will need to use that $\pi=\dot{\phi}$, and you will also need to know that $\int d^3x e^{i \vec{k}\cdot \vec{x}}=(2\pi)^3 \delta^3(k)$.)
\item Problem 3: Rederive equation \eqref{freevac} by evaluating the Euclidean path integral \eqref{epath}.  (Hint: look for a solution of the Euclidean equation of motion $\frac{d^2}{dt_E^2}\phi+\nabla^2\phi-m^2\phi=0$ with the right boundary conditions for the path integral; this is easiest in momentum space.  Then evaluate the Euclidean action on this solution, and convince yourself that this is all you need to evaluate the path integral.)
\item Problem 4: Compute the Unruh temperature
\be
T_{Unruh}=\frac{\hbar a}{2\pi k_B c}
\ee
 seen by a detector accelerating at $a=1\,g =9.8 m/s^2$.  Also compute the Unruh temperature seen by electrons as they orbit the proton in a hydrogen atom, and compare $k_B T$ to the ground state energy $13.6eV$.  Finally compute the Hawking temperature 
\be
T_{Hawking}=\frac{\hbar c^3}{8\pi k_B G M}
\ee
for a solar mass black hole ($M_\odot=2 \times 10^{30} kg$), as well as the supermassive black hole in the center of the Milky way galaxy ($M=10^6 M_\odot$).
\item Problem 5: Derive equation \eqref{norms}.
\item Problem 6: Derive the quadratic unitary integral in equation \eqref{Uints}.  If the experience leaves you feeling enthusiastic, derive the quartic one.  Then use the quartic result to fill in the derivation of Page's theorem \eqref{Pagethm}.
\item Problem 7: Derive the positivity of relative entropy $S(\rho|\sigma)=\mathrm{tr} \rho \log \rho-\mathrm{tr} \rho \log \sigma\geq 0$, and show that equality holds if and only if $\rho=\sigma$.  Use this result to derive \eqref{Ipos} and \eqref{triangle}, as well as the fact that $I_{AB}=0$ only if $\rho_{AB}=\rho_A\otimes\rho_B$.
\item Problem 8: Derive equation \eqref{cft2pt} from the conformal invariance of the vacuum and the symmetry transformation $\mathcal{O}$.  Also show that \eqref{sphere2pt} follows from \eqref{cft2pt} and \eqref{conftrans}.  You may need to look up more details about how primary operators transform, one good place is \cite{Aharony:1999ti}.
\item Problem 9: Derive equation \eqref{adsT}.
\item Problem 10: Come up with explicit polynomial-sized circuits to prepare the states \eqref{psiscott} and \eqref{psiscott2}, using the Hadamard and CNOT gates of appendix \ref{compsec}.
\item Problem 11: Derive equation \eqref{finalstateresult}.  If the indices are too annoying to keep track of you may find the diagrams explained in \cite{Bousso:2013uka} to be helpful.
\end{itemize}
\bibliographystyle{jhep}
\bibliography{bibliography}
\end{document}